\documentclass[a4paper,fleqn]{cas-dc}
\usepackage[numbers]{natbib}
\usepackage[thinlines]{easymat} % pour matrices partitionnees
\usepackage{amsmath,amsfonts,amssymb} % amsthm,amsxtra
\usepackage{indentfirst}
\usepackage{graphicx}
\usepackage{subeqnarray}
\usepackage{mdwlist}
\usepackage[percent]{overpic}
\usepackage{subfigure}
\usepackage{epstopdf}
\usepackage{rotating}
\usepackage{xcolor}
\usepackage{siunitx}
\usepackage{multirow}
\definecolor{amethyst}{rgb}{0.6, 0.4, 0.8}
\usepackage{geometry}
\usepackage{fleqn.clo}
\usepackage{hyperref}
\usepackage{cleveref}

\usepackage{caption}
\usepackage{pifont}
\definecolor{green2}{RGB}{0,128,0}
\usepackage[utf8]{inputenc}
\usepackage{booktabs,lipsum}
\usepackage[ruled, norelsize, linesnumbered]{algorithm2e}
\usepackage{algpseudocode}
\usepackage{multirow}

\renewcommand{\intextsep}{0pt} % espace vertical entre flottants et
\allowdisplaybreaks % Autorise la coupure de fin de page

\def\aujour{\number\day \space \ifcase\month\or
janvier\or f�vrier\or mars\or avril\or mai\or
juin\or juillet\or ao�t\or septembre\or octobre\or
novembre\or d�cembre\fi \space \number\year}

\def\cH{{\cal H}}

\def\cL{{\cal L}}

\newtheorem{remark}{Remark}

\def\C{{\setbox0=\hbox{$\displaystyle{\rm C}$}
        \hbox{\hbox to0pt{\kern 0.4\wd0\vrule height 0.95\ht0\hss}\box0}}}
\def\Q{{\setbox0=\hbox{$\displaystyle{\rm Q}$}%
    \hbox{\raise 0.2\ht0\hbox to0pt{\kern 0.4\wd0\vrule height
    0.85\ht0\hss}\box0}}} % ensemble des Rationnels --> � revoir
\def\R{\mathop{\rm I\mkern -3.5mu R}} % ensemble des R�els --> OK
\def\cH2{{\cal H}_2} % H2 calligraphi� (norme)

\def\cL2{\mathop{\mathcal L}_{2}} % espace carr�-int�grable
\def\cRH2{\mathop{\cal R \cal H}_2} % RH2 calligraphi� (norme)
\def\cRL2{\mathop{\cal R \cal L}_{2}} % espace carr�-int�grable

\DeclareMathOperator*{\der}{d}

\makeatletter
\DeclareRobustCommand\sfrac[1]{\@ifnextchar/{\@sfrac{#1}}
                                            {\@sfrac{#1}/}}
\def\@sfrac#1/#2{\leavevmode\kern.1em\raise.5ex
         \hbox{$\m@th\fontsize\sf@size\z@
                           \selectfont#1$}\kern-.1em
         /\kern-.15em\lower.25ex
          \hbox{$\m@th\fontsize\sf@size\z@
                            \selectfont#2$}}
\makeatother
\def\tsc#1{\csdef{#1}{\textsc{\lowercase{#1}}\xspace}}
\tsc{WGM}
\tsc{QE}
\tsc{EP}
\tsc{PMS}
\tsc{BEC}
\tsc{DE}
\begin{document}
\let\WriteBookmarks\relax
\def\floatpagepagefraction{1}
\def\textpagefraction{.001}
\shorttitle{Model-based versus model-free feeding control and water quality monitoring for fish growth tracking}
\shortauthors{F. Aljehani, I. N'Doye and T.-M. Laleg-Kirati}

% \title [mode = title]{Feeding control and water quality monitoring for fish growth tracking \textcolor{blue}{under different levels of unionized
% ammonia exposure} in aquaculture systems} 
\title [mode = title]{Model-based versus model-free feeding control and water quality monitoring for fish growth tracking in aquaculture systems}

%\title [mode = title]{Learning-based Optimal Control for Monitoring Fish Feeding and Water Quality in Aquaculture Systems} 

\tnotemark[1]

\tnotetext[1]{This work has been supported by the King Abdullah University of Science and Technology (KAUST), Base Research Fund (BAS/1/1627-01-01) to Taous Meriem Laleg.}

%\tnotetext[2]{The second title footnote which is a longer text matter
%   to fill through the whole text width and overflow into
%   another line in the footnotes area of the first page.}

\author[first]{Fahad Aljehani} 
\ead{fahad.aljehani@kaust.edu.sa}
\author[first]{Ibrahima N'Doye} 
\ead{ibrahima.ndoye@kaust.edu.sa}
\author[first,second]{Taous-Meriem Laleg-Kirati}
\ead{taousmeriem.laleg@kaust.edu.sa}
\address[first]{Computer, Electrical and Mathematical Sciences and Engineering Division (CEMSE), King Abdullah University of Science and Technology (KAUST), Thuwal 23955-6900, Saudi Arabia.}
\address[second]{National Institute  for Research in Digital Science and Technology (INRIA), Saclay, France.}                       
 
% \author[]{Fahad Aljehani}[]
                        
% %\cormark[1]
% %\fnmark[1]
% \ead{fahad.aljehani@kaust.com.sa}

% \author[]{Ibrahima N'Doye}[]

% \ead{ibrahima.ndoye@kaust.edu.sa}

% \author[]{Taous~Meriem Laleg-Kirati}[]

% \ead{taousmeriem.laleg@kaust.edu.sa}

% \address[]{Computer, Electrical and Mathematical Sciences and Engineering Division (CEMSE), King Abdullah University of Science and Technology (KAUST), Thuwal 23955-6900, Saudi Arabia.}

\cortext[cor1]{Corresponding author: T. M. Laleg-Kirati}

\begin{abstract}
\noindent Aquaculture systems can benefit from the recent development of advanced control strategies to reduce operating costs and fish losses and increase growth production efficiency, resulting in fish welfare and health. Monitoring the water quality and controlling the feeding are fundamental elements to balancing fish productivity and shaping fish's life history in the fish growth process. Currently, most fish feeding processes are conducted manually in different phases and rely on time-consuming and challenging artificial discrimination. The ability of the feeding control approach influences the fish growth and breeding through the feed conversion rate; hence, controlling these feeding parameters is crucial for enhancing fish welfare and minimizing the general fishery costs. On the other hand, the high concentration level of the environmental factors, such as a high ammonia concentration and pH level, affect the water quality, affecting fish's survival and mass death. Therefore, there is a critical need to develop control strategies to determine optimal, efficient, and reliable feeding and water quality monitoring processes. In this paper, we revisit the representative fish growth model describing the total biomass change by incorporating the fish population density and mortality. Since the measurement data of the total biomass and population from the aquaculture systems are limited and difficult to obtain, we validate the new dynamic population model with the individual fish growth data for tracking control purposes. We specifically focus on relative feeding as a manipulated variable to design traditional and optimal control to track the desired weight reference within the sub-optimal temperature and dissolved oxygen profiles under different levels of unionized ammonia exposure. Then, \textcolor{black}{we propose a Q-learning approach that learns an optimal feeding control policy from the simulated data of the fish growth weight trajectories while managing the ammonia effects. The proposed Q-learning feeding control prevents fish mortality and achieves good tracking errors of the fish weight under the different levels of unionized ammonia. However, it maintains a relative food consumption that potentially underfeeds the fish. Finally,} we propose an optimal algorithm that optimizes the feeding and water quality of the dynamic fish population growth process. We also show that the model predictive control decreases fish mortality and reduces food consumption in all different cases \textcolor{black}{of unionized ammonia exposure}.% by an average of $26.9\%$ compared to the bang-bang controller, $22.6\%$ compared to the PID controller. %\textcolor{blue}{Finally, we propose a Q-learning approach that learns an optimal feeding control policy from the simulated data of the fish growth weight trajectories while managing the ammonia effects. The proposed Q-learning feeding control prevents fish mortality. Compared to the model-based approaches, it achieves better tracking errors of the fish weight under the different levels of unionized ammonia exposure and penalizes the food consumption.}
\end{abstract}

\begin{keywords}
\sep Feeding control
\sep Water quality monitoring
\sep Fish growth tracking %\sep Control theory 
%\sep Aquaculture control systems 
 \sep Bioenergetic growth model 
 %\sep Model-based control \sep  
%\sep Reinforcement learning 
\sep Model predictive control
\sep \textcolor{black}{Reinforcement learning}
\sep \textcolor{black}{Q-learning}
%\sep Reinforcement learning based model predictive control
\end{keywords}

\maketitle
\section{Introduction}
\begin{figure}[!t]
 \vspace{0.5cm}
\centering
      \begin{overpic}[scale=0.17]{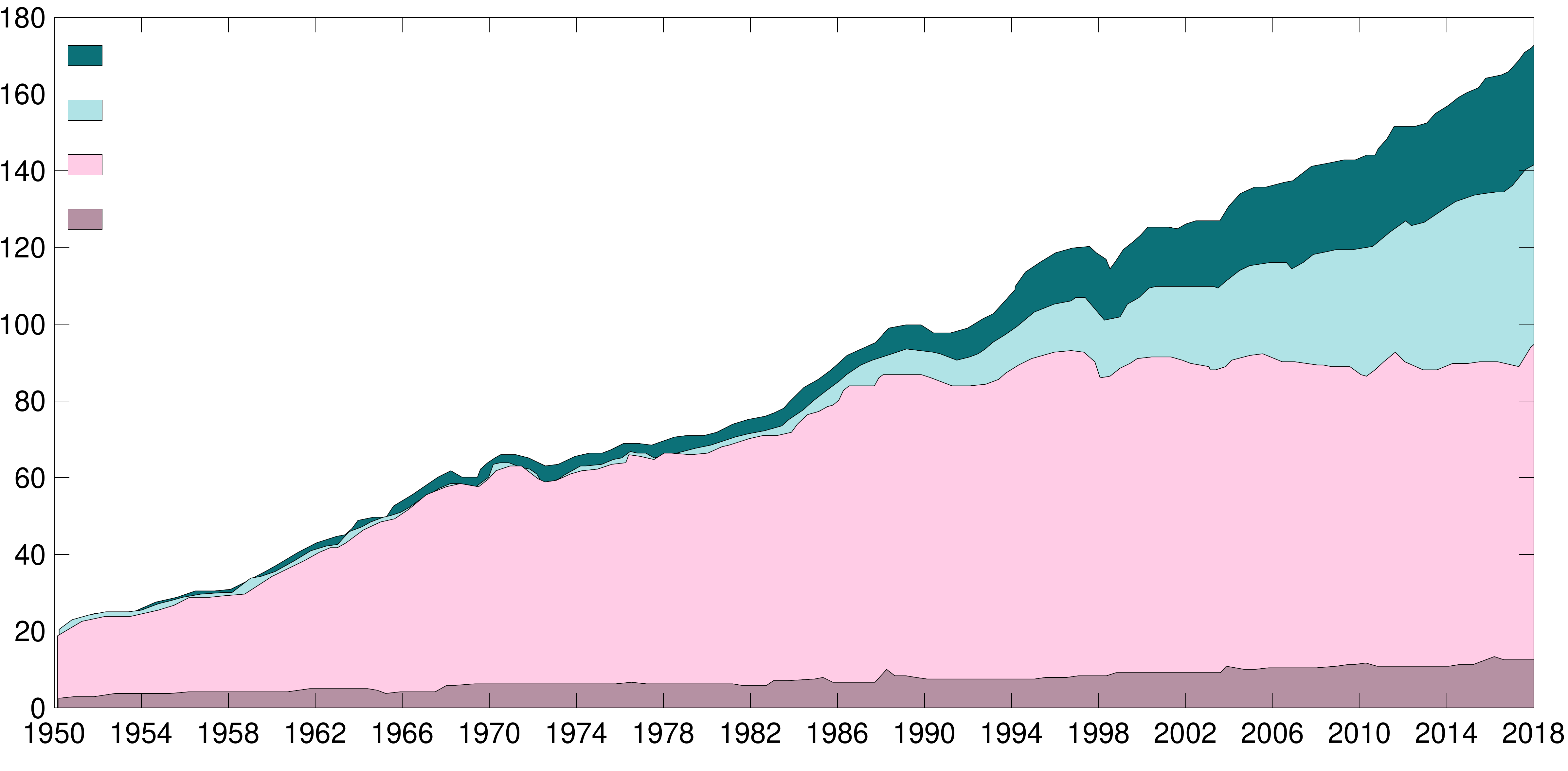}
            \put(9,45){\scriptsize Aquaculture - marine water}
    \put(9,41.5){\scriptsize Aquaculture - inland water}
     \put(9,38){\scriptsize Capture fisheries - marine water}
     \put(9,34.5){\scriptsize Capture fisheries - inland water}
     \put(-3,18){\scriptsize  \rotatebox{90}{Million tonnes}}
     \put(43,-2){\footnotesize {Time (year)}} 
      \end{overpic}
      \caption{Marine and aquaculture worldwide production}\label{roadmap} 
      \vspace{-0.25cm}
 \end{figure}
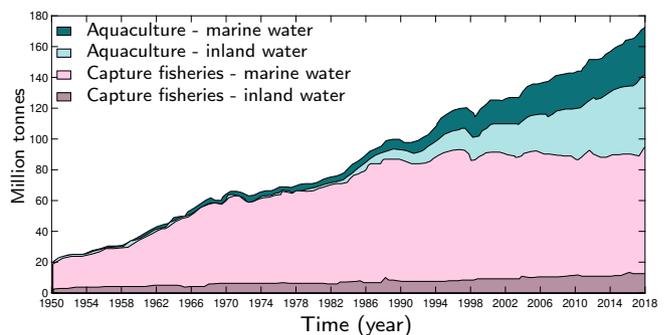

\noindent Aquaculture is one of the largest and fastest-growing food production sectors in the world and is likely to become the primary source of seafood in the future \cite{Fao:18}, as illustrated in Fig.\ref{roadmap}. As commercial fish production continues to increase, its impact and reliance on protein sources provided by ocean fisheries are likely to expand. To mitigate these impacts, adequate growth models are relevant for efficient aquaculture management, as they provide an optimized protocol for feeding and monitoring fish welfare throughout the grow-out cycle from stocking through harvesting \cite{Seg:16}. 

\noindent Growth is a biological process comprising many crucial processes and fish's life history \cite{FlM:21}. For aquaculture systems to achieve better control, tracking and predicting fish growth trajectory is vital. However, the tracking of growth trajectory poses challenges where environmental factors influence fish feed, such as unionized ammonia dissolved oxygen, salinity, water temperature, and light \cite{SHL:16}.

\noindent The feeding process and water quality monitoring are usually done manually, on-site, or analyzed in laboratories after the data is collected \cite{SHOLS:19}. This monitoring process delays the detection of abnormalities and the relevant control actions and is arduous to manage the costs and complexity of the cleaning, and the stabilization of the water quality \cite{Fo:18}. Depending on water quality measurements, an electric on-off actuator was implemented, and the objective was to design a bang-bang controller that switches the valves and pumps on-off \cite{LEE1995205}. However, this type of controller 
requires a set of desired points so the controller can meet them. As a result, undesirable responses like chattering phenomena are induced. Proportional-integral-derivative (PID) controllers have been examined experimentally in aquaculture for water quality and feeding. For automatic feeding, the feeding rate and time have been implemented using PID controller to enhance the fish growth \cite{Zain2013}. Whereas in \cite{Pedro2020}, the nitrate concentration in water was controlled by PID controller to track the desired reference. The design of closed-loop systems guarantees the best performance against requirements. However, PID may not be efficient for various fish growth tracking systems due to different management factors that account for the overfeeding specifications. Therefore, optimal control law that ensures a trade-off between optimal growth rate performance guarantees, minimization of the costs, and system complexity is relevant to increasing the overall performance.

\noindent Optimal control approaches using optimization techniques were developed to enhance aquaculture process economics. Many of these optimal control techniques enhance the plants aquaculture management and economics, considering the best harvesting time and the market variation  \cite{optiz1}, \cite{Optz2}. %However, these works do not examine all the growth model's biological variables and focus on designing a generic framework. 
Another line of work is based on the fish price and mortality effects to provide the best schedule of feeding \cite{optiz3}, \cite{Hea:95}. However, these works focus on designing a generic framework and do not examine all the growth model's biological variables. \textcolor{black}{To this end, they might lack to maintain fish growth tracking performance in the presence of model uncertainty, such as ammonia effects.} %Therefore, there is an essential need to optimize feeding schedules and enhance fish farming productivity \cite{NGPKB:00}.
\textcolor{black}{The learning-based method for fish growth, such as reinforcement learning (RL), can alleviate the ammonia exposure effects. The Q-learning algorithm for fish trajectory tracking is a model-free RL algorithm, which has been developed in \cite{CNMBL:22}. Specifically, the Q-learning algorithm controller aims to track the fish growth trajectory while managing model uncertainties and environmental factors.}

This paper proposes a fish population growth model that relies on the fish stocking density and mortality rate. The representative bioenergetic fish population growth model accounts for the biodiversity by expressing the total biomass change through the fish population density and mortality. This proposed model is a step beyond the results in \cite{CNMBL:21} limited to a single bio-energetic fish growth model and does not consider the effect of UIA on mortality. Since the measurement data of the total biomass and population from the aquaculture systems is limited and difficult to obtain, we validate the new dynamic population model with the individual fish growth data \cite{Yan:98}. We particularly zoom on the relative feeding as a manipulated variable to design traditional and optimal control to track the desired weight reference within the controlled temperature and dissolved oxygen (DO) profiles through different unionized ammonia (UIA) exposures. To reduce the mortality related to UIA exposure under varying levels of fish stocking density at which the fish growth rate tracks the desired growth reference, we propose an optimal \textcolor{black}{model-based} feeding and water quality controller that includes temperature, DO, and UIA and \textcolor{black}{a Q-learning approach developed in \cite{CNMBL:22} that only learns an optimal feeding control policy while managing the ammonia exposure.}

% \noindent The paper's outline is organized as follows: Section \ref{sub-bioenergetic} describes the individual and the new population dynamic fish growth models.  Section \ref{trad-adv} provides the main results of the traditional and optimal feeding control problem and water quality monitoring. Section \ref{Performance_assessments_section} highlights the performance assessments of the proposed controllers for controlling fish feeding along with numerical simulation tests. Thanks to the flexibility of the model-predictive framework, we provide the optimal feeding and water quality monitoring in which the temperature, dissolved oxygen, and unionized ammonia are controlled in Section \ref{MPC2}. \textcolor{blue}{Section \ref{RL} extends the Q-learning approach developed in \cite{CNMBL:22} to minimize growth trajectory tracking error while dealing with the different levels of ammonia exposure.} Finally, the paper summarizes the main contributions and some future works in Section \ref{conclusion}. 
\noindent The paper's outline is organized as follows: Section \ref{sub-bioenergetic} describes the individual and the new population dynamic fish growth models.  Section \ref{trad-adv} provides the main results of the traditional and optimal feeding control problem and water quality monitoring. Section \ref{Performance_assessments_section} highlights the performance assessments of the proposed controllers for controlling fish feeding along with numerical simulation tests. Thanks to the flexibility of the model-predictive framework, we provide the optimal feeding and water quality monitoring in which the temperature, dissolved oxygen, and unionized ammonia are controlled in Section \ref{MPC2}. Finally, the paper summarizes the main contributions and some future works in Section \ref{conclusion}. 

\section{Bioenergetic fish growth models}\label{sub-bioenergetic}
\noindent This section describes fish growth's individual and population dynamics. First, we validate the new dynamic population model with the individual fish growth data for control and monitoring purposes. Then, we highlight the correlation between ammonia exposure and mortality which underly fish growth performance.

\noindent The dynamic energy budget (DEB) model has recently provided the most complete and well-connected environmental variables effects for growth predictions by covering the entire life cycle of the fish organism \cite{Koo:12, LiS:08, MHGEJ:20}. 
The DEB model describes the metabolic processes of organisms in terms of energy. The description incorporates tools in the early organisms' life cycle which enables; to maintain beneficial knowledge before establishing new farms \cite{VLGTH:20,FGCG:14}, approximate the fish production and amount of food \cite{ChB:98}, and enhance the production by farming aquatic species in integrated fashion \cite{RSPFG:12}. In line with this, the (DEB) growth model has become central in the modeling and analysis of many fish species. It can be formulated as in the metabolic process, which is the difference between anabolism and catabolism, according to Ursin's work \cite{Urs:67}. 
\begin{itemize}
\item \noindent{\textbf{Individual fish growth models:}} The dynamics of the individual bioenergetic growth model is described in terms of fish biomass and fish population density as follows \cite{CNMBL:21,Yan:98}
\begin{equation}\label{sys1aa}
\!\!\!\!\!\!\!\!\frac{\der w}{\der t}= \underbrace{\Psi\big(f, T, DO\big)v(UIA)}_{\mbox{\scriptsize anabolism}} w^m - \underbrace{k(T)}_{\mbox{\scriptsize catabolism}} w^n
\end{equation}
where {$w$ is the individual fish weight, $\Psi\big(f, T, DO\big)$ $(\si{kcal^{1\!-\!m}}\si{day^{-1}})$ and $ v(UIA)$ are the coefficients of anabolism and  $k(T)$ (\si{kcal^{1-n}}\si{day^{-1}}) is the fasting catabolism coefficient formulated as
\begin{equation*}\label{eq1a}
\!\!\Psi\big(f, T, DO\big)= h\rho f b(1-a)\tau(T)\sigma(DO),
\end{equation*}
and 
\begin{equation*}
k(T)=k_{\mbox{\tiny min}}\exp\Big({j(T-T_{\mbox{\tiny min}})}\Big).
\end{equation*}
The model constitutes the effects of water quality parameters as temperature ($T$), dissolved oxygen ($DO$), and un-ionized ammonia ($UIA$) and feeding parameters as food availability \cite{Yan:98}. Nomenclature and the fish growth model parameters are summarized in Table \ref{para} \cite{Yan:98}. The effects of temperature $\tau(T)$, unionized ammonia $v(UIA)$, and dissolved oxygen $\sigma(DO)$ on food consumption and their formulations are found in Appendix \cite{Yan:98,CNMBL:21,Sink2010}.
%%%%%%%%%%%%%%%%%%%%%%%%%%%%%%%%%%%%%%%%%%%%%%%%%%%
\begin{table*}[!t]
\begin{center}
\line(1,0){480}
\end{center}
{\normalsize
\begin{equation}\label{sys1}
\left\{\begin{array}{llllllll}
   \displaystyle \frac{\der \xi}{\der t}=\underbrace{p_s\xi_i}_{\mbox{\scriptsize fish stocking density}}+ \underbrace{\Psi\big(f, T, DO\big)v(UIA)}_{\mbox{\scriptsize anabolism}}\xi^m - \underbrace{k(T)}_{\mbox{\scriptsize catabolism}} \xi^n-\underbrace{pk_1(UIA)\xi_a}_{\mbox{\scriptsize fish mortality}}\\
\displaystyle \frac{\der p}{\der t}=p_s-\mbox{INT}(pk_1)
\end{array}\right.
\end{equation}
}
\begin{center}
\line(1,0){480}
\end{center}
\vspace{0.5cm}
\end{table*}
%%%%%%%%%%%%%%%%%%%%%%%%%%%%%%%%%%%%%%%%%%%%%%%%%%%%
\item \noindent{\textbf{Population dynamic fish growth models:}} Population models are crucial to describe potential pathways for fish population dynamic recovery and the patterns in biodiversity to understand and predict the future of marine resources. The fish growth population models based on the DEB integrate the management variable, such as feeding, and environmental variables, like the temperature, dissolved oxygen, and ammonia, as effects for fish growth predictions and performance by covering the life cycle of the fish organism. Besides, these models capture essential information on environmental and biological changes in fishes, including mortality and fish stocking density   \cite{Hiddink2008,Jenkins2020}.
%Population dynamic is a significant factor in increasing fish growth because it captures essential information on environmental and biological changes in fishes, including mortality and biodiversity \cite{Hiddink2008,Jenkins2020}. 
The dynamic of the population dynamic bioenergetic growth models is described in terms of fish biomass and population density as follows by \eqref{sys1}. The states $\xi$ and $p$ are total fish biomass and fish number, respectively. $p_s$ and $\xi_i$ are stocking fish number and individual fish biomass during fish stocking, $k_1(UIA)$ is fish mortality coefficient, and $\xi_a$ is mean fish biomass which equals to $\xi$ divided by $p$. 
%\begin{equation*}
%\textcolor{red}{k_1= \text{see Appendix please. I updated the below paragraph with ref 115 }}
%\end{equation*}
The fish growth model \eqref{sys1} can be written in a compact form as follows
\begin{equation}\label{sys00}
\left\{\begin{array}{llllllll}
\displaystyle\frac{\der \xi}{\der t}= g\big(\xi,p, \textcolor{black}{\underbrace{f, T, DO,UIA}_{u}}, k_1\big)\\
    \displaystyle \frac{\der p}{\der t}=p_s-\mbox{INT}(pk_1)
\end{array}\right.
\end{equation}
where $\xi\in\mathbb{W}\subset\R$ denotes the state and \textcolor{black}{$u=[u_1,u_2,u_3,$ $u_4]^T$} is the input vector. \textcolor{black}{$u\in\mathbb{U}\subset\R^4$} describes the manipulated control input vector corresponds to the feeding rate, temperature, and dissolved oxygen, and unionized ammonia respectively. The set of admissible input values $\mathbb{U}$ is compact. 
The relative feeding rate $f=\dfrac{r}{R}$ is expressed in terms of the maximal daily ration $R$ and the daily ration $r$.}
\end{itemize}

% \section{Model calibration and water quality effects}\label{growth-model}

\subsection{Model simulation and validation}
\noindent This subsection aims to validate the proposed population growth model to the individual bioenergetic growth model proposed in \cite{Yan:98}. This step is essential to investigate the effectiveness of the proposed model and how it can reflect reality. Thus, one of the validation limitations is the lack of real data on total fish weight and population. This remains a challenge even for the individual fish since the normal procedure to measure the fish's weight is done manually. Due to these limitations, the validation process relies on the individual fish model (DEB) \cite{Yan:98} because it has been tested and validated experimentally.

\noindent The objective of the validation is to have a similar response of individual fish model when the population is selected to be $1$ in the proposed model given similar input profiles. The inputs, $f$, $T$, $DO$, and $UIA$, are generated within the range of their minimum and maximum as in Table~\ref{para} in Appendix. The initial weight for the proposed model and the individual fish (DEB) is selected to be $6.24$ [$\si{g \per fish}$]. And, the initials for the fish population are picked $1$, $5$, $10$, $50$, and $100$ [$\si{fish}$].

\begin{figure}[!t]
    \centering
    \subfigure[]{\includegraphics[width=0.4\textwidth]{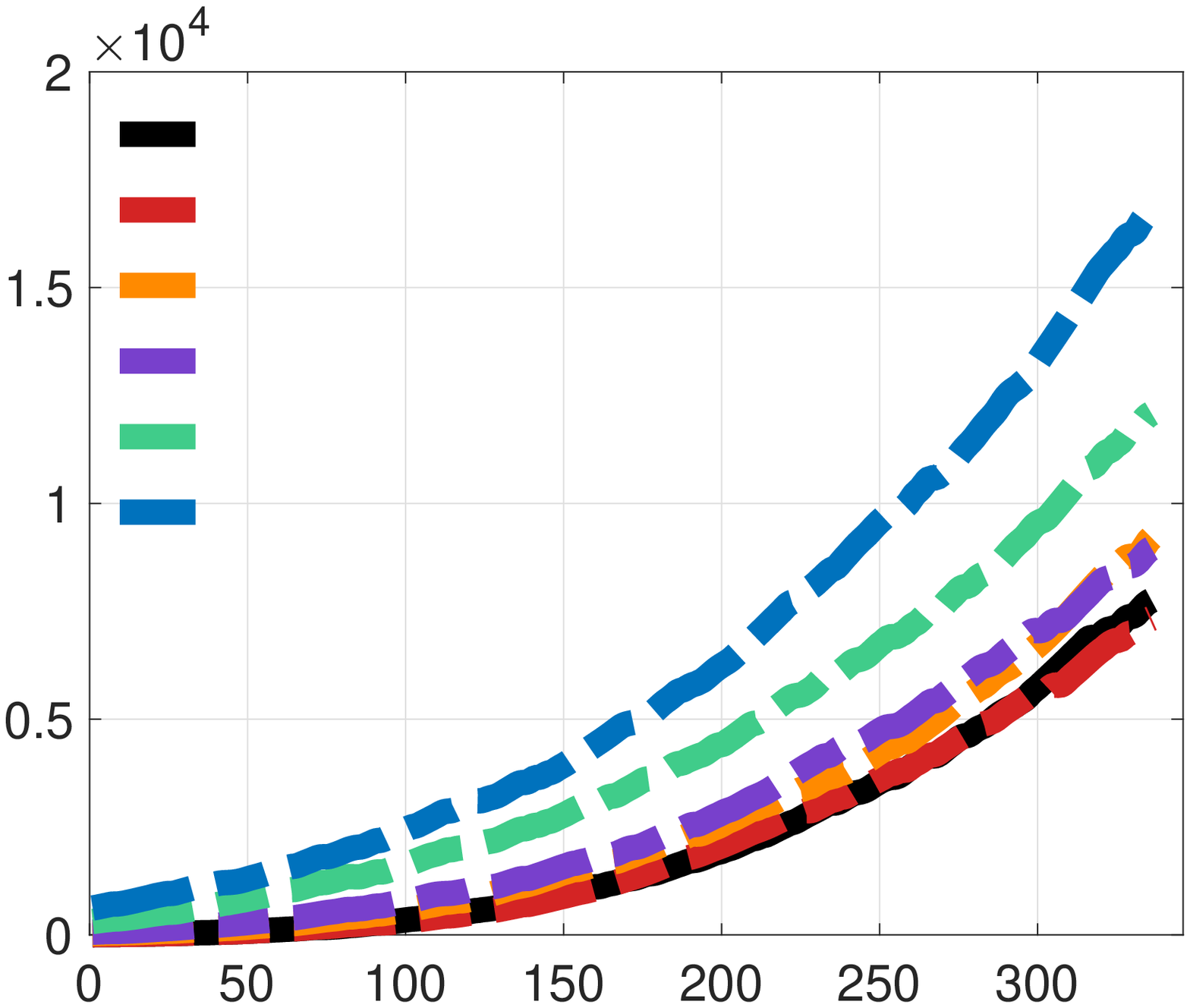}\put(-193,35){\scriptsize \rotatebox{90}{Total fish weight [$\si{g \per fish}$]}}\put(-130,-4){\scriptsize Culture period [day]}\put(-152,126){\scriptsize Simulated model}\put(-152,116){\scriptsize Population $=1$}\put(-152,105){\scriptsize Population $=5$}\put(-152,94){\scriptsize Population $=10$}\put(-152,84){\scriptsize Population $=50$}\put(-152,74){\scriptsize Population $=100$}}\vspace{-0.4cm}
    \subfigure[]{\includegraphics[width=0.4\textwidth]{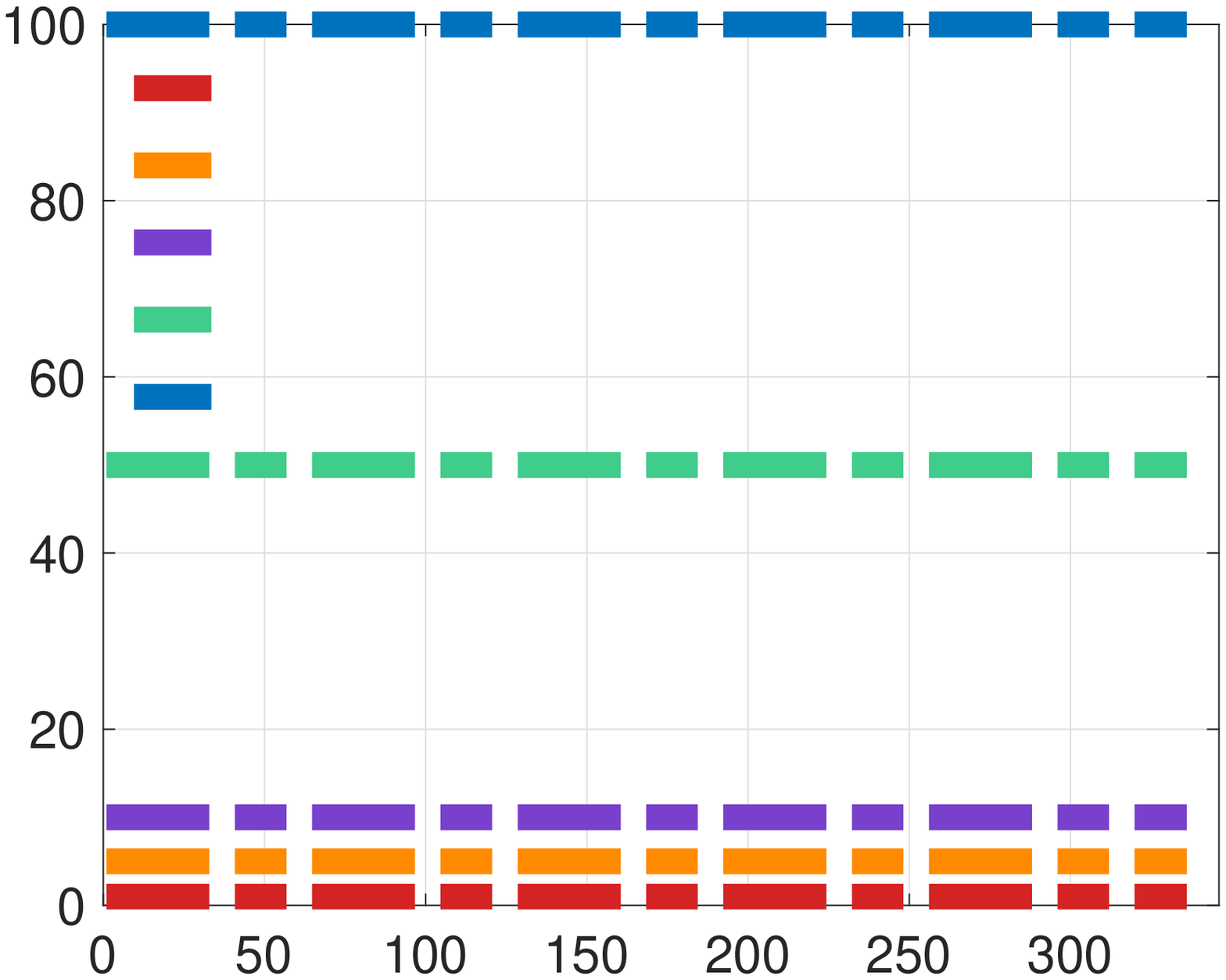}\put(-193,35){\scriptsize \rotatebox{90}{Fish population [$\si{fish}$]}}\put(-130,-4){\scriptsize Culture period [day]}\put(-152,126){\scriptsize Population $=1$}\put(-152,116){\scriptsize Population $=5$}\put(-152,105){\scriptsize Population $=10$}\put(-152,94){\scriptsize Population $=50$}\put(-152,84){\scriptsize Population $=100$}}
    \vspace{-0.35cm}
    \caption{ \footnotesize Validation results of the proposed model given similar input profiles. Figure (a) shows the total fish weight (dashed) compared to individual fish weight in \cite{Yan:98} (solid black). Figure (b) illustrates the population dynamic of the proposed model.}
    \label{Model_calibration}
    \vspace{0.1cm}
\end{figure}

\noindent Fig.~\ref{Model_calibration} illustrates the validation results of the proposed model where Fig.~\ref{Model_calibration}(a) shows the total fish weight. The solid black line represents the result of \cite{Yan:98} according to given the input profiles and the dashed red line is the result of the proposed model when the population is $1$. It is clear that the responses of both models have similar fish growth trajectories in the case of the individual population. While as expected, increasing the fish population increase the total fish in the proposed model. Fig.~\ref{Model_calibration}(b) shows population dynamic of the proposed model. The results of the population are not changing due to non-toxic feeding and water quality input factors. This leads us to investigate the effects of feeding and water quality factors in the proposed model.

\subsection{Feeding and water quality effects}
\begin{figure}[!t]
    \centering
    \subfigure[]{\includegraphics[width=0.235\textwidth]{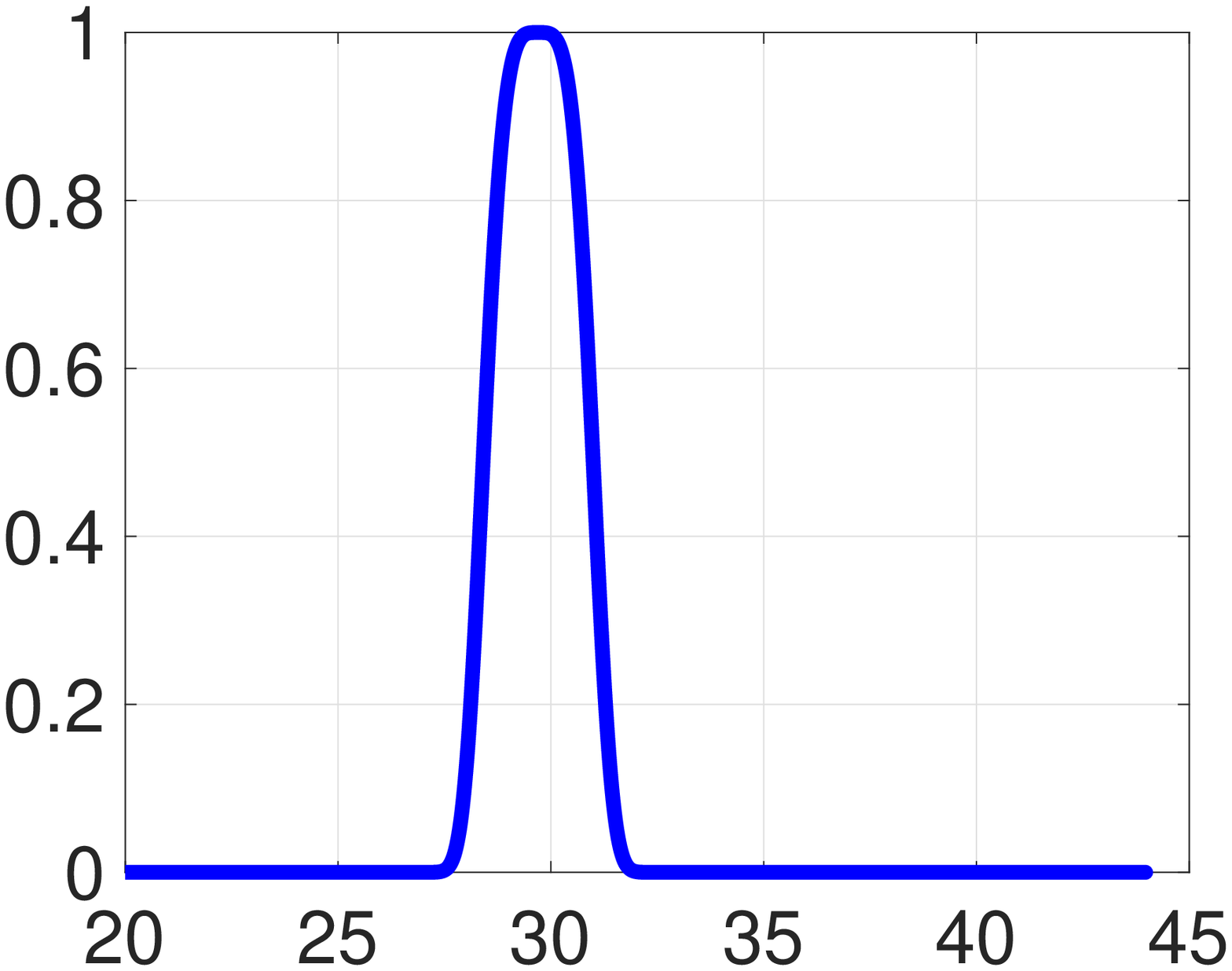}\put(-123,35){\scriptsize \rotatebox{90}{$\tau(T)$}}\put(-85,-7){\scriptsize Temperature [$^\circ C$]}}
    \subfigure[]{\includegraphics[width=0.235\textwidth]{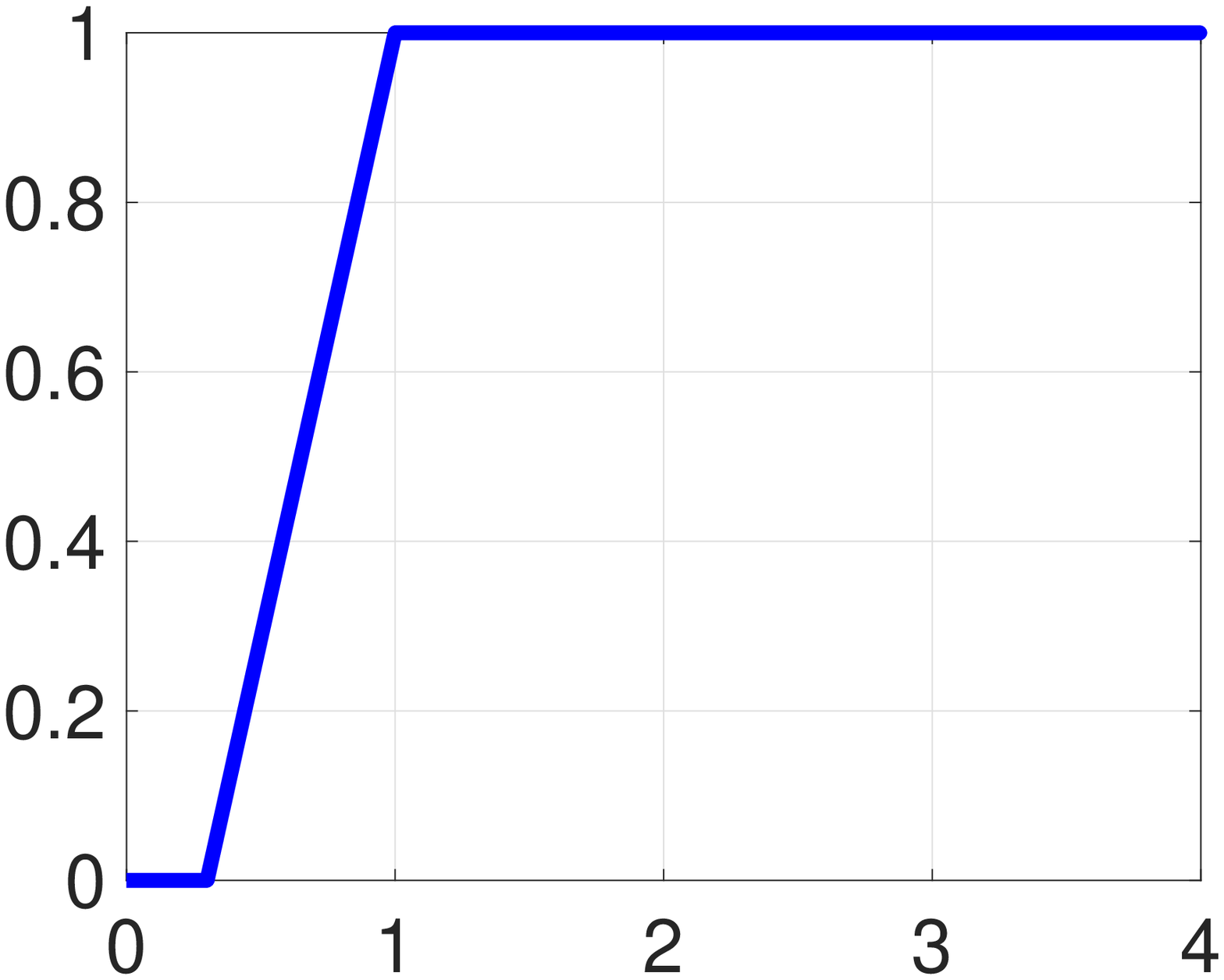}\put(-123,30){\scriptsize \rotatebox{90}{$\sigma(DO)$}}\put(-95,-7){\scriptsize Dissolved Oxygen [$\si{mg \per L}$] }} 
    \subfigure[]{\includegraphics[width=0.235\textwidth]{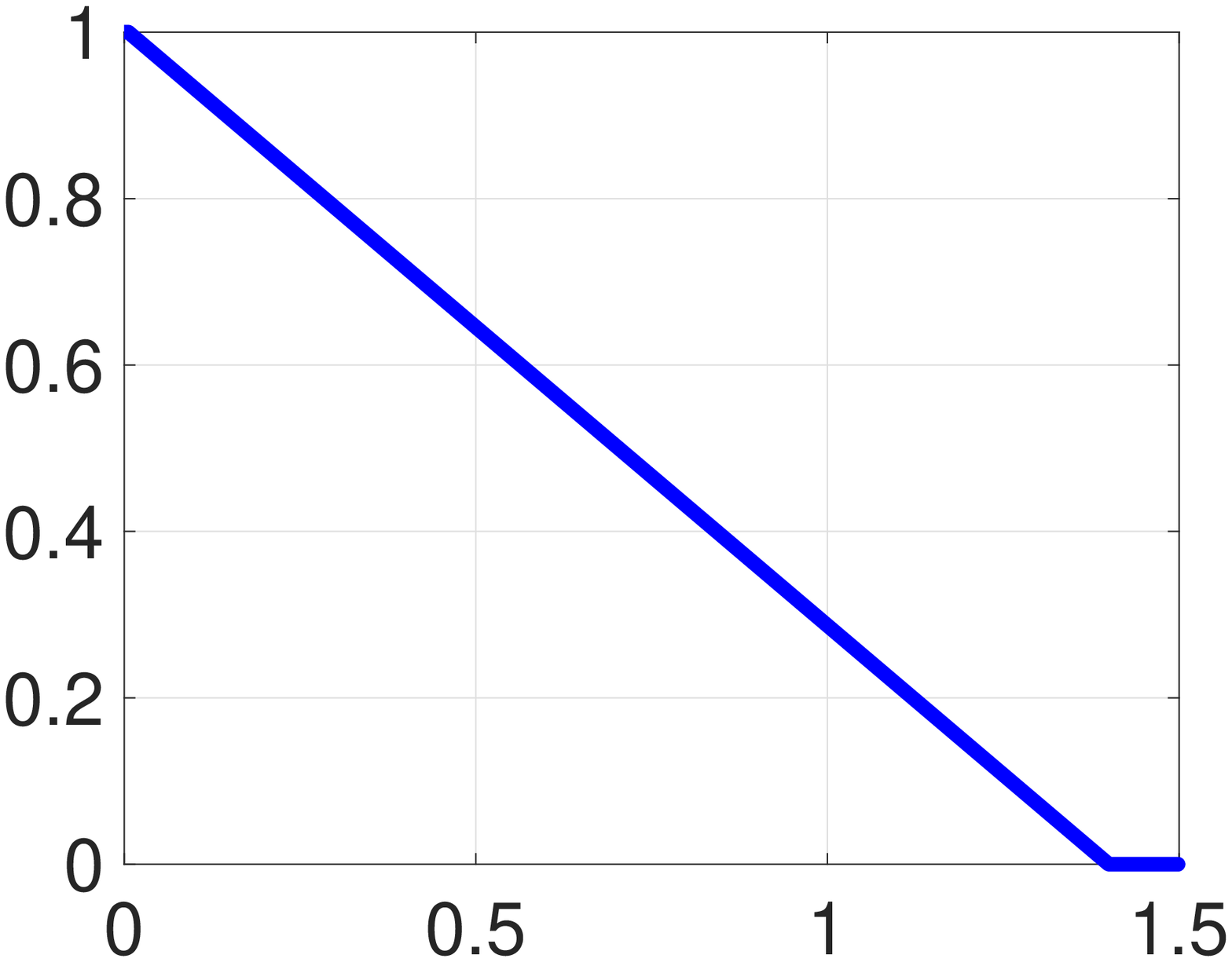}\put(-121,30){\scriptsize \rotatebox{90}{$v(UIA)$}}\put(-75,-7){\scriptsize UIA [$\si{mg \per L}$] }} 
    \vspace{-0.3cm}
    \caption{\footnotesize The effects of the water quality factors on fish food consumption. (a) $\tau (T)$ is the temperature factor ($0\leqslant \tau(T) \leqslant 1$, dimensionless), (b) $\sigma (DO)$ is the dissolved oxygen factor ($0\leqslant \sigma(DO) \leq 1$, dimensionless), and (c) $v(UIA)$ is the unionized ammonia factor ($0\leqslant v(UIA) \leqslant 1$, dimensionless).}
    \label{effect_env_factors}
\end{figure}

\noindent This subsection studies the effects of the feeding and water quality factors. The water quality effects are a function of the temperature, dissolved oxygen, and unionized in \eqref{sys00}. These functions appear in the system \eqref{sys1} as function of $\tau(T)$, $\sigma(DO)$, and $v(UIA)$. Figs.~\ref{effect_env_factors} illustrates the effect of the water quality factors on fish food consumption where Fig.~\ref{effect_env_factors}(a) corresponds to the effect of temperature on $\tau(T)$, Fig.~\ref{effect_env_factors}(b)  represents to the effect of dissolved oxygen on $\sigma(DO)$, and Fig.~\ref{effect_env_factors}(c) corresponds to the effect of unionized ammonia on $v(UIA)$.  

\begin{figure}[!t]
    \centering
    \subfigure[]{\hspace{1mm}\includegraphics[width=0.235\textwidth]{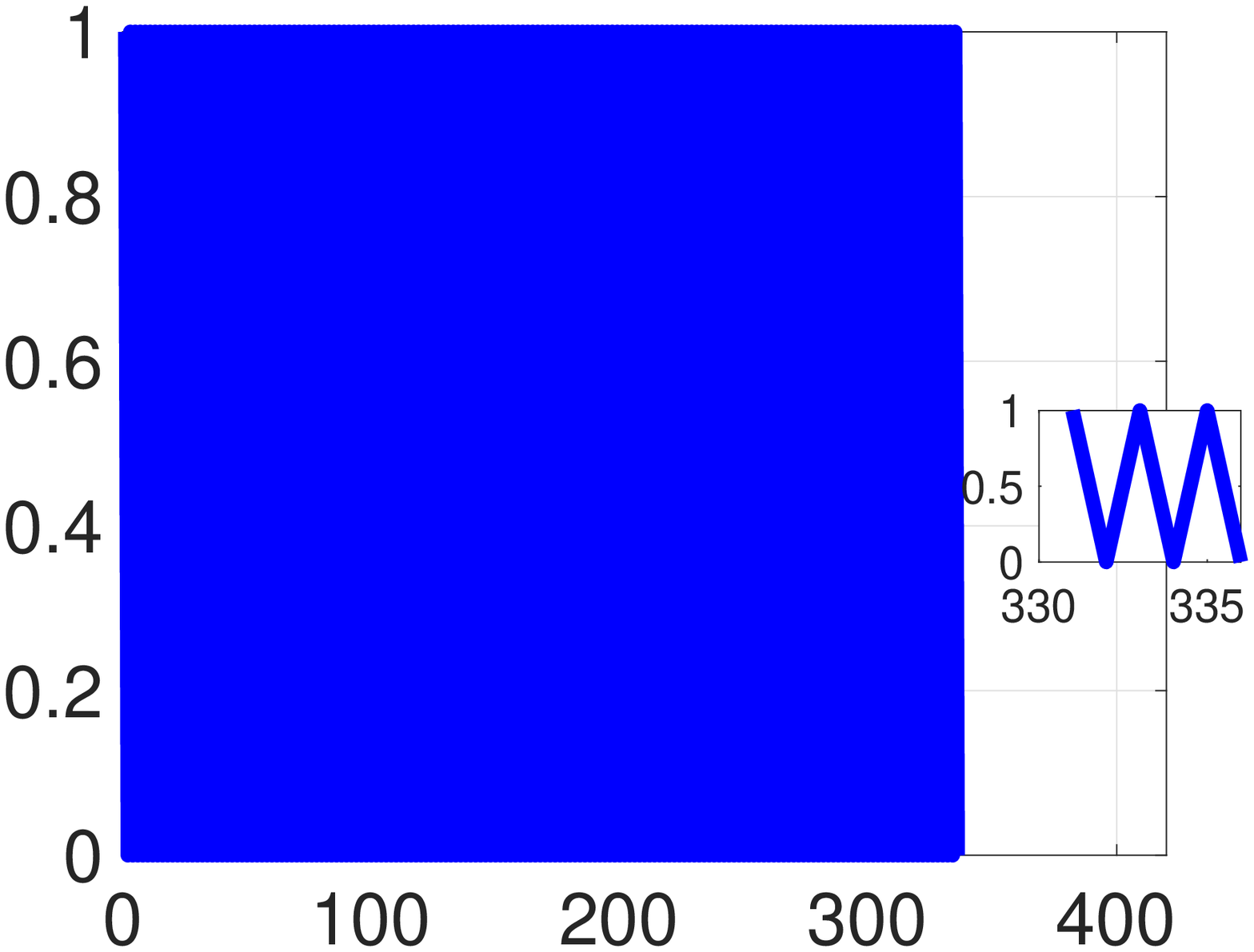}\put(-118,20){\scriptsize \rotatebox{90}{\textcolor{black}{Relative feeding}}}\put(-85,-7){\scriptsize \textcolor{black}{Culture period [day]}}}
    \subfigure[]{\includegraphics[width=0.235\textwidth]{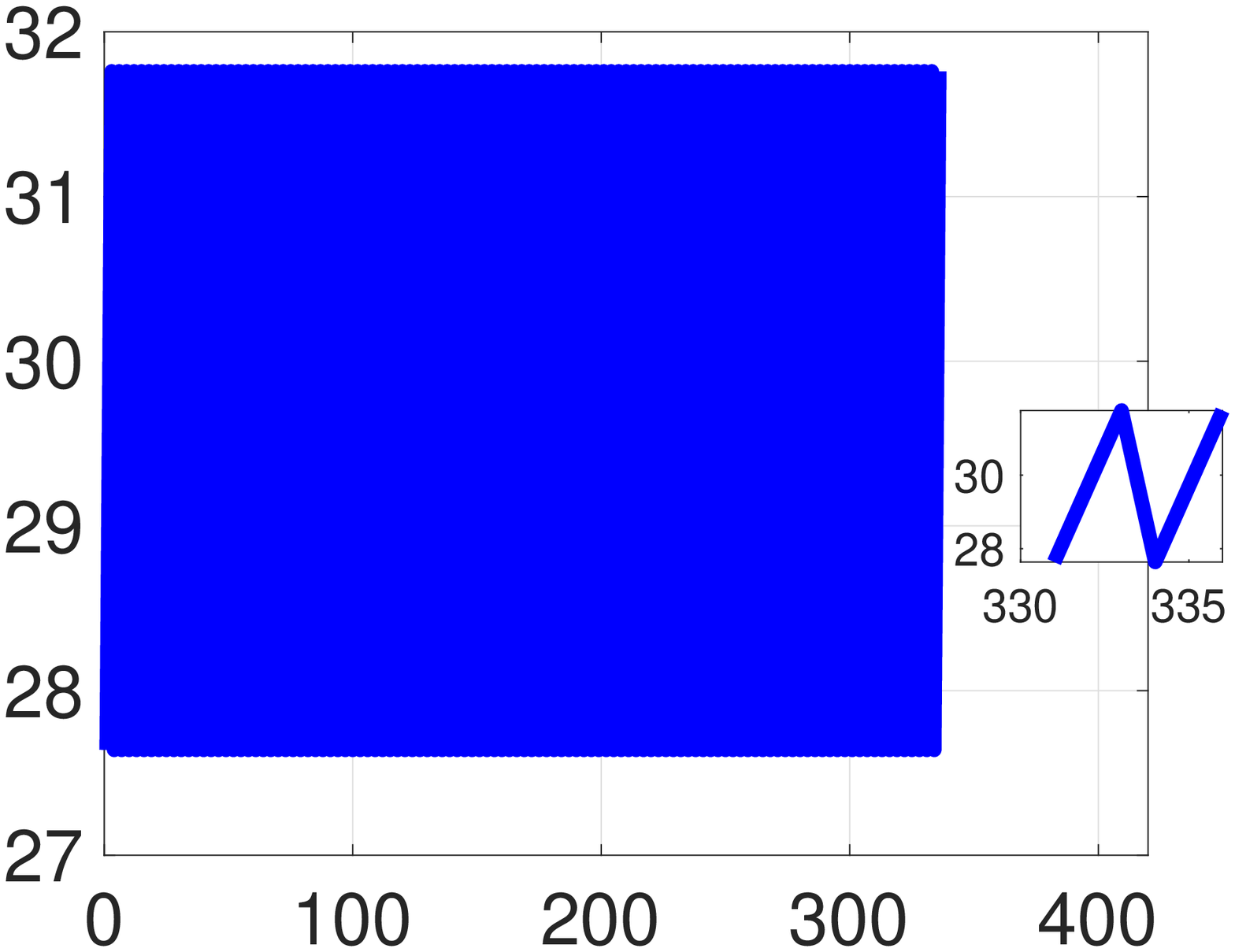}\put(-119,15){\scriptsize \rotatebox{90}{\textcolor{black}{Temperature [$^\circ C$]}}}\put(-85,-7){\scriptsize \textcolor{black}{Culture period [day]}}}
    \subfigure[]{\hspace{1.8mm}\includegraphics[width=0.23\textwidth]{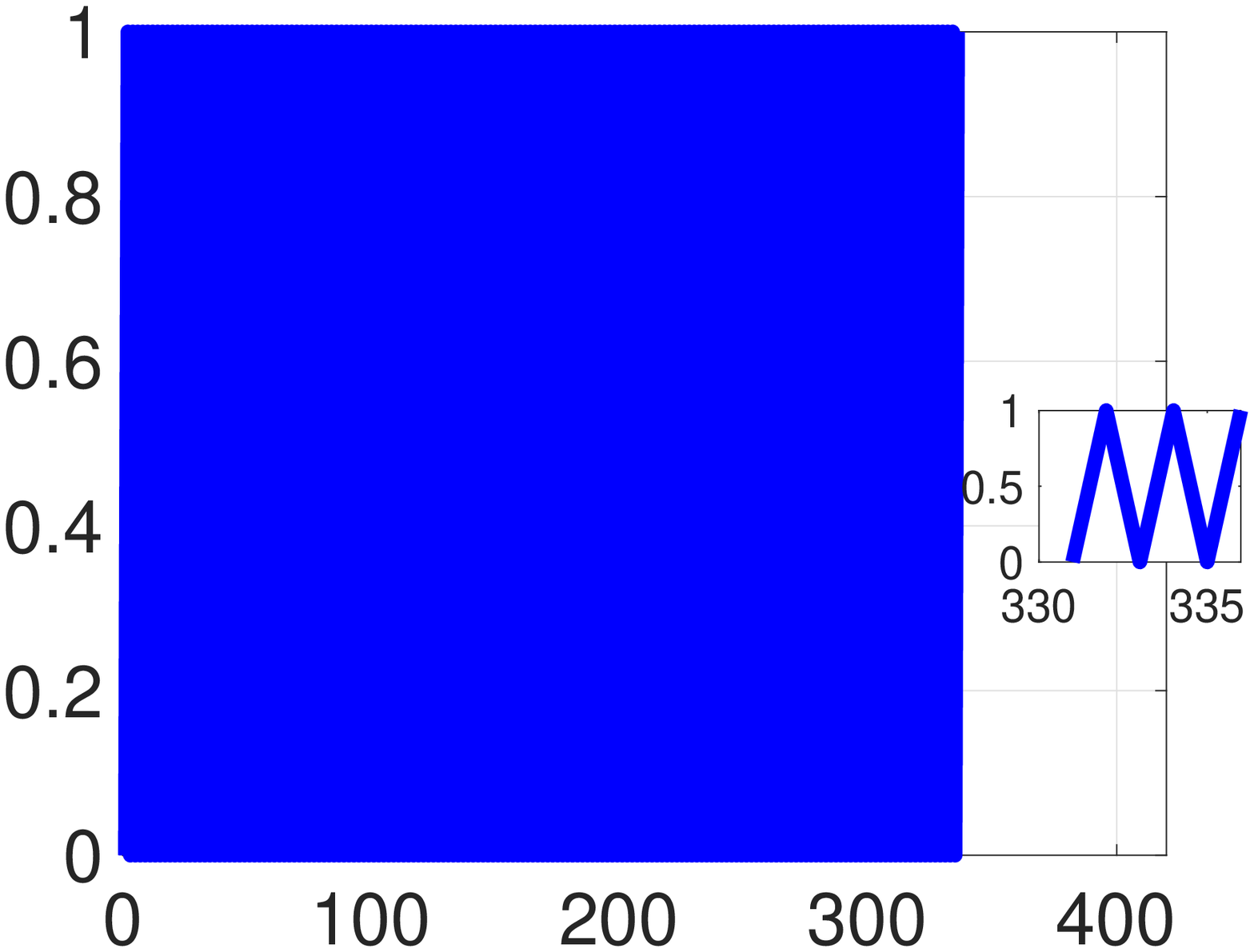}\put(-117,5){\scriptsize \rotatebox{90}{\textcolor{black}{Dissolved Oxygen [$\si{mg \per L}$]}}}\put(-85,-7){\scriptsize  \textcolor{black}{Culture period [day]}}}
    \subfigure[]{\includegraphics[width=0.235\textwidth]{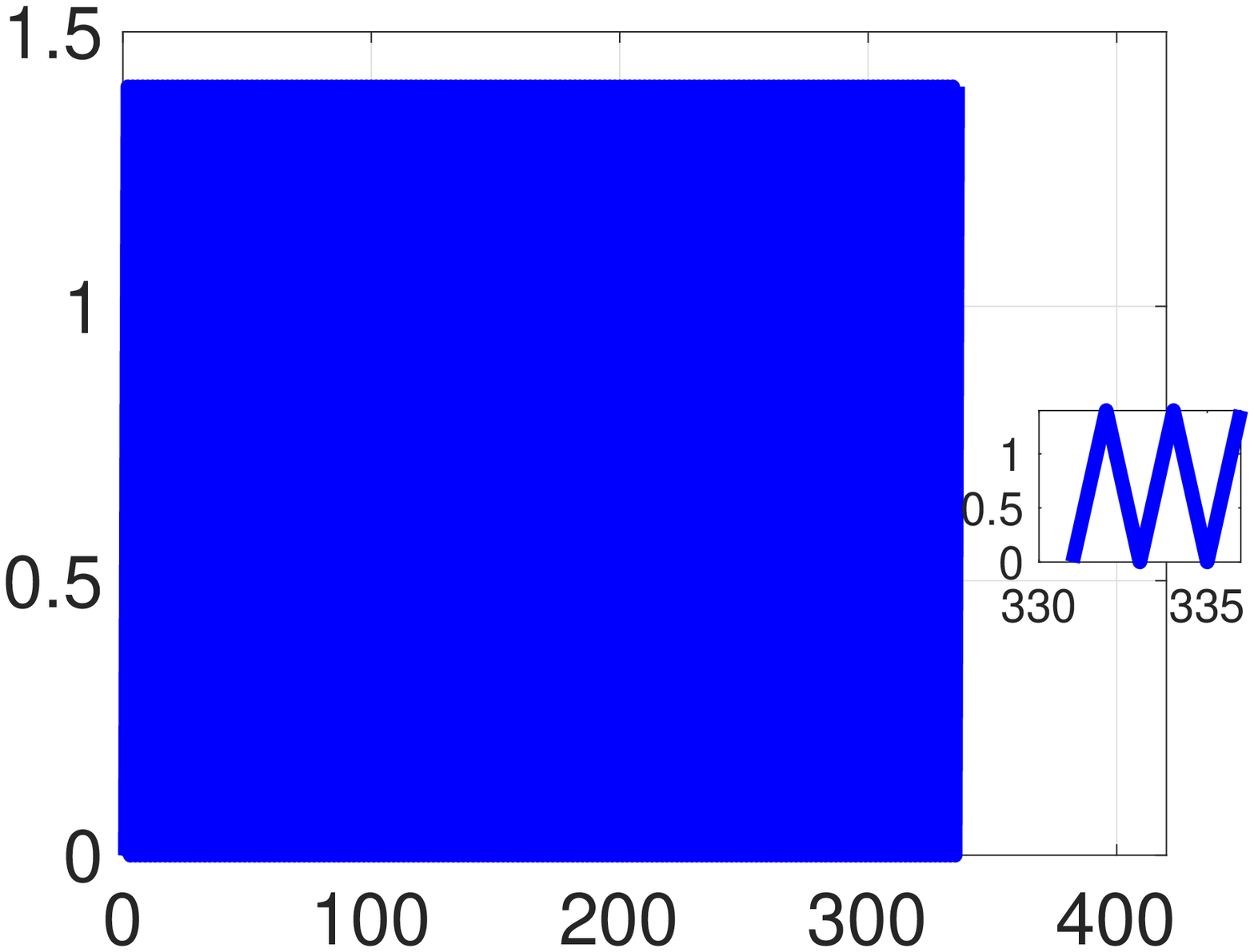}\put(-119,27){\scriptsize \rotatebox{90}{\textcolor{black}{UIA [$\si{mg \per L}$]}}}\put(-85,-7){\scriptsize \textcolor{black}{Culture period [day]}}} 
    \vspace{-0.3cm}
    \caption{\footnotesize \textcolor{black}{The generated input time-varying profiles for the sensitivity analysis. Figure (a) presents the relative feeding profile, (b) the temperature profile, (c) the dissolved oxygen profile, and (d) the unionized ammonia profile.}}
    \label{generated_input_profiles}
\end{figure}

\noindent We note that the optimal temperature, dissolved oxygen, and unionized ammonia result in maximum $\tau(T)$, $\sigma(DO)$, and $v(UIA)$, which are equal to $1$. In this case, the anabolism in \eqref{sys1} will be maximum, assuming the relative feeding $f=1$. With that being mentioned, it is worth investigating the effects of feeding and water quality factors on fish growth and population. In particular, five cases are considered as follows
\begin{itemize}
    \item Baseline case: The feeding and water quality are optimal, namely $f=1$, $\tau(T)=1$, $\sigma(DO)=1$, and $v(UIA)=1$   
    \item Time-varying relative feeding ($f$)
    \item Time-varying temperature ($T$)
    \item Time-varying dissolved oxygen ($DO$)
    \item Time-varying unionized ammonia ($UIA$)
\end{itemize}

\textcolor{black}{Figure~\ref{generated_input_profiles} illustrates the generated input time-varying profiles for each case. They are designed to fluctuate the functions $f$, $\tau(T)$, $\sigma(DO)$, and $v(UIA)$ between $0$ and $1$.  }Table~\ref{sensitivety_analysis} summarizes the performance results of the water quality and feeding factors. Feeding and water quality factors indeed affect the total fish weight compared to the baseline case. Particularly, unionized ammonia (UIA) directly affects the fish population, resulting in the death of $9$ fish. From these observations, UIA is a dominant factor, and it is worth investigating the responses of various control methods to different UIA exposure.

\begin{table}[!ht]
%\vspace{1.5cm}
\caption{Sensitivity analysis of the effect on feeding and water quality factor}
\begin{center}
\begin{tabular}{| c || c  | c |}
  \hline
\bf{Cases} & ~~\bf{Fish weight $\si{[g]}$}~~ & ~~\bf{Population $\si{[fish]}$}~~\\
\hline
\mbox{Maximum case} & $7922.9$& $10/10$\\
\mbox{Time-varying} $f$ & $4549.2$& $10/10$\\
\mbox{Time-varying} $T$ & $1220.15$& $10/10$\\
\mbox{Time-varying} $DO$ & $4759.2$& $10/10$\\
~~\mbox{Time-varying} $UIA$~~ & $0.066$ & $1/10$\\
\hline
\end{tabular}
\end{center}
\label{sensitivety_analysis}
\end{table}

%\newpage
%\section{Traditional and optimal feeding controllers for monitoring fish population growth dynamics: a case study under ammonia exposure}\label{trad-adv}
\section{Fish population growth tracking under ammonia exposure}\label{trad-adv}
\noindent This section studies the model-based feeding controllers for monitoring fish population growth under different UIA profiles. Indeed, ammonia exposure remains a significant concern in the fish population growth model and the possible effects on fish health and survival in aquaculture systems. Besides, the high ammonia accumulation directly results in fish mortality and increases potential signs of stress resulting in behavioral responses and disease resistance. Therefore, it is crucial to analyze the fish growth responses under the different traditional and optimal controllers through the UIA exposure to monitor the value of UIA to enhance fish farming productivity. We start with the classical controllers, as the bang-bang and proportional-integral-derivative (PID) controllers; then, we move to the optimal control, namely model predictive control \textcolor{black}{and Q-learning}. 

\subsection{Bang-bang control}\label{bang}
The Bang-bang controller is an on-off controller. It is considered one of the simplest controllers yet widely utilized in various market devices. It has been used to regulate water quality and automate feeding in aquaculture systems. For instance, the bang-bang controller was implemented to track the desired set-point of dissolved oxygen by switching the aerator on or off in \cite{Francis2019}. The mathematical representation for tracking the desired set-point can be given as follows 
\begin{equation}\label{bang-bang cont}
u^j(t) = \left\{\begin{array}{lll}
   \displaystyle \mbox{on} \quad \quad \mbox{if} \ \ \ \ e^j(t) > 0,\\
\displaystyle \mbox{off}  \quad \quad \mbox{if} \ \ \ \ e^j(t)\leqslant 0,
\end{array}\right.
\end{equation}
where $u^j(t)$ is the input action to the system,  $j$ is the number of control variables and $e^j(t)=\mathcal{X}_d^j(t)-\mathcal{X}^j(t)$ is the error between desired references $\mathcal{X}_d^j(t)=[T_d(t),$ $DO_d(t),$ $f_d(t),$ $UIA_d(t)]$ and output measurements $\mathcal{X}^j(t)=[T(t),$ $DO(t),$ $f(t),$ $UIA(t)]$. For instance, consider the temperature to be the controllable parameter $T$, then the heater will turn on if $T_{\mbox{\scriptsize desired}}- T> 0 \implies T<T_{\mbox{\scriptsize desired}}$ or turn off if $T_{\mbox{\scriptsize desired}} - T \leqslant 0 \implies T\geqslant T_{\mbox{\scriptsize desired}}$. The given conditions in the bang-bang controller are not restricted to tracking desired references but can also be set according to duty cycles. This enables the design of an automated feeding system. 
In the sequel, we monitor the feeding using the bang-bang controller to track the desired fish growth as follows
\begin{equation}\label{bang-bang feeding}
f = \left\{\begin{array}{lll}
   \displaystyle 1 \quad \quad \mbox{if} \ \ \ \ w_d-w > 0,\\
\displaystyle 0.1  \quad \quad \mbox{if} \ \ \ \ w_d-w \leqslant 0,
\end{array}\right.
\end{equation}
where $w = \frac{FB}{FP}$ is the mean weight of fish under the assumption that it is measurable, and $w_d$ is the desired fish weight. In the case that $f=1$, the provided food quantity is $10\%$ of the fish's mean body weight. On the other hand, when $f=0.1$, the provided food quantity is $1\%$ of the fish's mean body weight. Using this structure of relative feeding allows {the fish to be fed daily.}

\subsection{Proportional-Integral-Derivative (PID) controller}\label{pid}
\noindent   Proportional-Integral-Derivative (PID) controllers have been examined experimentally in aquaculture for water quality and feeding. For automatic feeding, the feeding rate and time have been implemented using a PID controller to enhance the fish growth \cite{Zain2013}. Whereas in \cite{Pedro2020}, the nitrate concentration in water was controlled by a PID controller to track the desired reference. Assuming that the objective is to drive temperature, dissolved oxygen, and feeding rate to desired references, then the measurements are $\mathcal{X}^j(t)$, and the references are $\mathcal{X}_d^j(t)$. PID controller is calculated from the error feedback as follows
\begin{equation}
    u^j(t) = K_p^j e^j(t) + K_i^j \int e^j(t) \,dt + K_d^j \frac{d e^j(t)}{d t},
\end{equation}
where $u^j(t)$ is the input action to the system, $e^j(t)=\mathcal{X}_d^j(t)-\mathcal{X}^j(t)$ is the feedback or tracking error, $K_p^j$, $K_i^j$, and $K_d^j$ are the gains of proportional, integral, derivative, respectively. These gains are tunable parameters tuned with trial and error.

\noindent  Similar to the bang-bang controller, the manipulated variable is the relative feeding, and the PID formulation is given as follows 
\begin{equation}\label{PID feeding}
    f = K_p e + K_i \int e \,dt + K_d \frac{d e}{d t},
\end{equation}
where $e = w_d-w$ is the tracking error.

\subsection{Optimal feeding under different UIA profiles (MPC$^1$)}\label{advance} 
\noindent Optimal control strategies are relevant to maximize efficiency growth rate while incurring the cost of wasted food due to overfeeding that adversely impacts water quality. Subsequently, these advanced control methods can handle disturbance attenuation of external factors such as time-varying environmental factors. In the case of the fish growth application, the closed-loop PID-type controllers and the bang-bang controller do not satisfy the desirable features such as optimizing the feed conversion ratio, which minimizes the feed while maximizing the predicted growth state.

%\subsubsection{Optimal feeding}\label{MPC1}
\noindent The fish population growth reference tracking problem is formulated similarly to \cite{CNMBL:21} as a minimization with a finite-time prediction horizon as follows 
%  \begin{subeqnarray}\label{mpc1_tracking}
% \!\!\!\!\!\!\!\!\!\!\!\!\!\!\!\!\displaystyle   \min_{{u \in \mathcal{U}(\varepsilon)}}\!\!J\!=&&\!\!\!\!\!\!\!\!\!\!\!\!\int_{t_k}^{t_{k+N}}\!\! \Bigg (\Big \| \frac{\tilde{w}(\tau)-w_d(\tau)}{w_d(\tau)} \Big \|^2 + \lambda \Big \| f(\tau) \Big \|^2 \Bigg )\der \tau\!\!\!\! \label{mpc1_tracking_a}\\  
% &&\!\!\!\!\!\mbox{s.t}\quad \dot{\tilde{w}}(t)= g\big(\tilde{w}(t),f(t)\big) \label{mpc1_tracking_b}\\
% &&\!\!\!\!\! f_{\mbox{\scriptsize min}}\leqslant f(t) \leqslant f_{\mbox{\scriptsize max}}, \quad \forall t \in[t_k,\, t_{k+N}]\label{mpc1_tracking_c}\\
% &&\!\!\!\!\! \Delta f(t_k)= f(t_k)-f(t_{k-1})\label{mpc1_tracking_d}\\
%  & &\!\!\!\!\! w_0 \leqslant \tilde{w}(t) \leqslant w_{\mbox{\scriptsize end}} , \quad \forall t \in[t_k,\, t_{k+N}]\label{mpc1_tracking_e} \\
% &&\!\!\!\!\! \tilde{w}(t_k)=w(t_k), \quad \tilde{w}(0) =w(t_0) \label{mpc1_tracking_f},
% \end{subeqnarray}

\begin{subequations}\label{mpc1_tracking}
\begin{align}
& \min_{{u \in \mathcal{U}(\varepsilon)}} J = \int_{t_k}^{t_{k+N}} \Bigg (\Big \| \frac{\tilde{w}(\tau)-w_d(\tau)}{w_d(\tau)} \Big \|^2 \notag \\
& ~~~~~~~~~~~~~~~~~~~~~~~~~~~~~~~~~~~~~~~~~~~~~~+\lambda \Big \| f(\tau) \Big \|^2 \Bigg ) \,d\tau \label{mpc1_tracking_a} \\  
& \text{s.t.}\quad \dot{\tilde{w}}(t)= g\big(\tilde{w}(t),f(t)\big) \label{mpc1_tracking_b} \\
& f_{\text{min}}\leqslant f(t) \leqslant f_{\text{max}}, \quad \forall t \in[t_k,\, t_{k+N}] \label{mpc1_tracking_c} \\
& \Delta f(t_k)= f(t_k)-f(t_{k-1}) \label{mpc1_tracking_d} \\
& w_0 \leqslant \tilde{w}(t) \leqslant w_{\text{end}} , \quad \forall t \in[t_k,\, t_{k+N}] \label{mpc1_tracking_e} \\
& \tilde{w}(t_k)=w(t_k), \quad \tilde{w}(0) =w(t_0) \label{mpc1_tracking_f}
\end{align}
\end{subequations}
% \begin{subequations}\label{mpc1_tracking}
% \begin{align}
% & \min_{{u \in \mathcal{U}(\varepsilon)}} J = \int_{t_k}^{t_{k+N}} \Bigg (\Big \| \frac{\tilde{w}(\tau)-w_d(\tau)}{w_d(\tau)} \Big \|^2 \notag \\
% & \phantom{\min_{{u \in \mathcal{U}(\varepsilon)}} J = \int_{t_k}^{t_{k+N}} \Bigg (}+\lambda \Big \| f(\tau) \Big \|^2 \Bigg ) \,d\tau \label{mpc1_tracking_a} \\  
% & \text{s.t.}\quad \dot{\tilde{w}}(t)= g\big(\tilde{w}(t),f(t)\big) \label{mpc1_tracking_b} \\
% & f_{\text{min}}\leqslant f(t) \leqslant f_{\text{max}}, \quad \forall t \in[t_k,\, t_{k+N}] \label{mpc1_tracking_c} \\
% & \Delta f(t_k)= f(t_k)-f(t_{k-1}) \label{mpc1_tracking_d} \\
% & w_0 \leqslant \tilde{w}(t) \leqslant w_{\text{end}} , \quad \forall t \in[t_k,\, t_{k+N}] \label{mpc1_tracking_e} \\
% & \tilde{w}(t_k)=w(t_k), \quad \tilde{w}(0) =w(t_0) \label{mpc1_tracking_f}
% \end{align}
% \end{subequations}

where $N$ is the prediction horizon, and $\lambda$ is the {weight} parameter. Equation~\eqref{mpc1_tracking_a} defines the objective function that minimizes the normalized difference error of tracking while minimizing the relative feeding over the prediction horizon. The constraint in \eqref{mpc1_tracking_b} is the growth model \eqref{sys00} used to predict the evolve $\tilde{w}(t)$ by bounded $f(t)$ in \eqref{mpc1_tracking_c} and initial condition in \eqref{mpc1_tracking_e}.
$w_0$ and $w_{\mbox{\scriptsize end}}$ are the desired initial and maximal fish weight constraints, respectively.

\textcolor{black}{
\subsection{Q-learning based optimal feeding control under ammonia exposure}\label{RL}
}

\noindent \textcolor{black}{Reinforcement learning (RL) technology has demonstrated the ability to learn the optimal feeding control policy for the fish growth rate for aquaculture in our recent work \cite{CNMBL:22}. Additionally, it has been applied to recognize the wave size to determine whether to continue or stop through an automatic fish-feeding system \cite{HCHL:22} and facilitate the feeding control process in \cite{KII:20}. RL is applied to fish robotic research and schooling navigation areas \cite{YCWTZ:21, YWYYZ:21, VNK:18}. However, few studies have assessed fish-feeding control based on the RL methodology in monitoring and optimizing fish growth rate production for aquaculture systems. In line with this, we implement the Q-learning algorithm to solve the fish growth tracking problem under different levels of ammonia exposure. Q-learning algorithm optimizes a lookup table iteratively. This Q-table is usually built from a record of feeding actions related to fish weight, and population states \cite{CNMBL:22}. Then, Q-learning algorithm maps the set of fish weight and population system states, $\mathbf{S}$, to the set of feeding actions, $\mathbf{A}$, such that the following reward, $r_t$ \cite{CNMBL:22}
\begin{equation*}
r_t\Big(s_t,a_t\Big)= -\left[\Big(\dfrac{ w(s_t)- w^d(t)}{w^d(t)} \Big)^2 + \lambda \Big(f\Big)^2 \right],
\end{equation*} 
minimizes the fish growth tracking error deviation and penalizes the feed ration under the sub-optimal temperature and dissolved oxygen profiles for the required action
\begin{equation*}
\textbf{a}=\max_{a'} Q\Big(s,a'\Big), \quad a\in \mathbf{A}, s\in\mathbf{S}
\end{equation*} 
where $f$ is the feeding rate, $\lambda$ is a positive regularization term to assess the feeding input preference, $w(s_t)$ is the fish weight at the state $s_t$ and  $w^d(t)$ is the desired reference live-weight growth trajectory.
At each time $t$, the temporal difference of the Q-learning algorithm uses sampling experiences to update the action-value function from the aquaculture environment's response according to the following equation \cite{CNMBL:22}
\begin{align*}
&Q\Big(s_{t},a_{t}\Big)\!\leftarrow\!Q\Big(s_{t},a_{t}\Big) \notag \\ &+\alpha\left[ r_t\Big(s_t,a_t\Big)\!+\!\gamma \max _{a} Q\Big(s_{t+1},a_{t+1}\Big)\!-\!Q\Big(s_{t},a_t\Big)\right],
\end{align*}
where $Q\Big(s_{t},a_{t}\Big)$ is the value function of the state-action pair $\Big(s_t,a_t\Big)$ at each time $t$,  $r_t\Big(s_t,a_t\Big)$ is the corresponding reward, $\gamma$ is the discount factor and $\alpha$ is the learning rate.}

%%%%%%%%%%%%%%%%%%%%%%%%%%%%%%%%%%%%%%%%%%%%%%%%%%%%%%%%%%%%%%%%%%%%%%%%%%%%%%%%

\section{Performance assessments of the different approaches for controlling fish feeding}\label{Performance_assessments_section}
\noindent  This section aims to design and implement classical controllers such as bang-bang and PID, and \textcolor{black}{optimal controllers such as model predictive control and Q-learning} to monitor the fish population growth model under different unionized ammonia (UIA) profiles. In this numerical simulation, the individual fish weight is extracted from experimental data in \cite{Dampin2012} for Nile Tilapia (Oreochromis niloticus). We specifically zoom on the relative feeding as a manipulated variable to design traditional and optimal control to track the desired weight reference within the controlled temperature and DO profiles under different levels of unionized ammonia (UIA) exposure.

\subsection{Case~1: UIA is maintained constant under its~critical value}\label{UIA1}
This case compares bang-bang, PID, MPC$^1$, \textcolor{black}{and Q-learning} controllers when UIA profile is maintained constant under its critical value. The temperature and dissolved oxygen (DO) profiles are controlled around the optimal temperature $T_{\mbox{\scriptsize opt}} = 29.7 \ ^\circ C$ and above the critical value $DO_{\mbox{\scriptsize min}} = 1$, respectively. %

\subsection{Case~2: UIA varies under its critical value}\label{UIA2}
\noindent This case aims to test the controllers by controlling the relative feeding using a time-varying UIA profile under its critical value. The temperature and dissolved oxygen (DO) profiles are maintained similar to case 1.

\subsection{Case 3: UIA varies under its critical value with a spike}\label{UIA3}
\noindent The last case considers identical profiles of case 2. Thus, one spike is added to the UIA profile to investigate the responses of the three controllers in the presence of a disturbance. Indeed, this case shows how the UIA directly affects fish mortality and is sensitive to the fish population growth mortality while reducing the food quantity.

\noindent In what follows, \textcolor{black}{we tuned the PID gains to achieve a good compromise between fish growth tracking error and food consumption rate performances. Then, after several trial and error tests, we selected the gains} to be $K_p=0.1$, $K_i=12$, and $K_d=0.01$. We also apply a saturation block when $f>1 \implies f = 1$ and $f<0.1 \implies f = 0.1$ which means $10\%$ or less than $1\%$ to prevent exceeding the boundary of the relative feeding. In the MPC$^1$ framework, the prediction horizon and regularization parameter are selected as $N\!=\!6$, and $\lambda=0.002$.
\begin{figure*}[!t]
    \centering %Figures/y_bang_bang_updated.eps
    \subfigure[]{\includegraphics[width=0.32\textwidth]{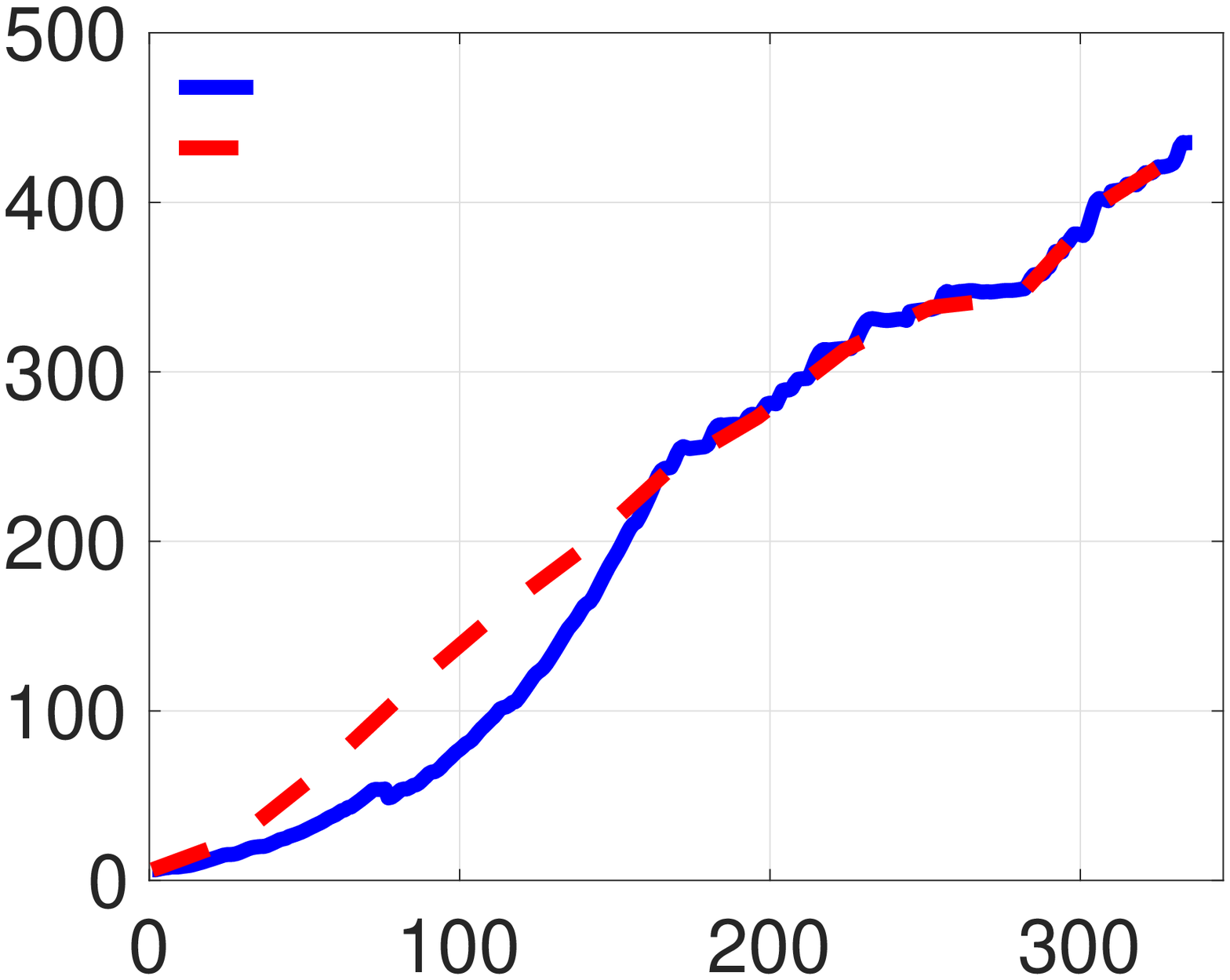}\put(-165,15){\scriptsize \rotatebox{90}{\textcolor{black}{Individual weight [$\si{g \per fish}$]}}}\put(-110,-7){\scriptsize \textcolor{black}{Culture period [day]}}\put(-123,102){\scriptsize Bang-Bang}\put(-123,95){\scriptsize Experimental}}
    \subfigure[]{\includegraphics[width=0.32\textwidth]{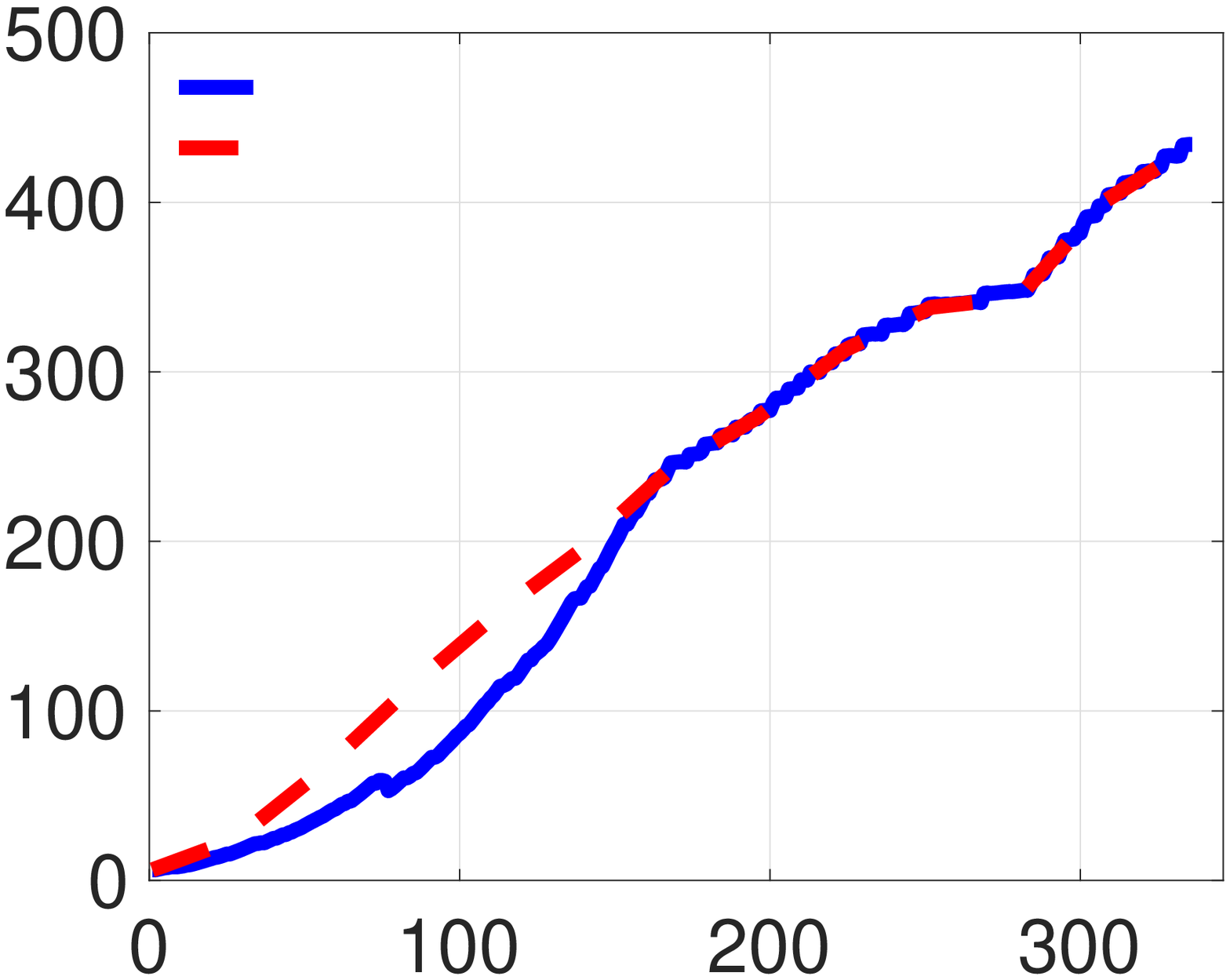}\put(-165,15){\scriptsize \rotatebox{90}{\textcolor{black}{Individual weight [$\si{g \per fish}$]}}}\put(-110,-7){\scriptsize \textcolor{black}{Culture period [day] }}\put(-123,102){\scriptsize PID}\put(-123,95){\scriptsize Experimental}} 
    \subfigure[]{\includegraphics[width=0.32\textwidth]{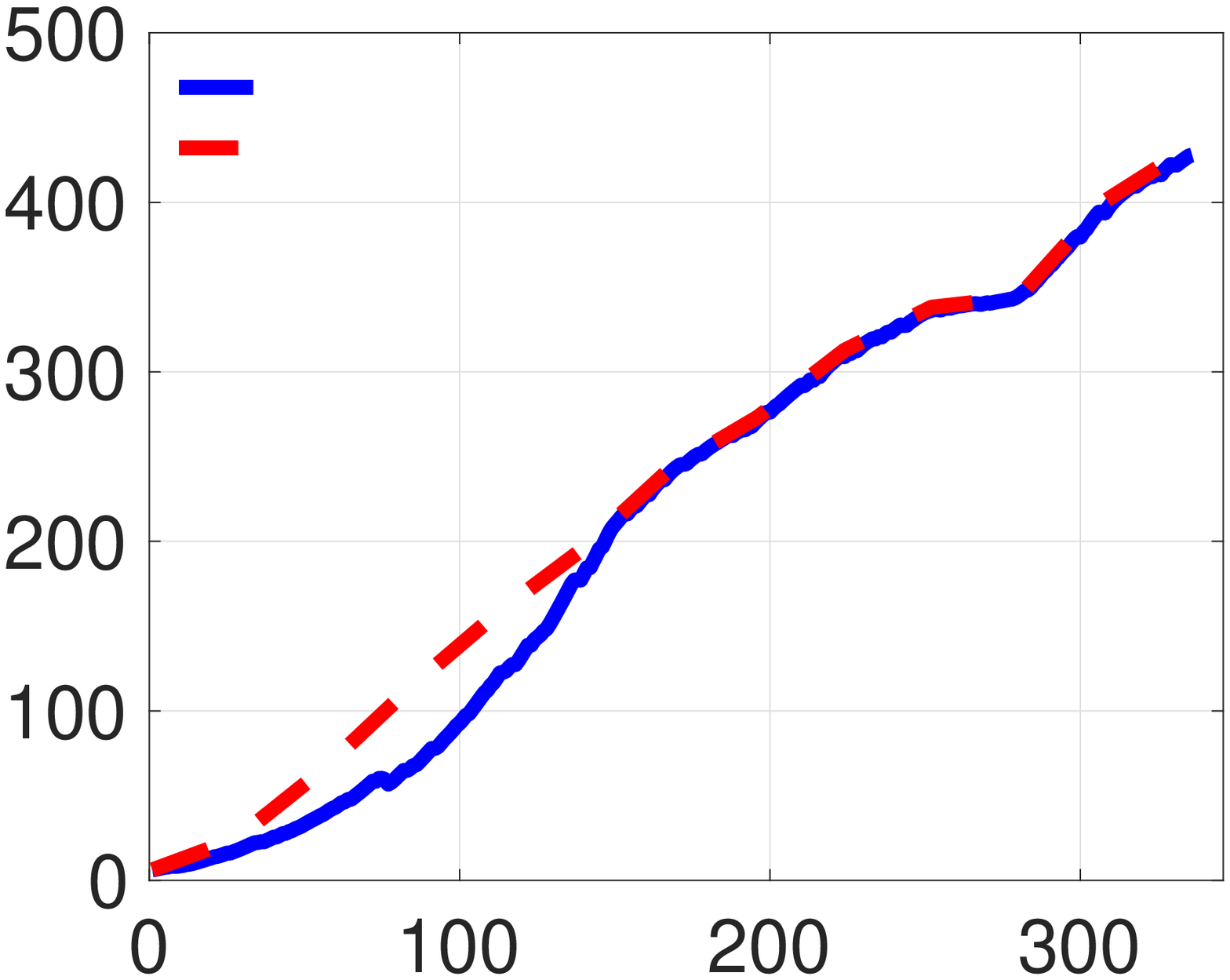}\put(-165,15){\scriptsize \rotatebox{90}{\textcolor{black}{Individual weight [$\si{g \per fish}$]}}}\put(-110,-7){\scriptsize \textcolor{black}{Culture period [day]} }\put(-123,102){\scriptsize MPC$^1$}\put(-123,95){\scriptsize Experimental}}
    \subfigure[]{\includegraphics[width=0.32\textwidth]{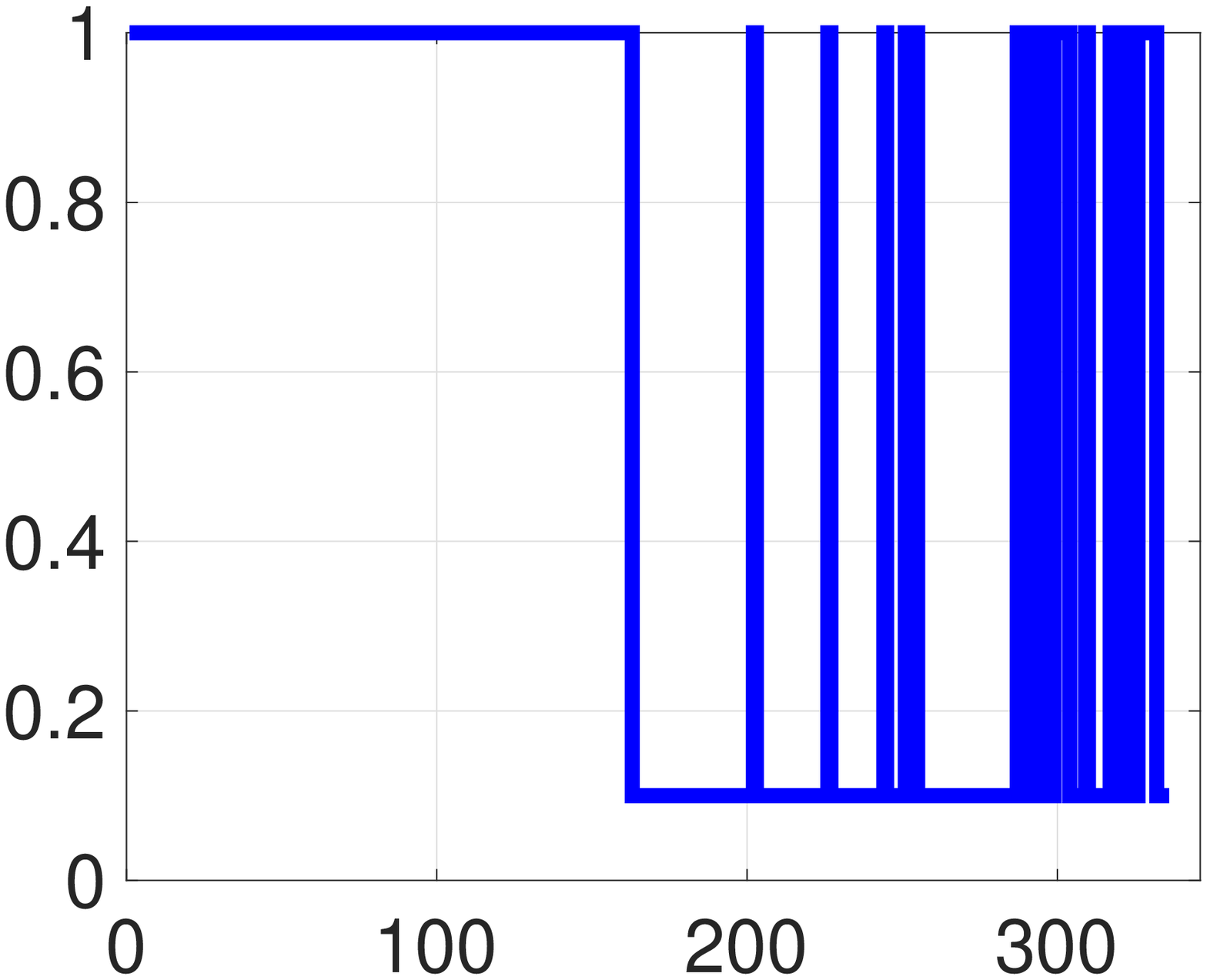}\put(-110,-7){\scriptsize \textcolor{black}{Culture period [day]}}\put(-162,32){\scriptsize \rotatebox{90}{\textcolor{black}{Relative feeding}}}} 
    \subfigure[]{\includegraphics[width=0.32\textwidth]{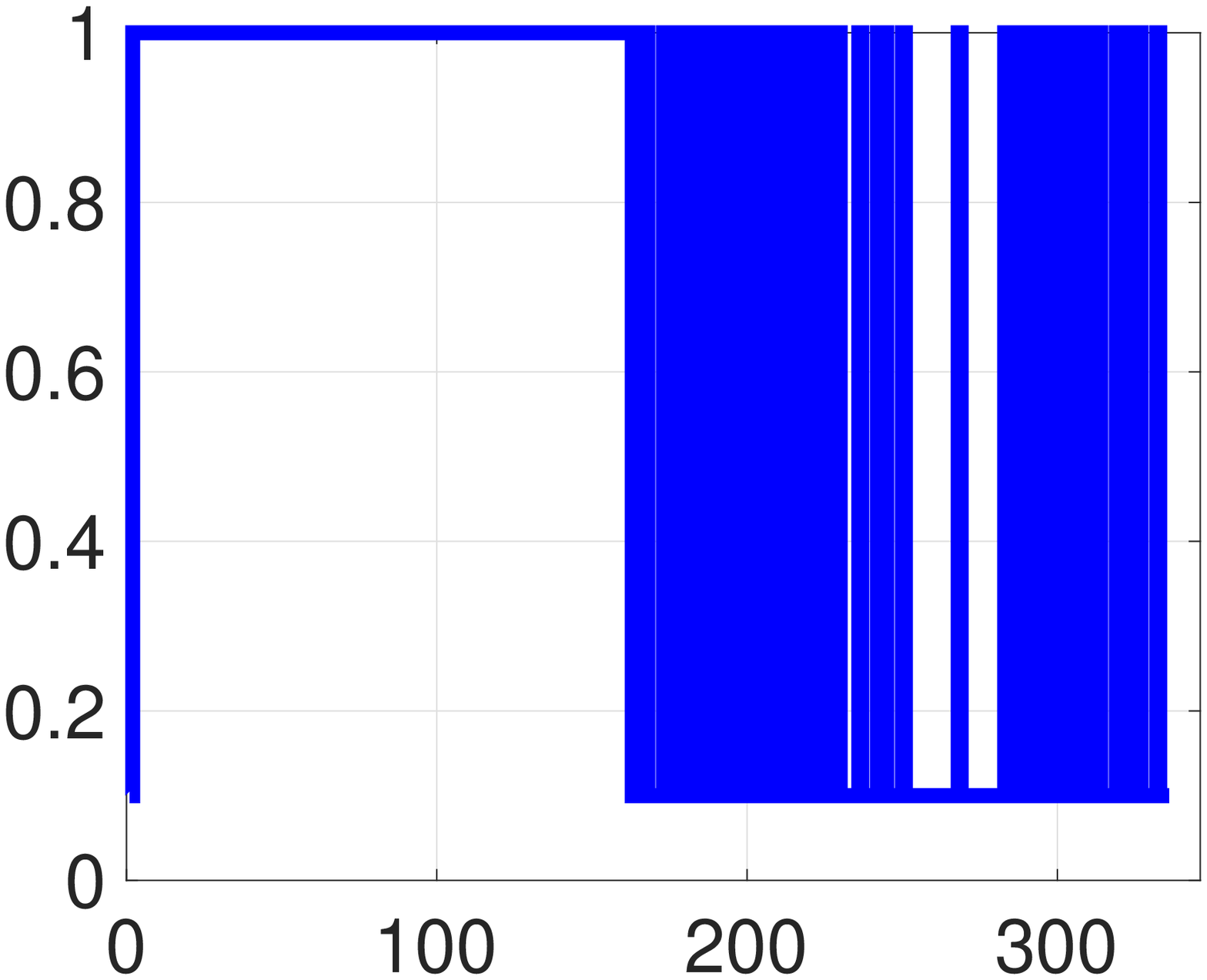}\put(-110,-7){\scriptsize \textcolor{black}{Culture period [day]}}\put(-162,32){\scriptsize \rotatebox{90}{\textcolor{black}{Relative feeding}}}} 
    \subfigure[]{\includegraphics[width=0.32\textwidth]{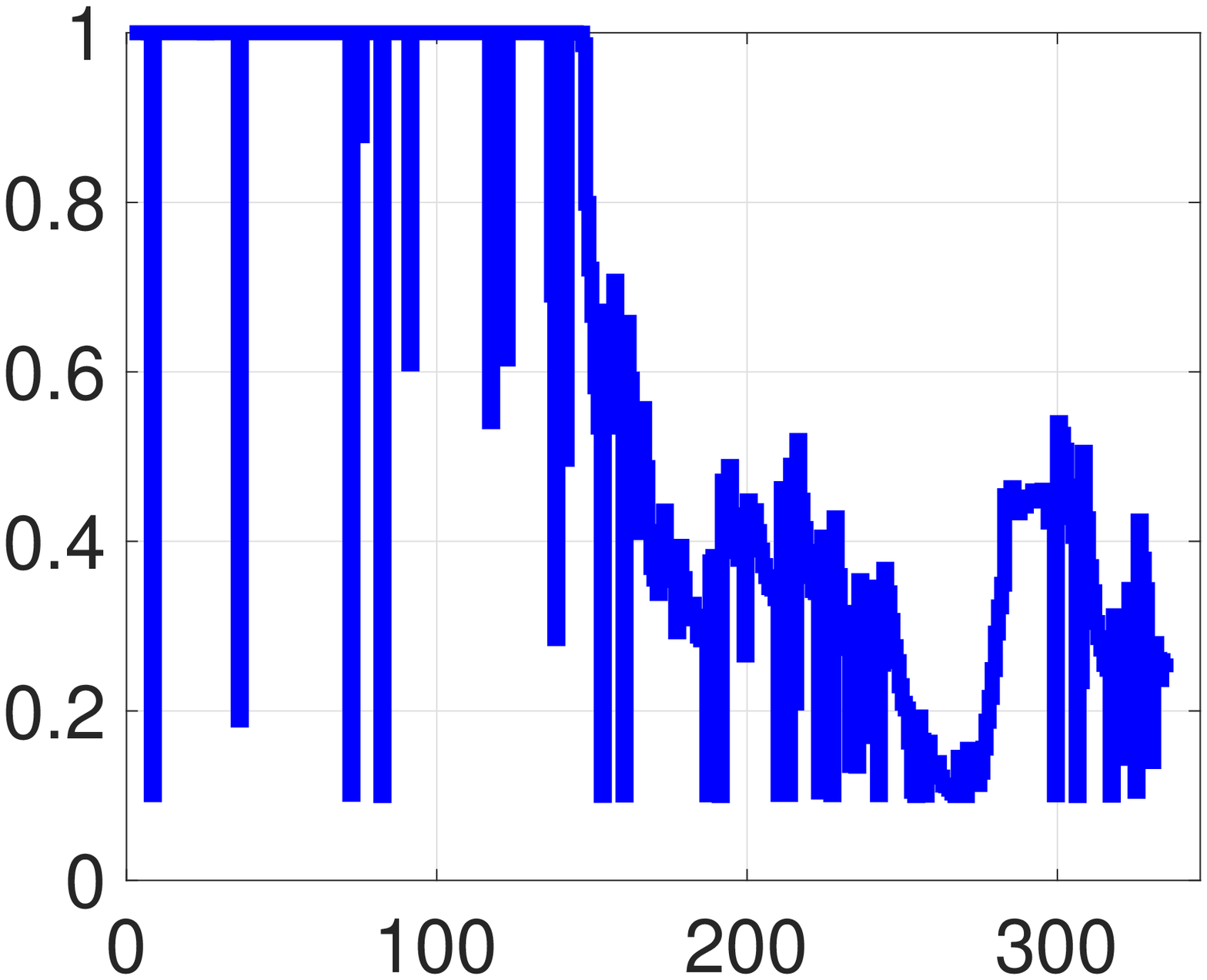}\put(-110,-7){\scriptsize \textcolor{black}{Culture period [day]}}\put(-162,32){\scriptsize \rotatebox{90}{\textcolor{black}{Relative feeding}}}} \vspace{-0.35cm}
    \caption{\footnotesize \textcolor{black}{The result of case 3 when the fish population size is equal to $10$ and UIA is varied under UIA$_{critical}$ within a spike. Figures (a), (b), and (c) illustrate fish individual growth weight tracking for Bang-Bang, PID, and MPC$^1$ controllers, respectively. Figures (d), (e), and (f) show the result of the manipulated variable (relative feeding) for Bang-Bang, PID, and MPC$^1$ controllers, respectively.}
    }
    \label{SpikeUIA}
\end{figure*}

\subsection{Results and discussion}\label{discussion}

\noindent{\textbf{\underline{Case 1:}}}
\noindent The three controllers bang-bang, PID and MPC$^1$ perform well in tracking and reach the final fish weight. Table~\ref{Performance_assessment_1} summarizes the performance assessment of all the controllers, including all cases, food consumption, RMSE, and fish mortality. The RMSE of the bang-bang controller is $14.109\%$, PID is $15.627\%$, and MPC$^1$ is $14.878\%$. 
Besides, the relative feeding of bang-bang and PID controllers are $f=1$ till the day $\approx135$. In contrast, MPC$^1$ seeks to track the desired fish weight while minimizing the relative feeding. 
To summarize this case, the tracking performance of the three controllers is quite similar (see Table \ref{Performance_assessment_1}). However, MPC$^1$ reduces the total food consumption by $17.47\%$ and $18.73\%$ compared to bang-bang and PID, respectively, even though the bang-bang RMSE is less than MPC$^1$.

\noindent{\textbf{\underline{Case 2:}}} 
From Table \ref{Performance_assessment_1}, a similar conclusion can be extracted as in case 1. The bang-bang and PID maximize the relative feeding for the first $\approx 135$ days while MPC$^1$ solves for optimal relative feeding in those days. 
Besides, MPC$^1$ reduces the total food consumption by $25.65\%$ and $7.26\%$ compared to bang-bang and PID, respectively.

\noindent{\textbf{\underline{Case 3:}}}
This case considers a spike in the UIA profile above its critical value. The resulting spike leads to one fish's mortality, indicating the need for further investigation of the three controllers' performances in Fig.~\ref{SpikeUIA}. The tracking weight performances of the bang-bang, PID, and MPC$^1$ are illustrated in Figs.~\ref{SpikeUIA}(a), (b), and (c), respectively. Moreover, the effect of the spike can be recognized on the day at $75$ when the trajectories decreased. The RMSE of the bang-bang controller is $27.72\%$, PID controller $22.817\%$, and MPC$^1$ controller is $19.914\%$. Figs.~\ref{SpikeUIA}(d), (e), and (f) show the result of the relative feeding of bang-bang, PID, and MPC$^1$ controllers, respectively. The relative feeding response of the MPC$^1$ controller is sufficient because it focuses on minimizing the relative feeding and decreases the RMSE under the presence of a spike. %

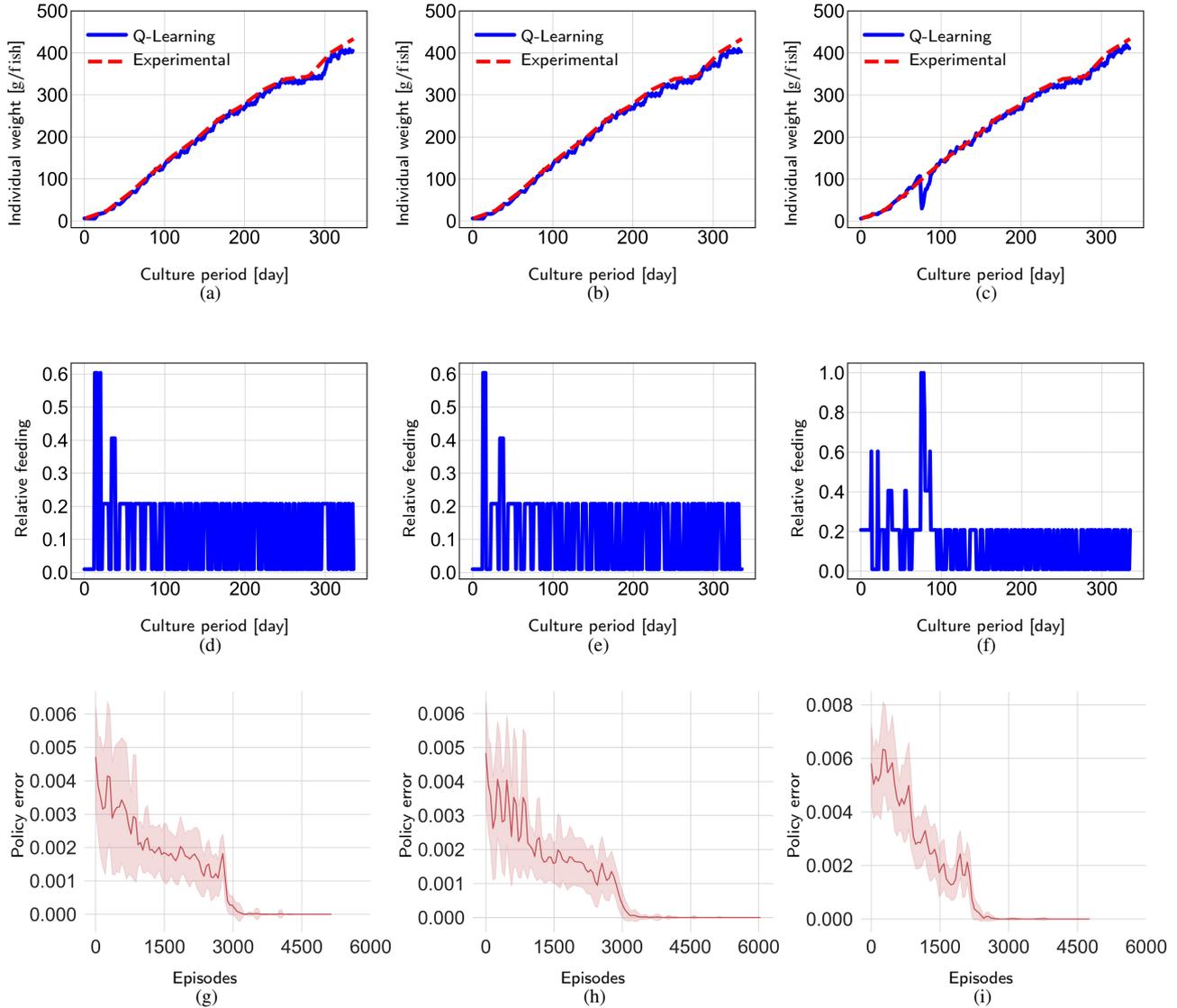
\begin{figure*}[!t]
    \centering %Figures/y_bang_bang_updated.eps
    \subfigure[]{\includegraphics[width=0.32\textwidth]{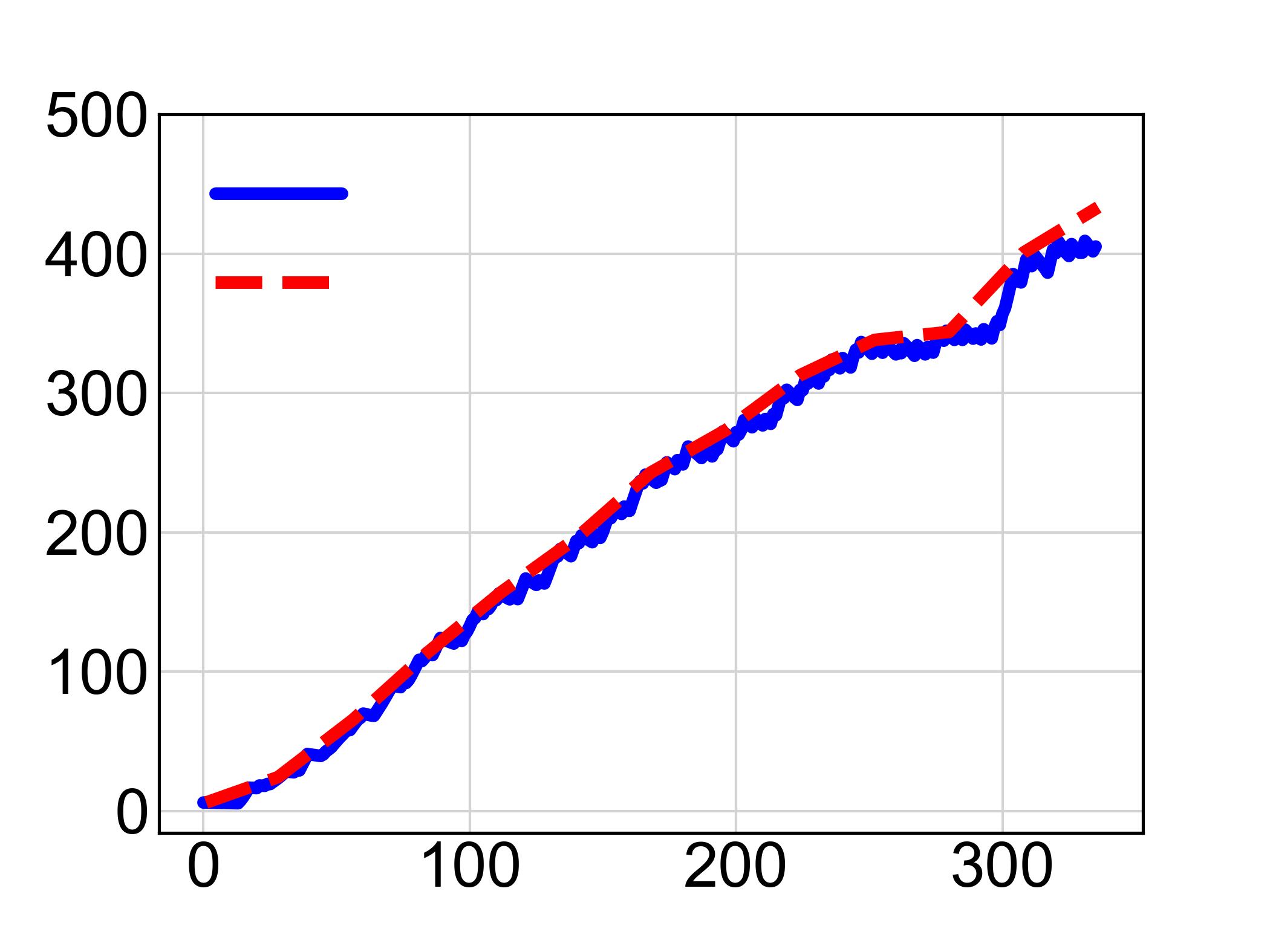}\put(-165,15){\scriptsize \rotatebox{90}{\textcolor{black}{Individual weight [$\si{g \per fish}$]}}}\put(-110,-7){\scriptsize \textcolor{black}{Culture period [day]}}\put(-113,92){\scriptsize Q-Learning}\put(-113,82){\scriptsize Experimental}}
    \subfigure[]{\includegraphics[width=0.32\textwidth]{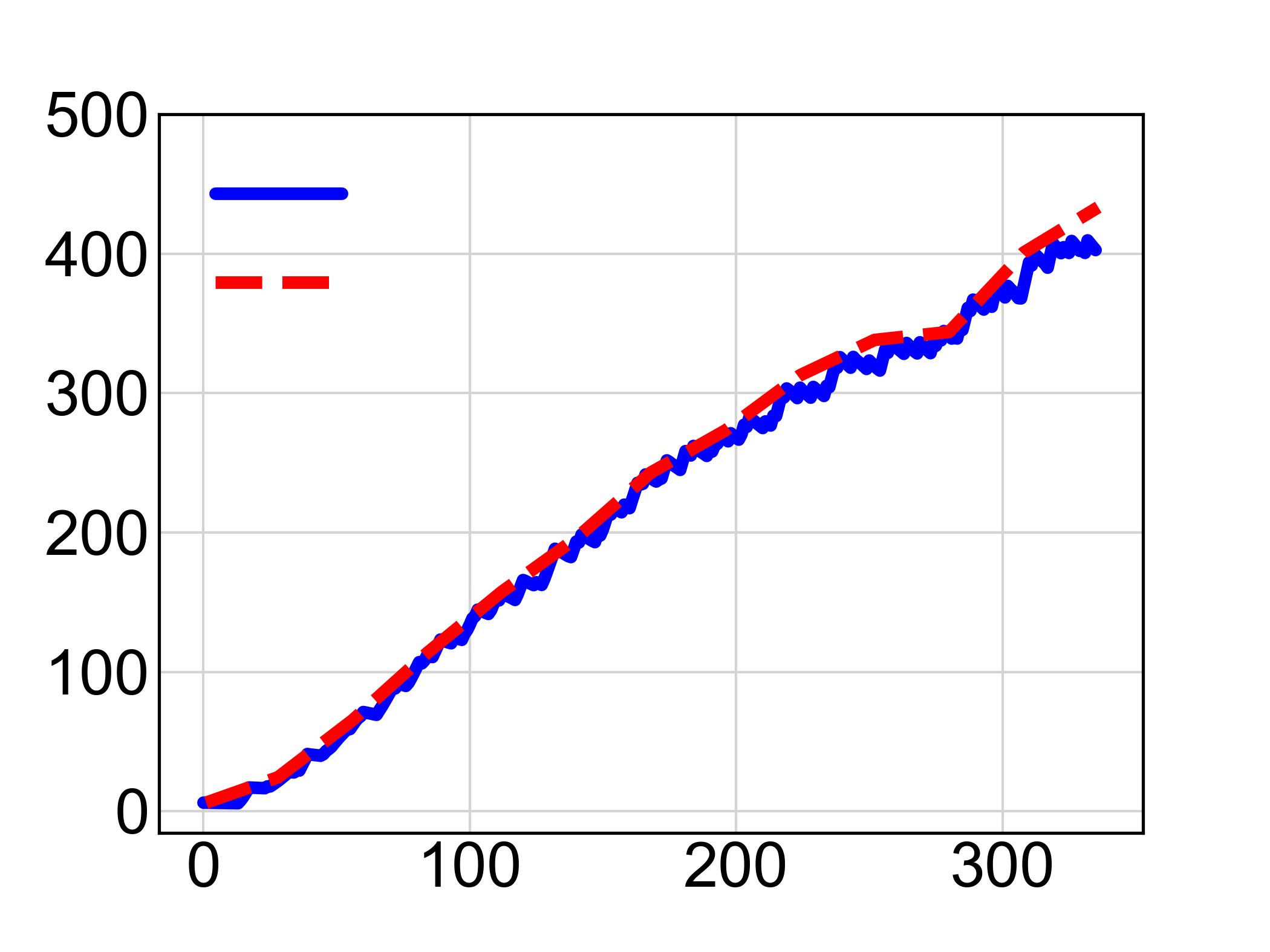}\put(-165,15){\scriptsize \rotatebox{90}{\textcolor{black}{Individual weight [$\si{g \per fish}$]}}}\put(-110,-7){\scriptsize \textcolor{black}{Culture period [day] }}\put(-113,92){\scriptsize Q-Learning}\put(-113,82){\scriptsize Experimental}} 
    \subfigure[]{\includegraphics[width=0.32\textwidth]{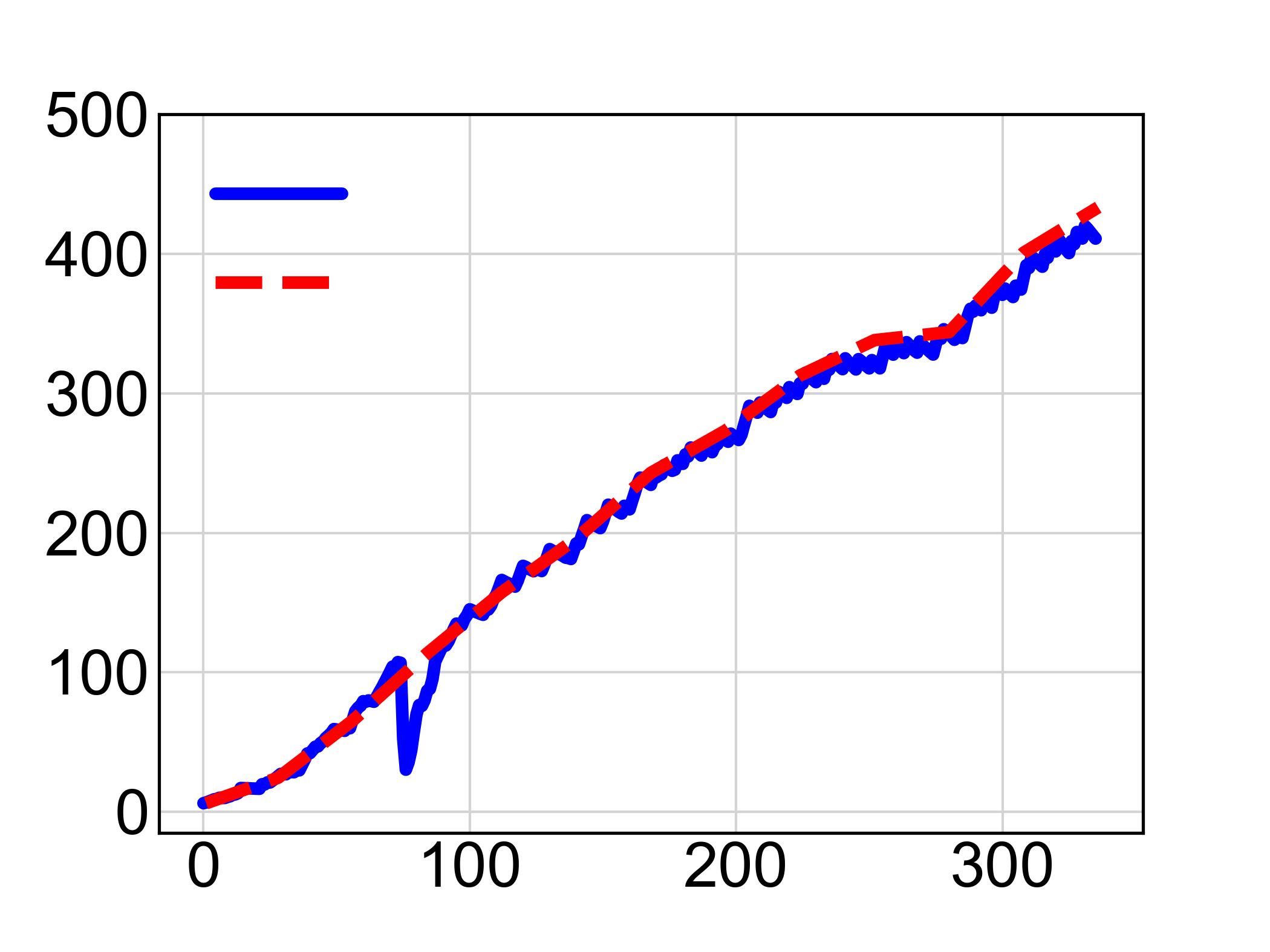}\put(-165,15){\scriptsize \rotatebox{90}{\textcolor{black}{Individual weight [$\si{g \per fish}$]}}}\put(-110,-7){\scriptsize \textcolor{black}{Culture period [day]} }\put(-113,92){\scriptsize Q-Learning}\put(-113,82){\scriptsize Experimental}}
    \subfigure[]{\includegraphics[width=0.32\textwidth]{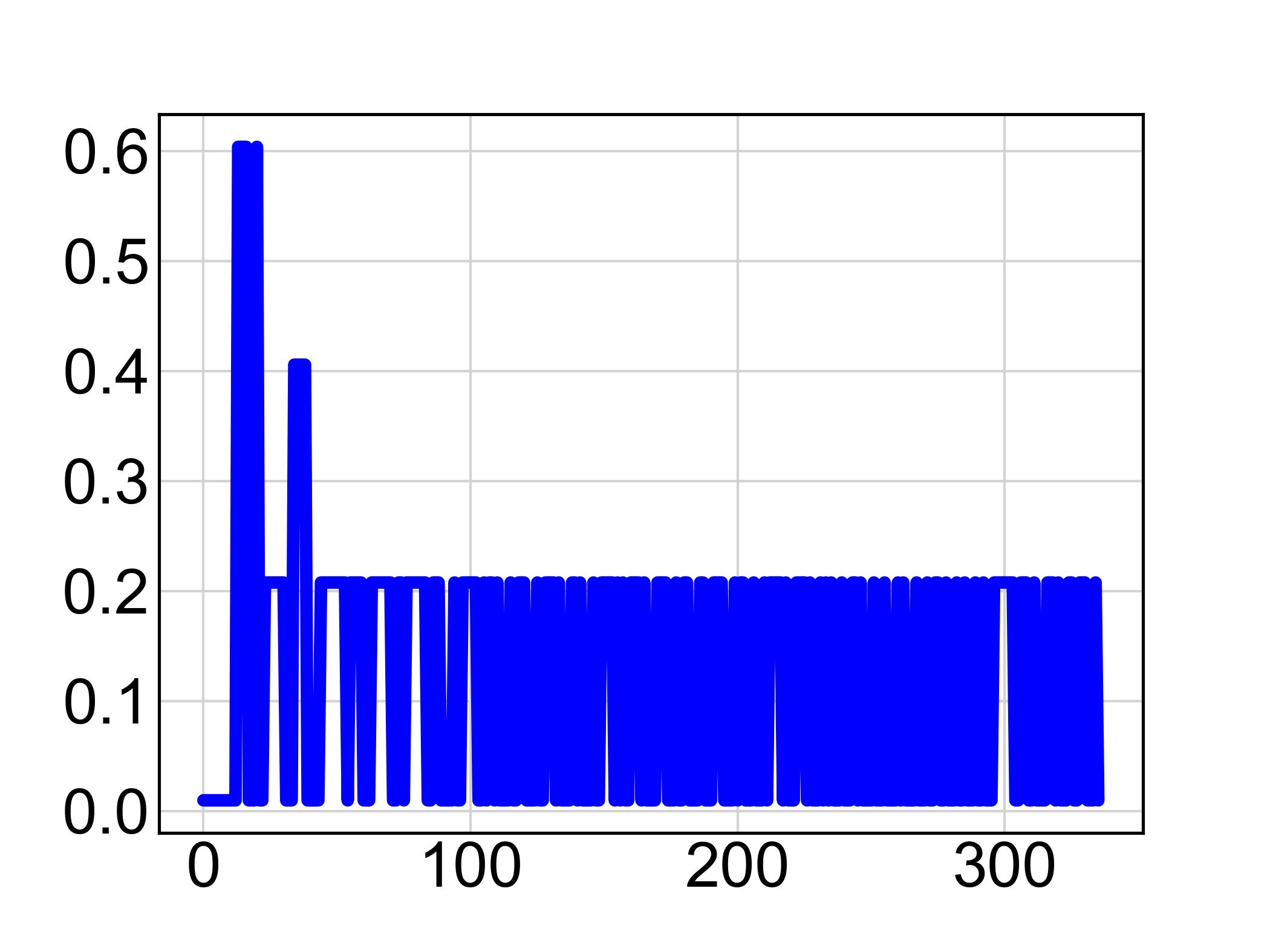}\put(-110,-7){\scriptsize \textcolor{black}{Culture period [day]}}\put(-162,32){\scriptsize \rotatebox{90}{\textcolor{black}{Relative feeding}}}} 
    \subfigure[]{\includegraphics[width=0.32\textwidth]{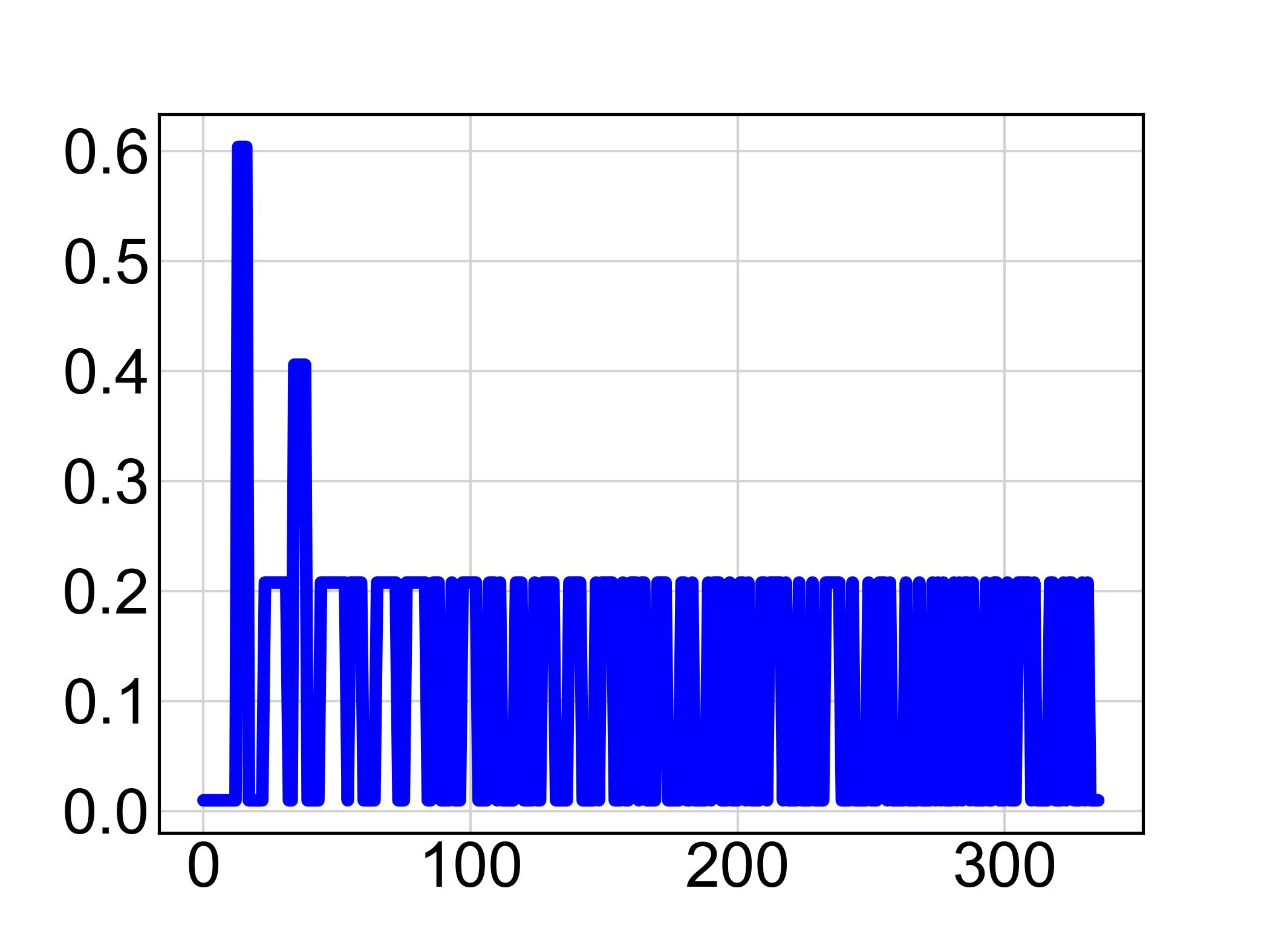}\put(-110,-7){\scriptsize \textcolor{black}{Culture period [day]}}\put(-162,32){\scriptsize \rotatebox{90}{\textcolor{black}{Relative feeding}}}} 
    \subfigure[]{\includegraphics[width=0.32\textwidth]{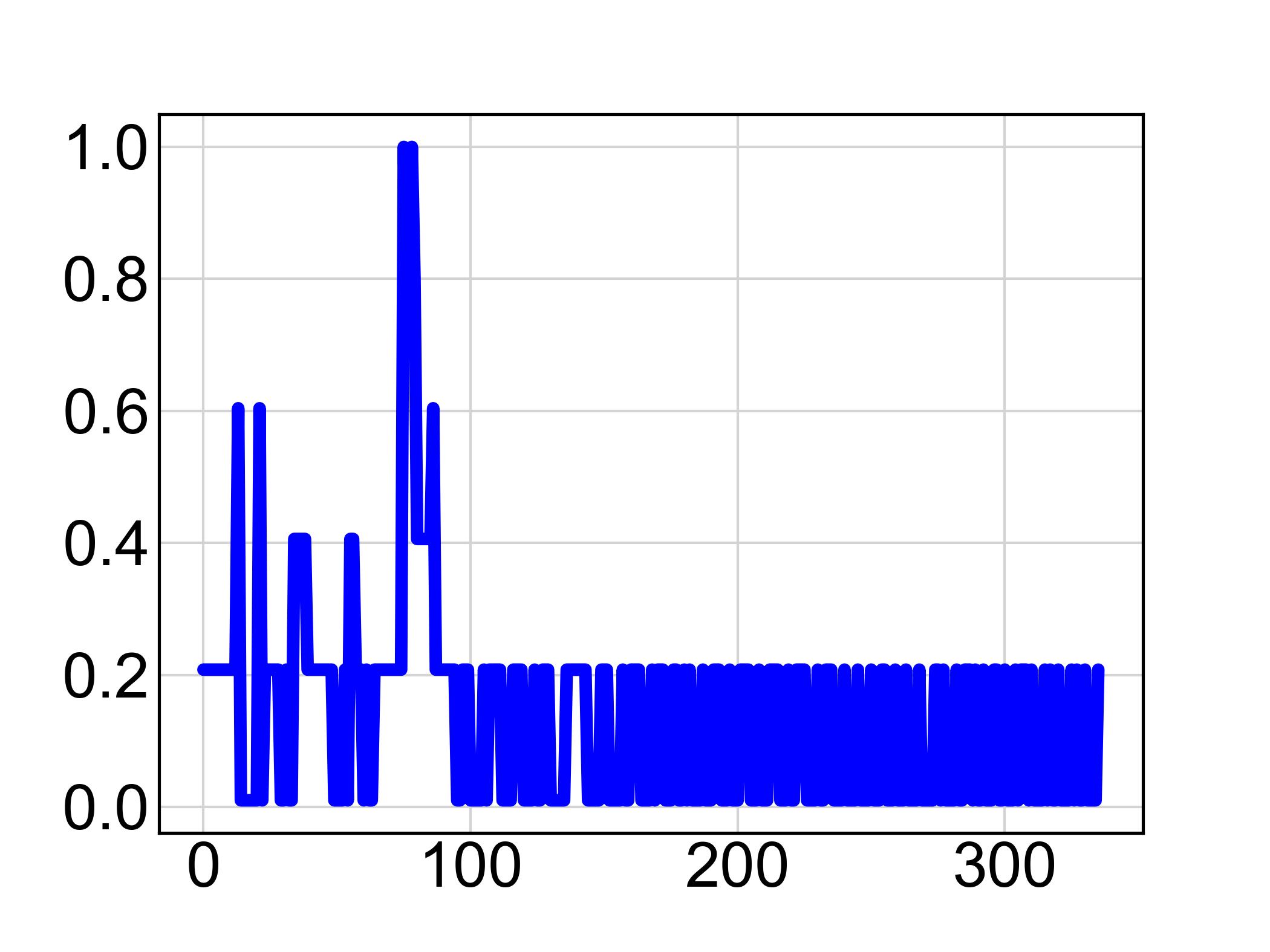}\put(-110,-7){\scriptsize \textcolor{black}{Culture period [day]}}\put(-162,32){\scriptsize \rotatebox{90}{\textcolor{black}{Relative feeding}}}} 
    \subfigure[]{\includegraphics[width=0.32\textwidth]{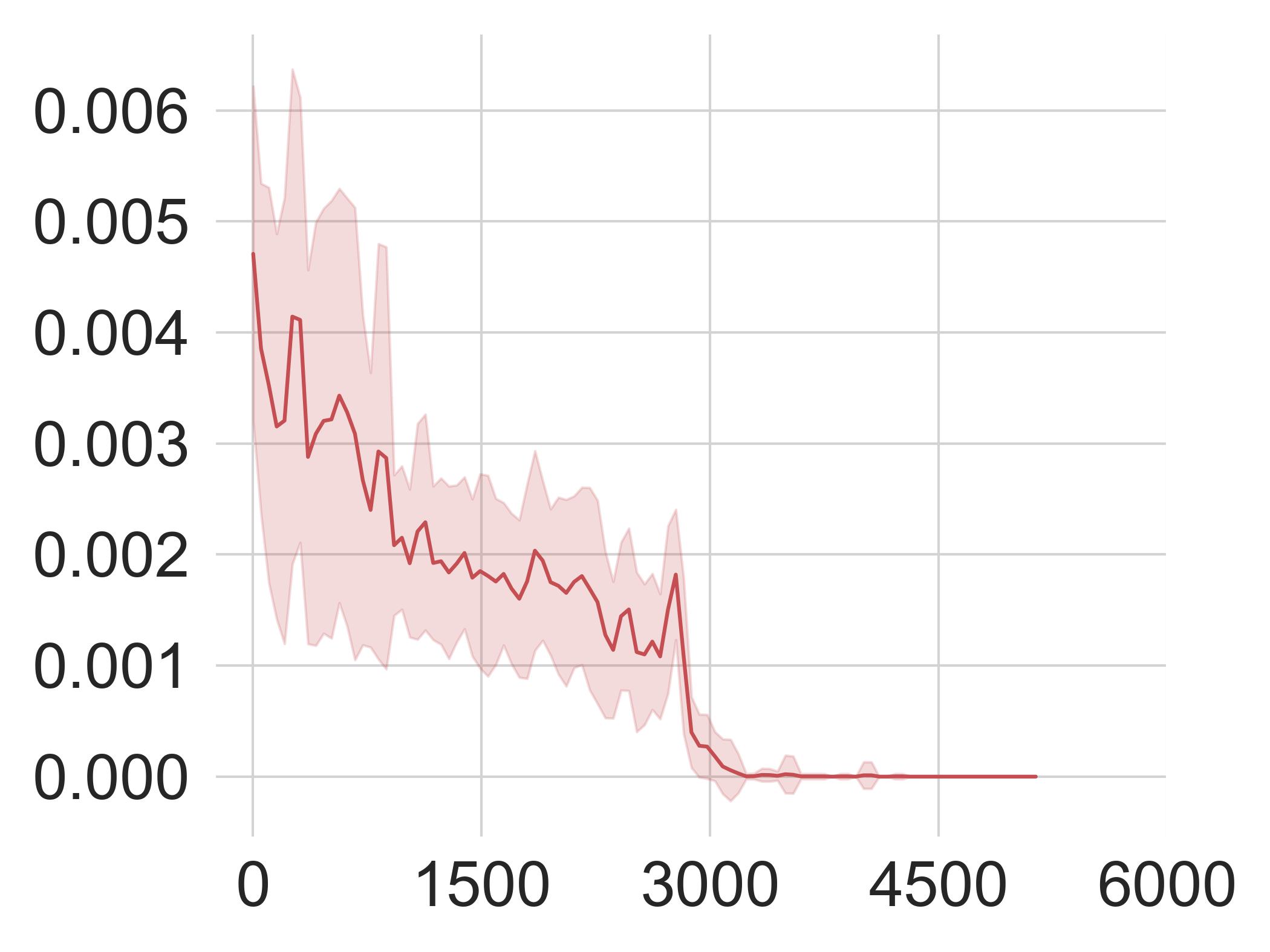}\put(-95,-7){\scriptsize \textcolor{black}{Episodes}}\put(-162,45){\scriptsize \rotatebox{90}{\textcolor{black}{Policy error}}}} 
    \subfigure[]{\includegraphics[width=0.32\textwidth]{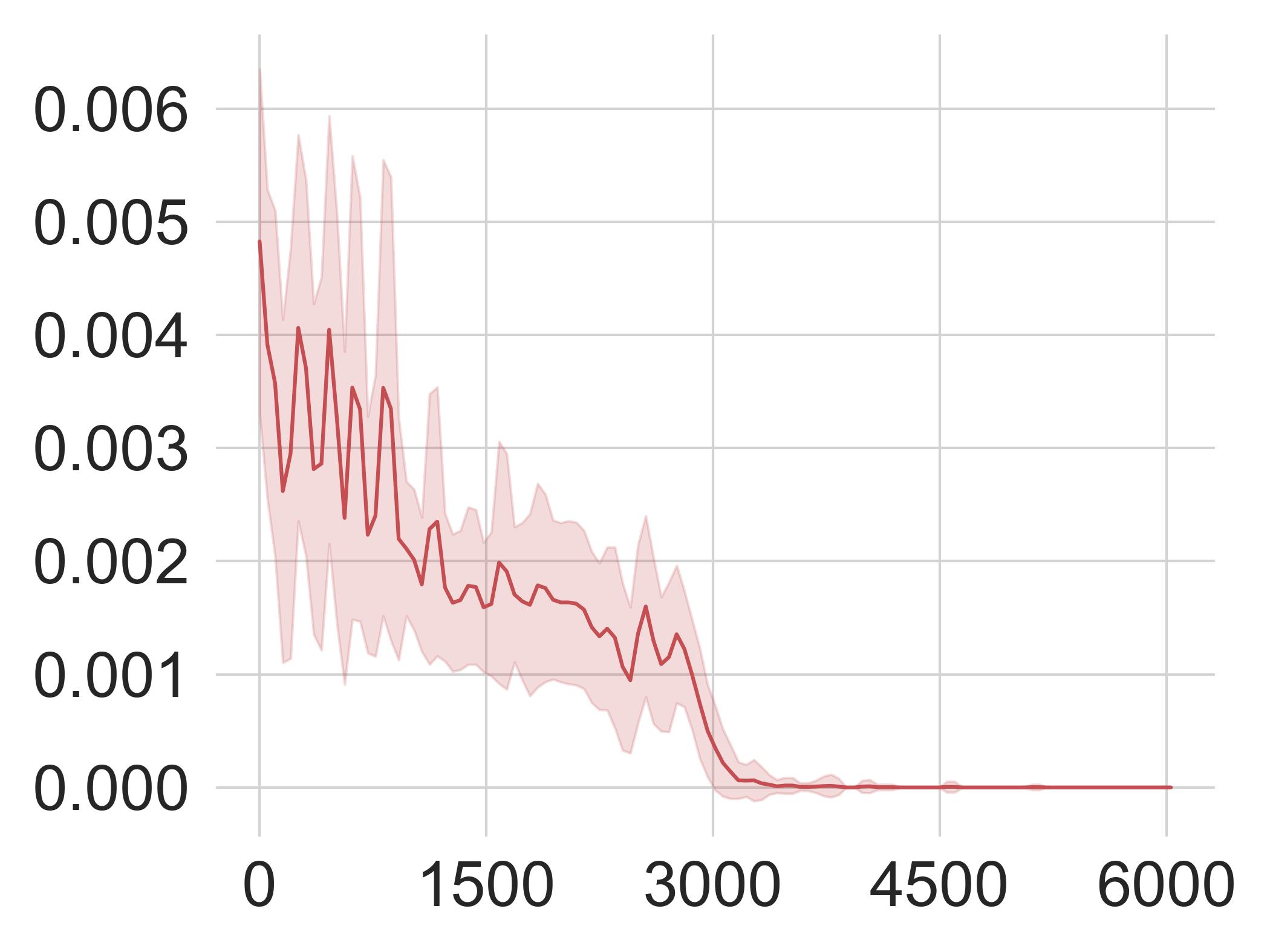}\put(-95,-7){\scriptsize \textcolor{black}{Episodes}}\put(-162,45){\scriptsize \rotatebox{90}{\textcolor{black}{Policy error}}}} 
    \subfigure[]{\includegraphics[width=0.32\textwidth]{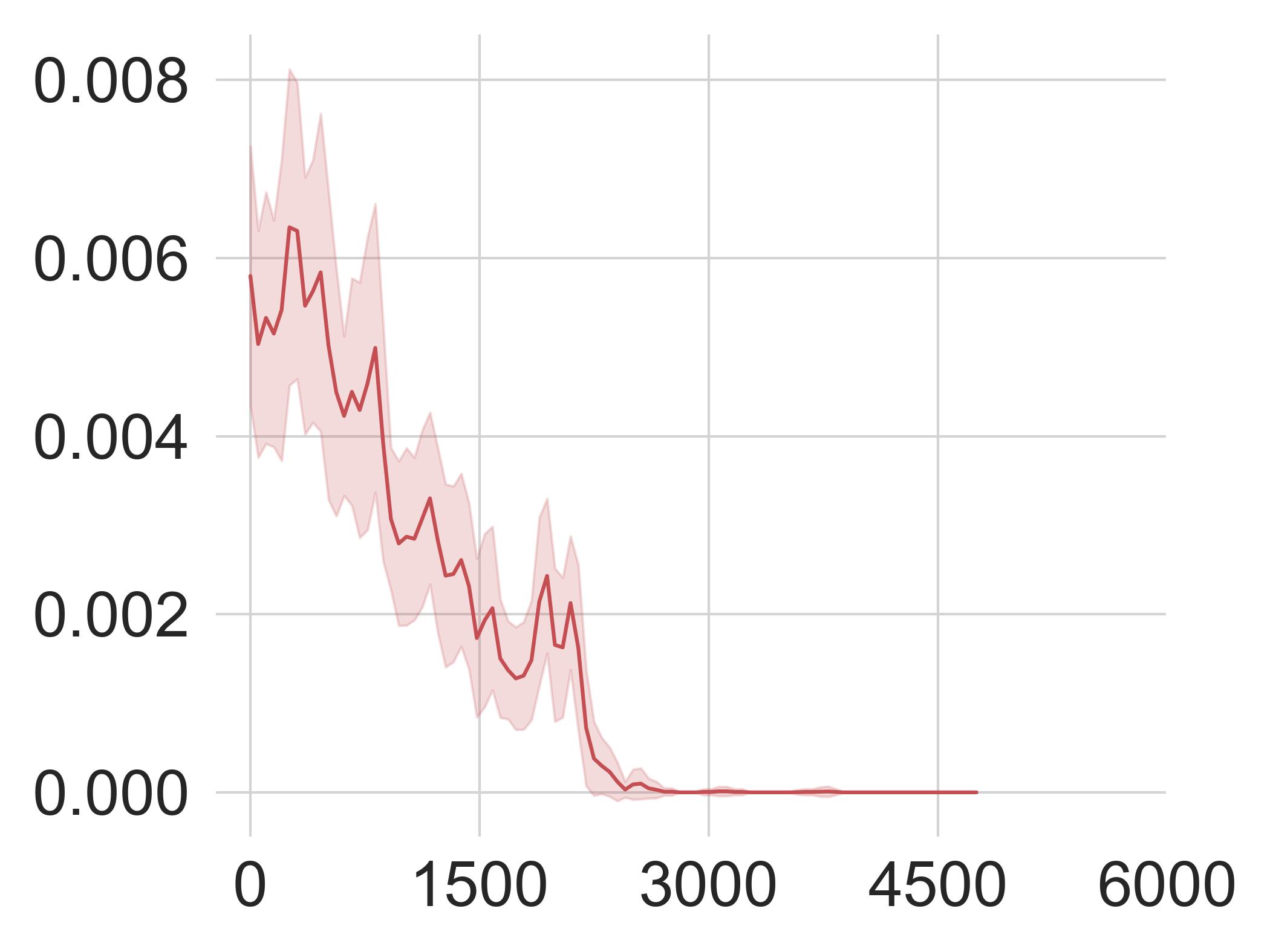}\put(-95,-7){\scriptsize \textcolor{black}{Episodes}}\put(-162,45){\scriptsize \rotatebox{90}{\textcolor{black}{Policy error}}}} \vspace{-0.35cm}
    \caption{\footnotesize \textcolor{black}{Performance results of Q-Learning under the three cases when the fish population is $10$. Figures (a), (b), and (c) illustrates  fish individual growth weight tracking for case 1, 2, and 3, respectively. Figures (d), (e), and (f) show the result of the manipulated variable (relative feeding) for case 1, case 2, and case 3, respectively. Figures (g), (h), and (i) present the policy improvement during the training episodes.}
    }
    \label{QL}
\end{figure*}

\noindent \textcolor{black}{On the other hand, Figure~\ref{QL} illustrates the fish growth tracking performance results using Q-Learning when the fish population is $10$ under different levels of unionized ammonia (UIA) exposure. Figs.~\ref{QL} (a), (b), and (c) show the fish individual growth weight tracking for cases 1, 2, and 3, respectively. Figs. (d), (e) and (f) show the result of the relative feeding variable for case 1, case 2, and case 3, respectively. Figs.~\ref{QL}  (g), (h), and (i) present the policy improvement during the training episodes. We note that the proposed Q-learning feeding control prevents fish mortality to some degree compared to the model-based approaches and achieves better tracking errors of the fish weight for all the different levels of unionized ammonia exposures. However, Q-learning provides a feeding policy and maintains a relative food consumption that potentially underfeeds the fish.}
\textcolor{black}{
\begin{remark}
% $\lambda=0.1$ and
The discretization of system (\ref{sys00}) for Q-learning is sensitive to convert continuous dynamics to finite sets. Using a variable resolution grid scheme, introduced in \cite{Kushner1992}, improves the performance of Q-learning.% where $\Delta t= 7$ days and $\lambda=0.1$. 
\end{remark}
}
  \begin{table*}[!t]
\caption{\textcolor{black}{Performance assessment of different controllers for the three cases (UIA constant, varying, and varying with a spike)}}
\textcolor{black}{
\begin{center}
\begin{tabular}{| c || c  | c | c | c | c |}
  \hline
\bf{Cases} & ~~\bf{Controller}~~ & ~~\bf{Fish mortality}~~ & ~~~~\bf{RMSE}~~~~ & ~~\bf{Food consumption [$\si{g}$]}~~ \\ 
\hline
Case 1& Bang-Bang & 0/10 & $14.109\%$  & $3479.7$\\
 &  PID & 0/10 & $15.627\%$ & $3245.5$\\
 & MPC$^1$ & 0/10 & $14.878\%$ & $2871.9$\\ 
  & \textcolor{black}{Q-learning} & \textcolor{black}{0/10} & \textcolor{black}{$11.71\%$} & \textcolor{black}{$718.28$}\\
%& RL+MPC &  &  & \\
 \hline
~~Case 2~~& ~~Bang-Bang~~ & 0/10 & $15.645\%$ & $4054$\\
 & PID & 0/10 & $18.385\%$ & $3249.8$\\
 & MPC$^1$ & 0/10 & $17.69\%$ & $3013.8$\\
  & \textcolor{black}{Q-learning} & \textcolor{black}{0/10} & \textcolor{black}{$10.98\%$} & \textcolor{black}{$668.06$}\\
%& RL+MPC &  &  & \\
 \hline
Case 3& ~Bang-Bang & 1/10 & $27.72\%$ & $2835.4$\\
 &  PID& 1/10 & $22.817\%$ & $3291$\\
 & MPC$^1$ & 1/10 & $19.914\%$ & $2928.5$\\
  & \textcolor{black}{Q-learning} & \textcolor{black}{0/10} & \textcolor{black}{$12.09\%$} & \textcolor{black}{$706.99$} \\
%& RL+MPC$^1$ &  &  & \\
\hline
Case 3& ~Bang-Bang & 4/25 & $56.428\%$ & $4607.5$\\
 &  PID& 4/25 & $57.799\%$ & $4106$\\
 & MPC$^1$ & 4/25 & $57.0144\%$ & $3728.6$\\
   & \textcolor{black}{Q-learning} & \textcolor{black}{0/25} & \textcolor{black}{$14.74\%$} & \textcolor{black}{$979.85$}\\
%& RL+MPC$^1$ &  &  & \\
 \hline
\end{tabular}
\end{center}
\label{Performance_assessment_1}
\vspace{0.5cm}
}
\end{table*}

\section{Optimal feeding and water quality monitoring (MPC$^2$)}\label{MPC2}
\noindent As presented above, a high concentration of exposure to UIA that is reflected by a spike in case 3 results in fish mortality by only controlling the feeding along with the density. To fill this gap in fish mortality related to UIA exposure under conditions of fish stocking density at which the fish growth rate tracks the desired growth reference, we design an optimal feeding and water quality monitoring that includes the temperature, dissolved oxygen (DO), and UIA in the objective function of MPC$^1$. 
We formulate this second model predictive control (called MPC$^2$) as follows. 
%  \begin{subeqnarray}\label{mpc2_tracking1}
% &&\!\!\!\!\!\!\!\!\!\!\!\!\!\!\!\!\displaystyle   \min_{{u \in \mathcal{U}(\varepsilon)}}\!\!J\!=\!\!\int_{t_k}^{t_{k+N}}\!\! \Bigg (\Big \| \frac{\tilde{w}(\tau)-w_d(\tau)}{w_d(\tau)} \Big \|^2 \!+\! \lambda_1 \Big \| f(\tau) \Big \|^2 \notag \\
% &&~~~~~+\lambda_2 \Big \| T(\tau)-T_{d} \Big \|^2+\lambda_3 \Big \| DO(\tau) - DO_{d} \Big \|^2 \notag \\
% && ~~~~~+\lambda_4 \Big \| \frac{UIA(\tau)-UIA_d}{UIA_d} \Big \|^2\Bigg )\der \tau\!\!\!\! \label{mpc2_tracking_a1}\\  
% &&\!\!\!\!\!\mbox{s.t}\quad \dot{\tilde{w}}(t)= g\big(\tilde{w}(t),u(t)\big) \label{mpc2_tracking_b1}\\
% &&\!\!\!\!\! u_{\mbox{\scriptsize min}}\leqslant u(t) \leqslant u_{\mbox{\scriptsize max}}, \quad \forall t \in[t_k,\, t_{k+N}]\label{mpc2_tracking_c1}\\
% &&\!\!\!\!\! \Delta u(t_k)= u(t_k)-u(t_{k-1})\slabel{mpc2_tracking_d1}\\
%  & &\!\!\!\!\! w_0 \leqslant \tilde{w}(t) \leqslant w_{\mbox{\scriptsize end}} , \quad \forall t \in[t_k,\, t_{k+N}]\label{mpc2_tracking_e1} \\
% &&\!\!\!\!\! \tilde{w}(t_k)=w(t_k), \quad \tilde{w}(0) =w(t_0) \slabel{mpc2_tracking_f1}
% \end{subeqnarray}
\begin{subequations}\label{mpc2_tracking1}
\begin{align}
& \min_{{u \in \mathcal{U}(\varepsilon)}} J = \int_{t_k}^{t_{k+N}} \Bigg (\Big \| \frac{\tilde{w}(\tau)-w_d(\tau)}{w_d(\tau)} \Big \|^2 + \lambda_1 \Big \| f(\tau) \Big \|^2 \notag \\
& ~~~~~+\lambda_2 \Big \| T(\tau)-T_{d} \Big \|^2+\lambda_3 \Big \| DO(\tau) - DO_{d} \Big \|^2 \notag \\
& ~~~~~+\lambda_4 \Big \| \frac{UIA(\tau)-UIA_d}{UIA_d} \Big \|^2\Bigg ) \,d\tau \label{mpc2_tracking_a1} \\  
& \text{s.t.}\quad \dot{\tilde{w}}(t)= g\big(\tilde{w}(t),u(t)\big) \label{mpc2_tracking_b1} \\
& u_{\text{min}}\leqslant u(t) \leqslant u_{\text{max}}, \quad \forall t \in[t_k,\, t_{k+N}] \label{mpc2_tracking_c1} \\
& \Delta u(t_k)= u(t_k)-u(t_{k-1}) \label{mpc2_tracking_d1} \\
& w_0 \leqslant \tilde{w}(t) \leqslant w_{\text{end}} , \quad \forall t \in[t_k,\, t_{k+N}] \label{mpc2_tracking_e1} \\
& \tilde{w}(t_k)=w(t_k), \quad \tilde{w}(0) =w(t_0) \label{mpc2_tracking_f1}
\end{align}
\end{subequations}

where the prediction horizon is selected as $N=5$, and the regularization parameters $\lambda_1=0.001$, $\lambda_2=0.2$, $\lambda_3=0.5$, and $\lambda_4=0.5$ are appropriately chosen to balance the objective function~\eqref{mpc2_tracking_a1}. $T_d$, $DO_d$ and $UIA_d$ are the desired references.
\begin{figure*}[!t]
    \centering
    \subfigure[]{\includegraphics[width=0.32\textwidth]{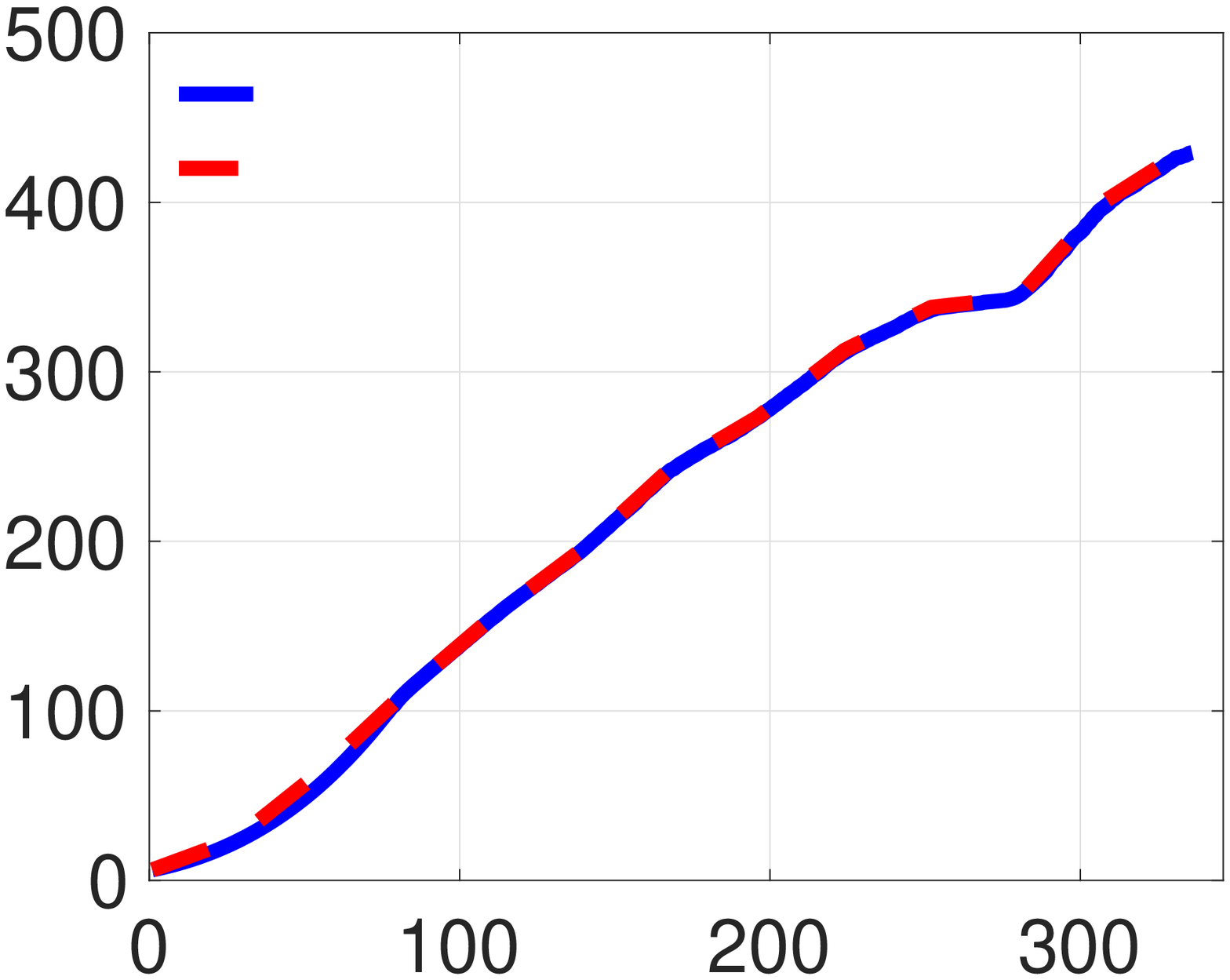}\put(-165,15){\scriptsize \rotatebox{90}{\textcolor{black}{Individual weight [$\si{g \per fish}$]}}}\put(-110,-7){\scriptsize \textcolor{black}{Culture period [day]}}\put(-123,102){\scriptsize MPC$^2$}\put(-123,93){\scriptsize Experimental}}
    \subfigure[]{\includegraphics[width=0.32\textwidth]{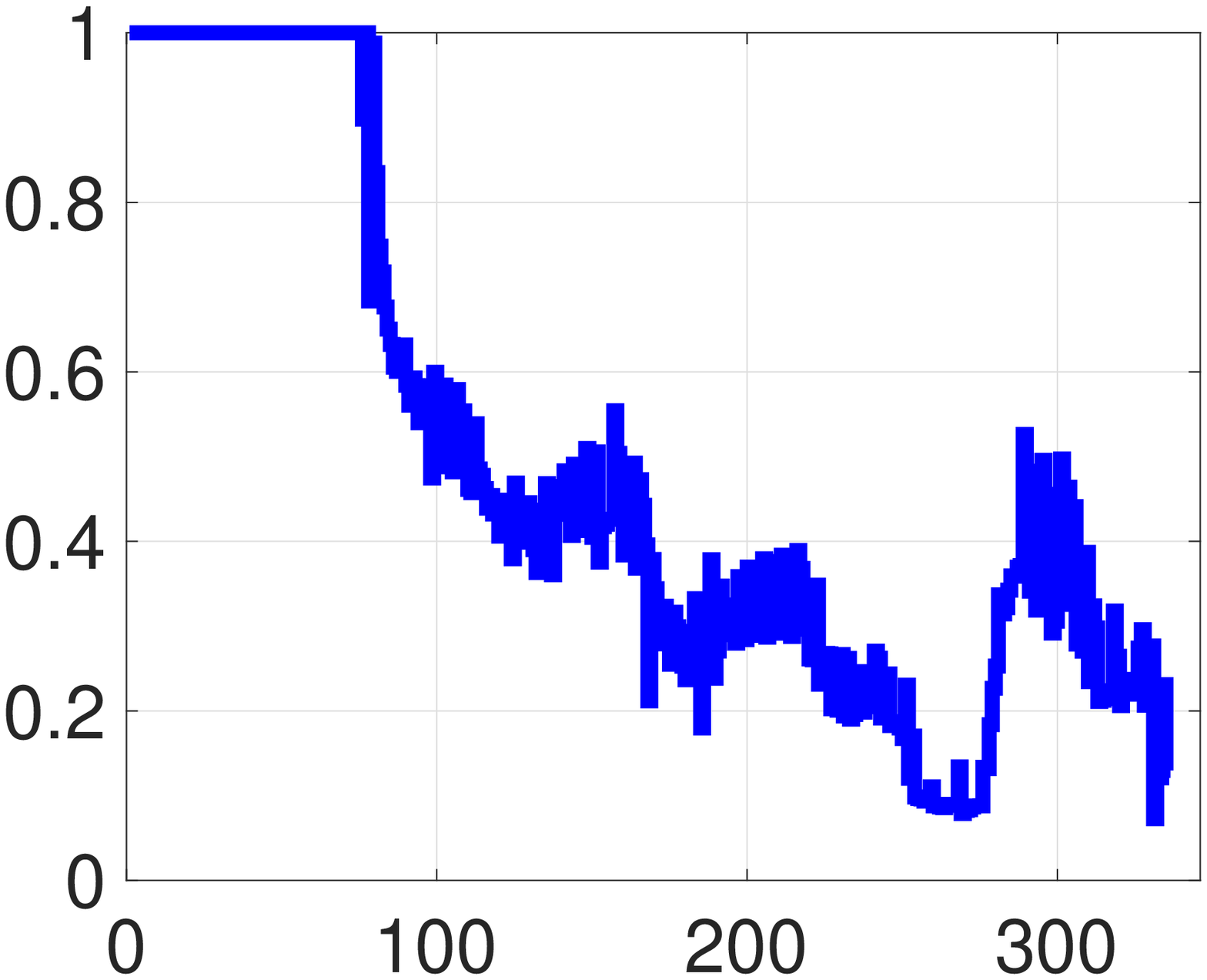}\put(-165,32){\scriptsize \rotatebox{90}{\textcolor{black}{Relative feeding}}}\put(-110,-7){\scriptsize \textcolor{black}{Culture period [day]}}} 
    \subfigure[]{\includegraphics[width=0.32\textwidth]{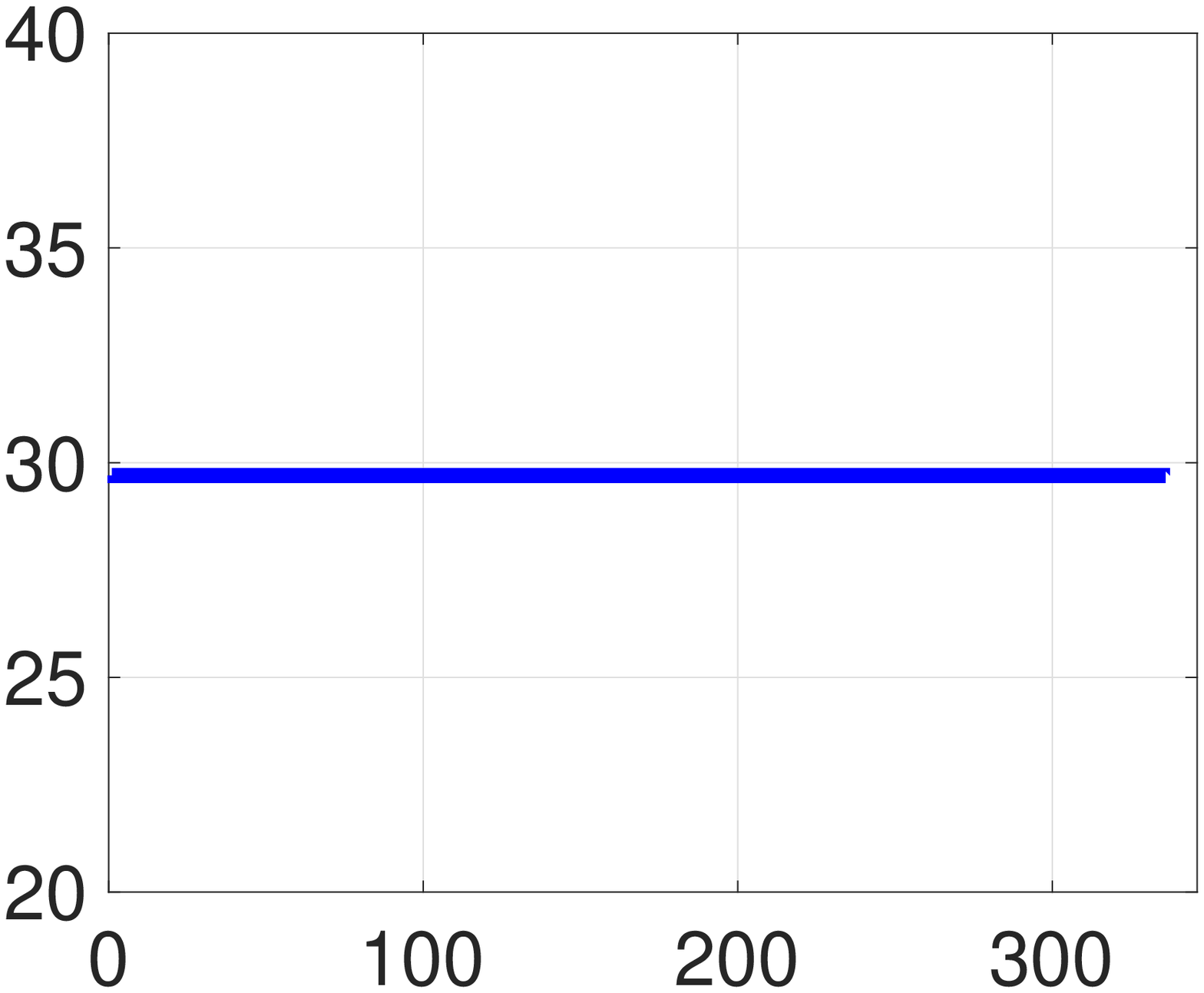}\put(-165,30){\scriptsize \rotatebox{90}{\textcolor{black}{Temperature [$^\circ C$]}}}\put(-110,-7){\scriptsize \textcolor{black}{Culture period [day]}}}
    \subfigure[]{\hspace{1cm}\includegraphics[width=0.32\textwidth]{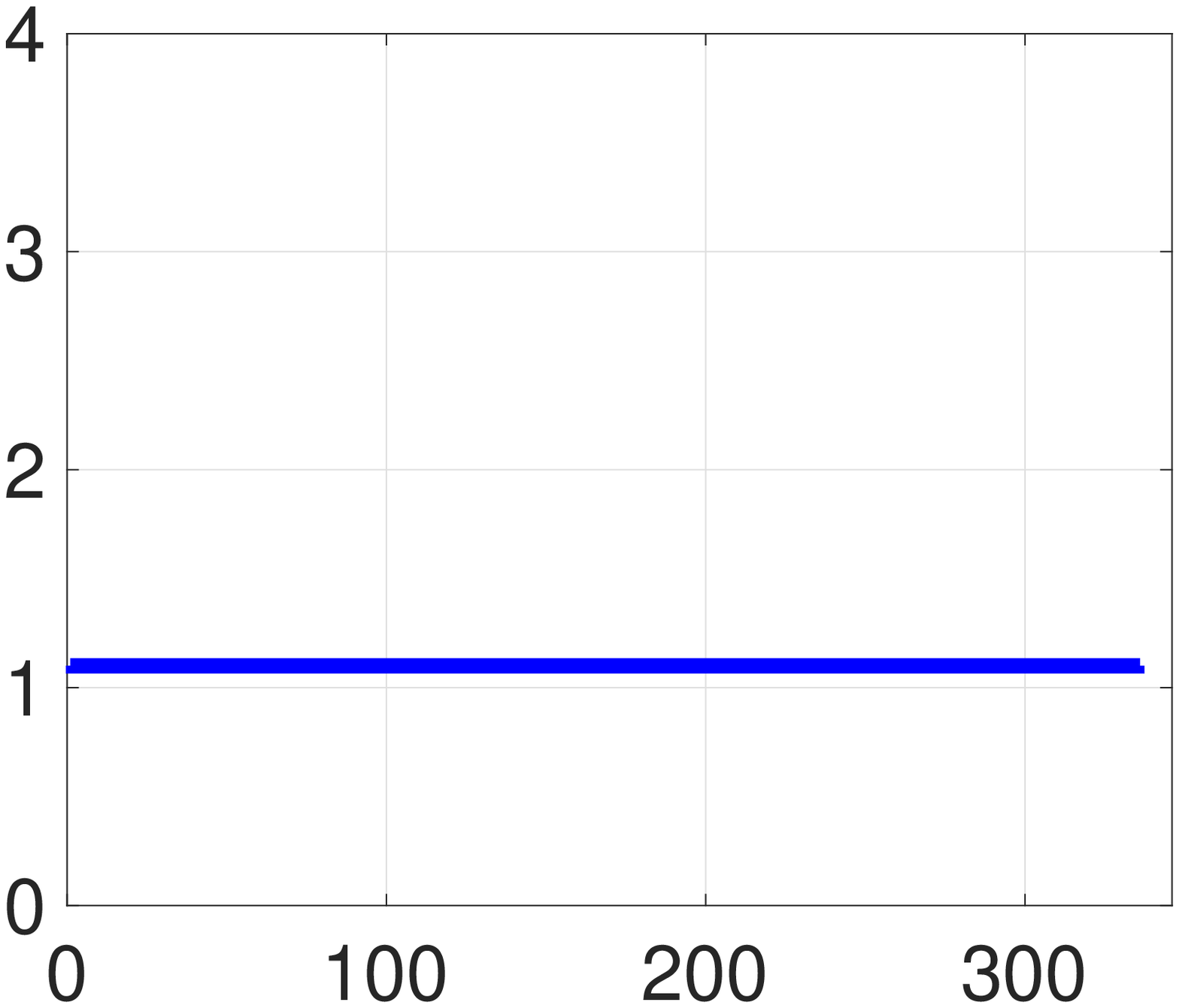}\put(-110,-7){\scriptsize \textcolor{black}{Culture period [day]}}\put(-160,20){\scriptsize \rotatebox{90}{\textcolor{black}{Dissolved oxygen [$\si{mg \per L}$]}}}} 
    \subfigure[]{\includegraphics[width=0.32\textwidth]{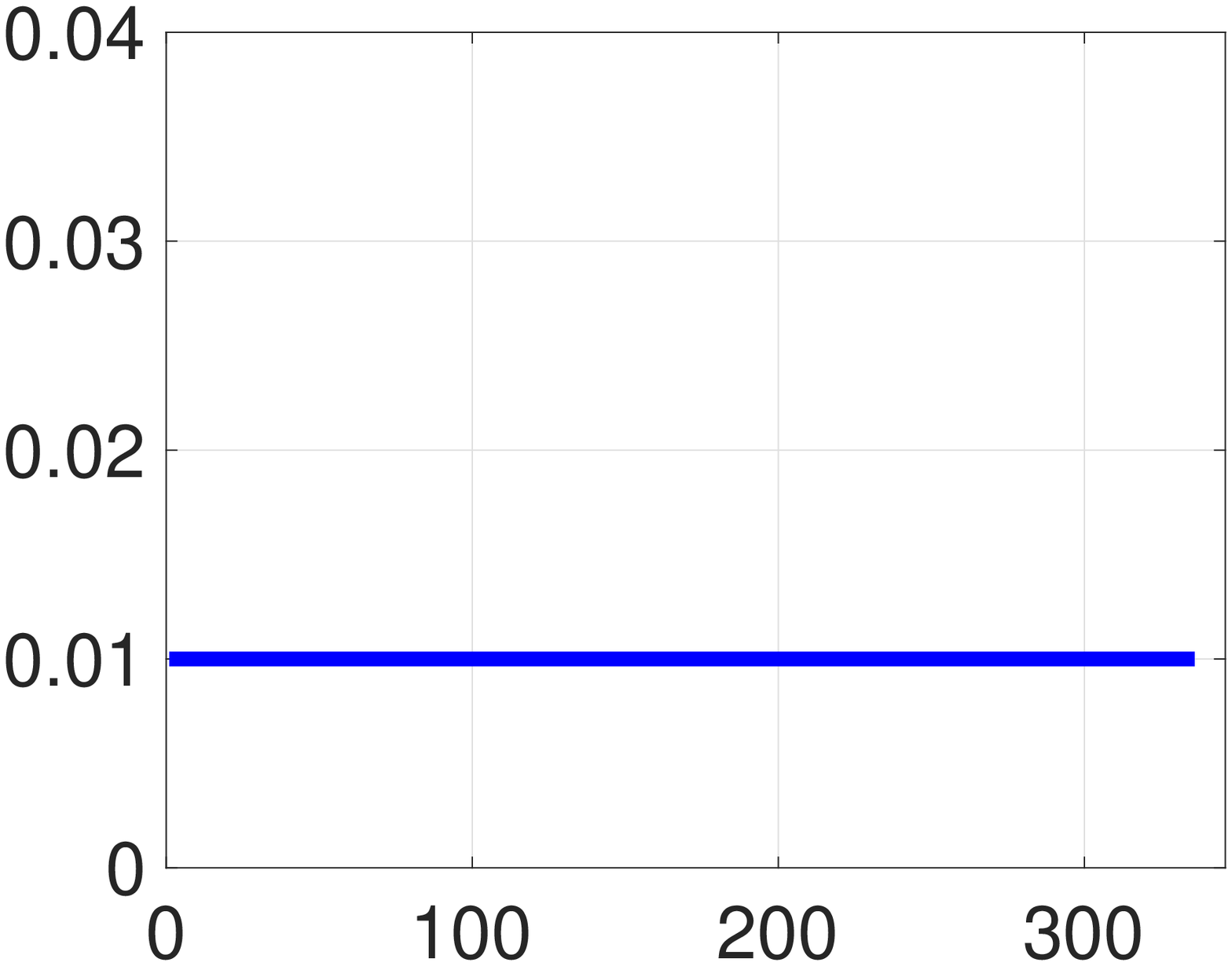}\put(-110,-7){\scriptsize \textcolor{black}{Culture period [day]}}\put(-166,12){\scriptsize \rotatebox{90}{\textcolor{black}{Unionized ammonia [$\si{mg \per L}$] }}}} 
    \subfigure[]{\includegraphics[width=0.32\textwidth]{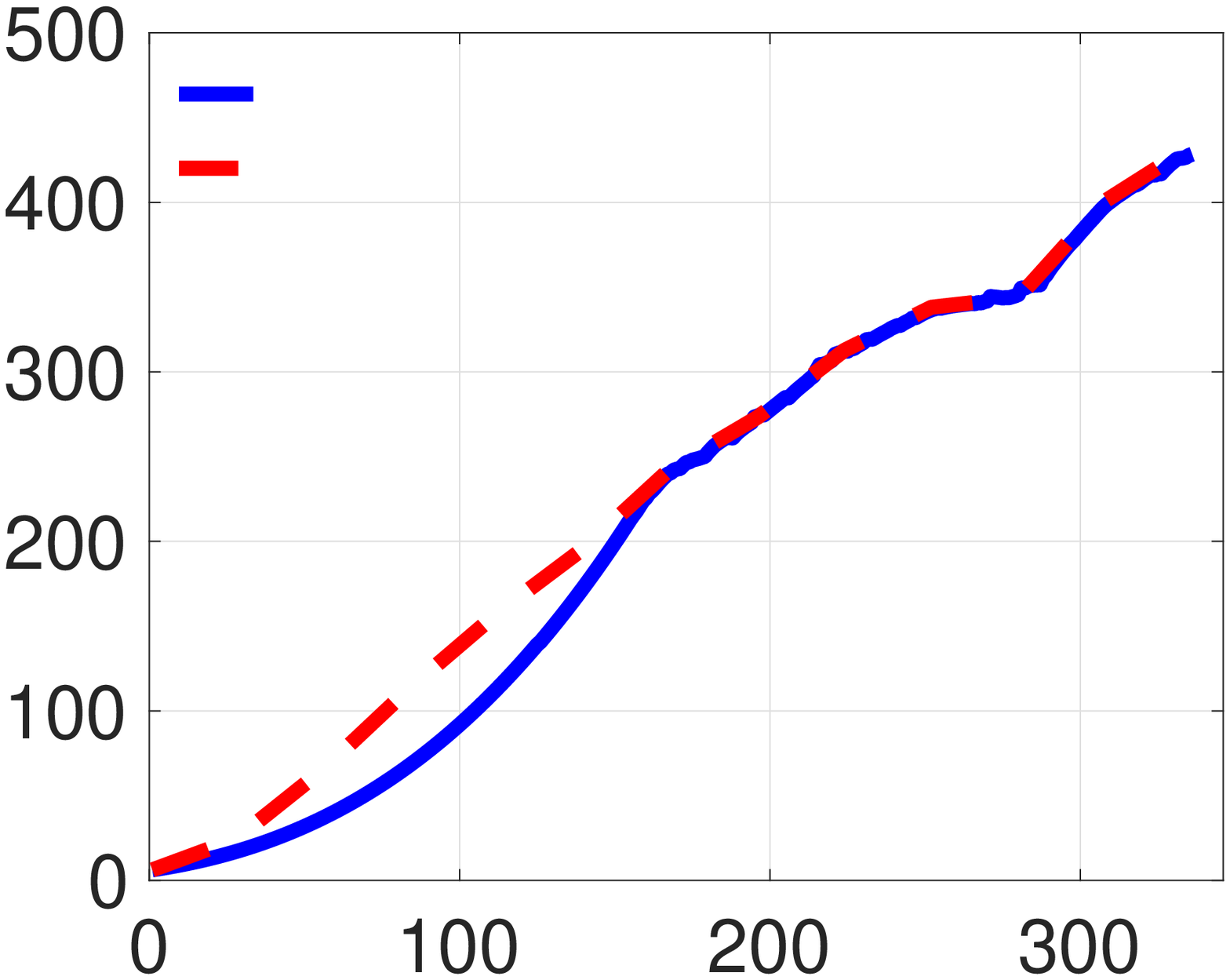}\put(-165,15){\scriptsize \rotatebox{90}{\textcolor{black}{Individual weight [$\si{g \per fish}$]}}}\put(-110,-7){\scriptsize \textcolor{black}{Culture period [day]}}\put(-123,102){\scriptsize MPC$^2$}\put(-123,93){\scriptsize Experimental}}
    \subfigure[]{\includegraphics[width=0.32\textwidth]{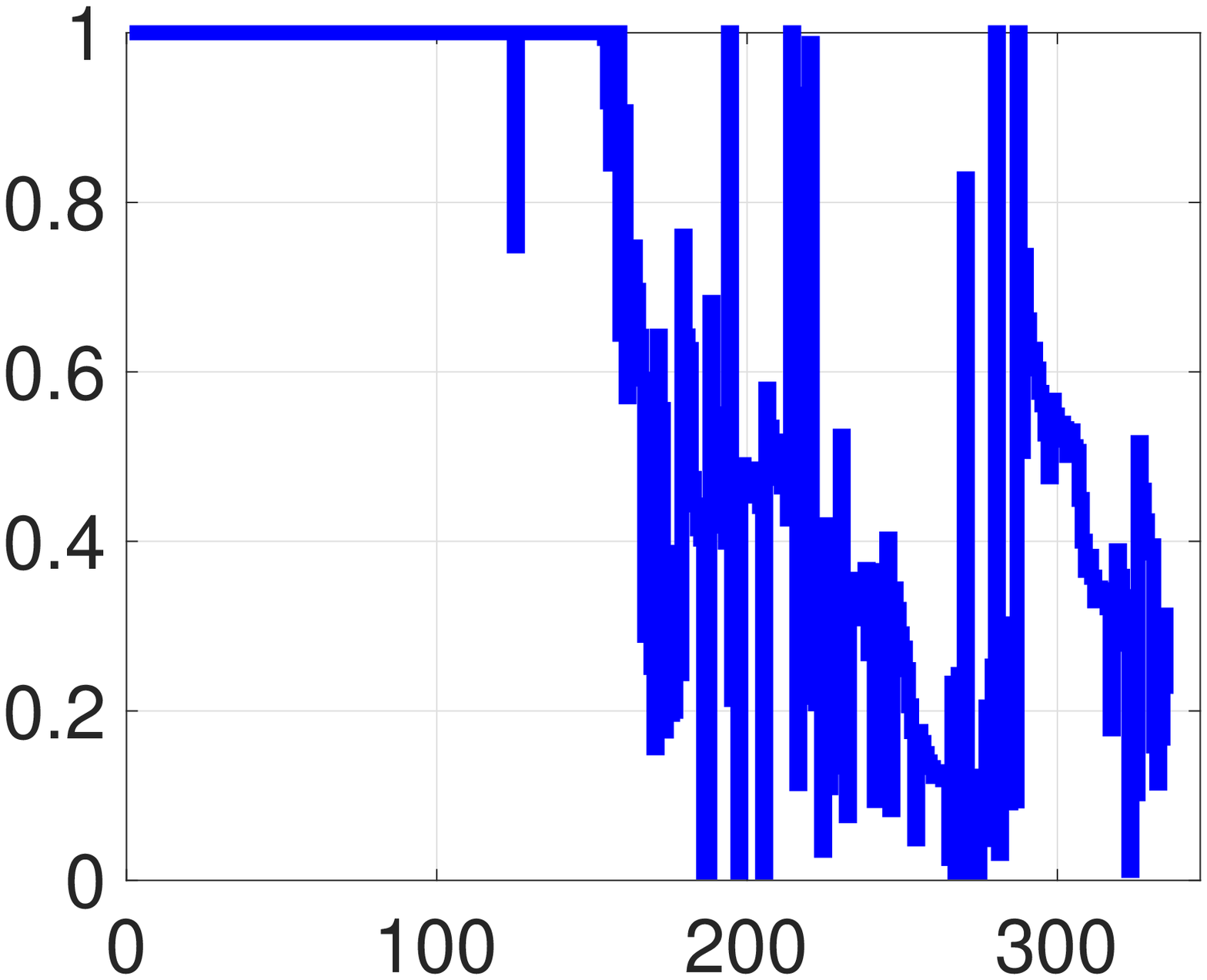}\put(-165,32){\scriptsize \rotatebox{90}{\textcolor{black}{Relative feeding}}}\put(-110,-7){\scriptsize \textcolor{black}{Culture period [day]}}} 
    \subfigure[]{\includegraphics[width=0.32\textwidth]{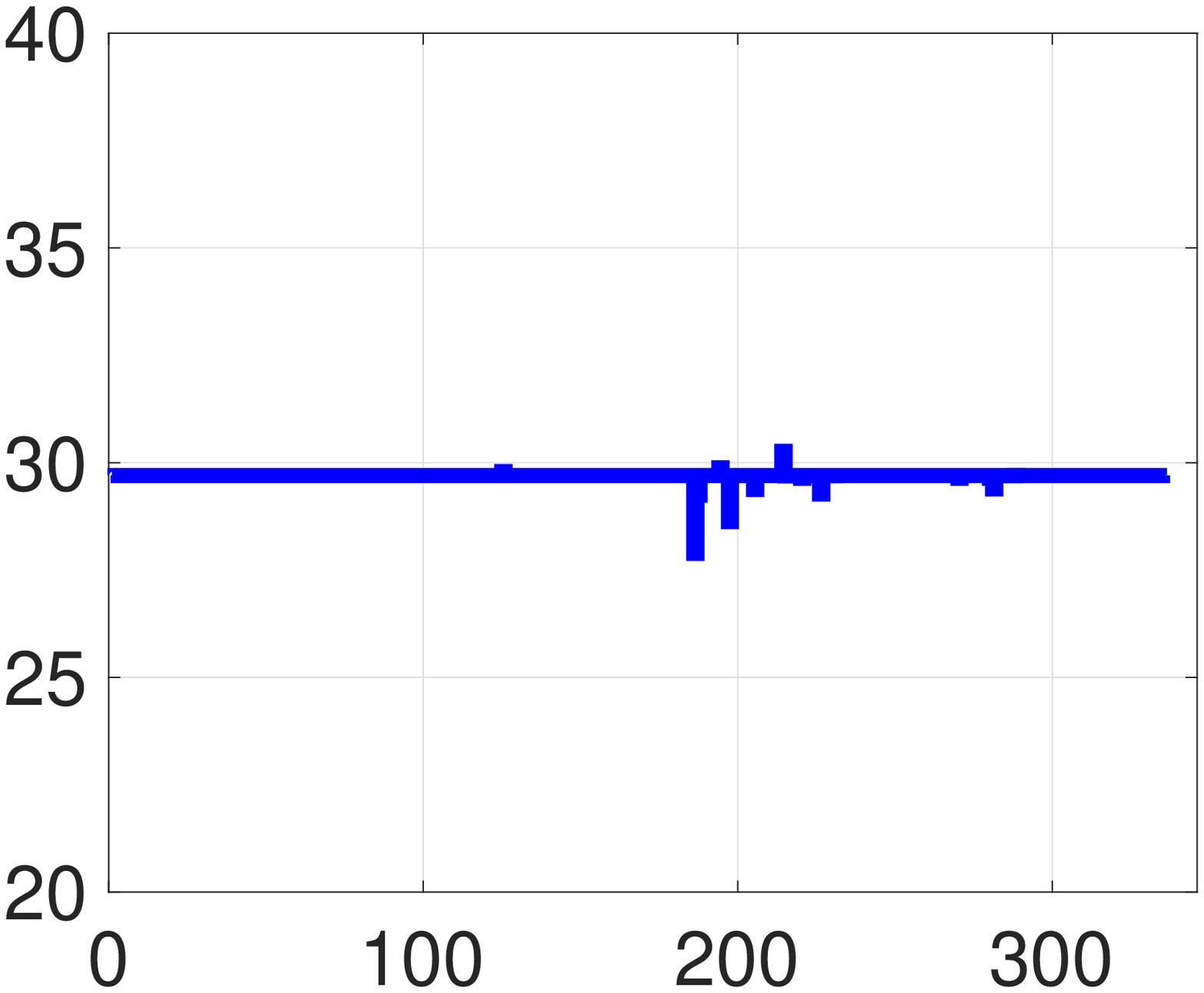}\put(-165,30){\scriptsize \rotatebox{90}{\textcolor{black}{Temperature [$^\circ C$]}}}\put(-110,-7){\scriptsize \textcolor{black}{Culture period [day]}}}
    \subfigure[]{\hspace{1cm}\includegraphics[width=0.32\textwidth]{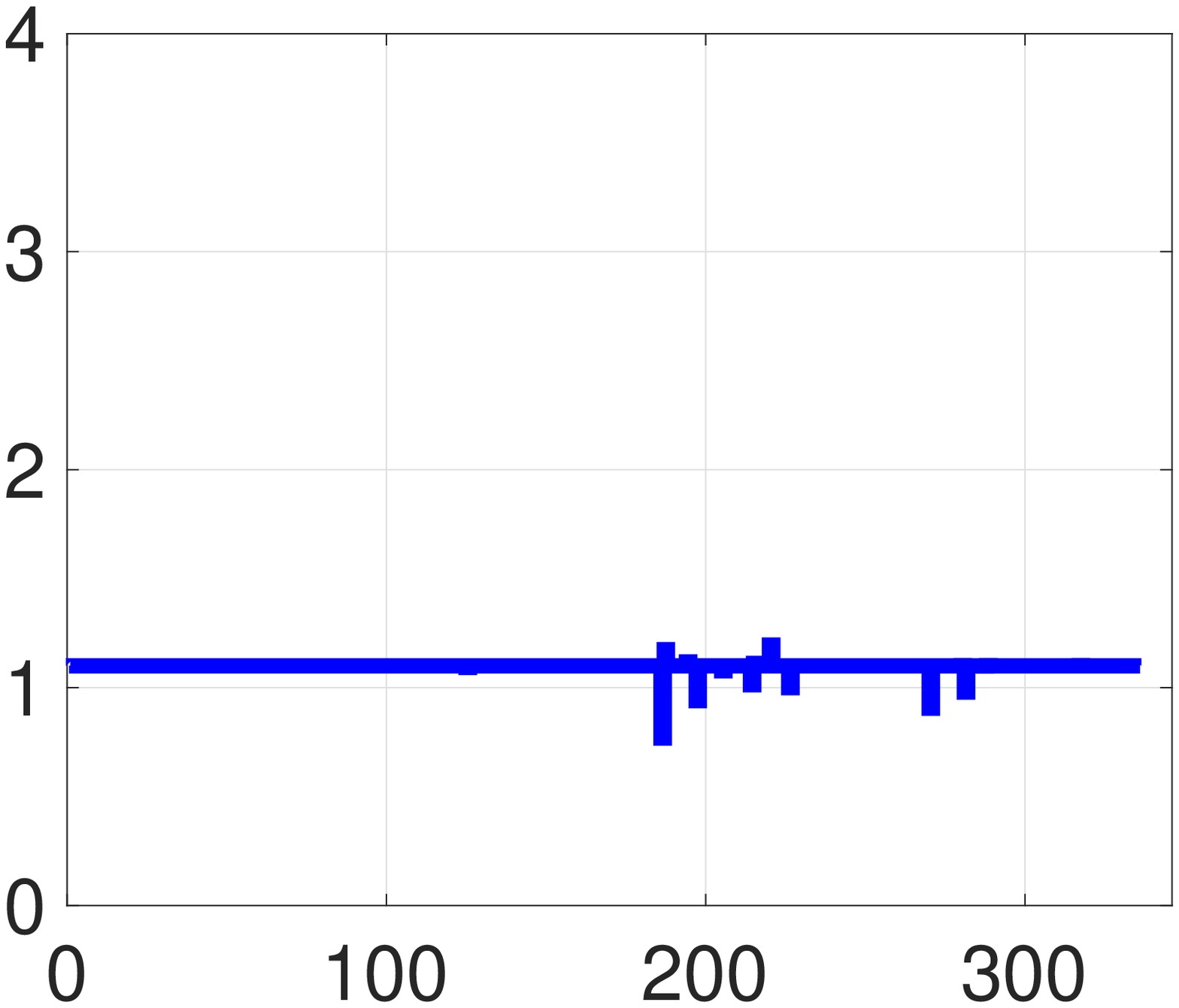}\put(-110,-7){\scriptsize \textcolor{black}{Culture period [day]}}\put(-160,20){\scriptsize \rotatebox{90}{\textcolor{black}{Dissolved oxygen [$\si{mg \per L}$]}}}} 
    \subfigure[]{\includegraphics[width=0.32\textwidth]{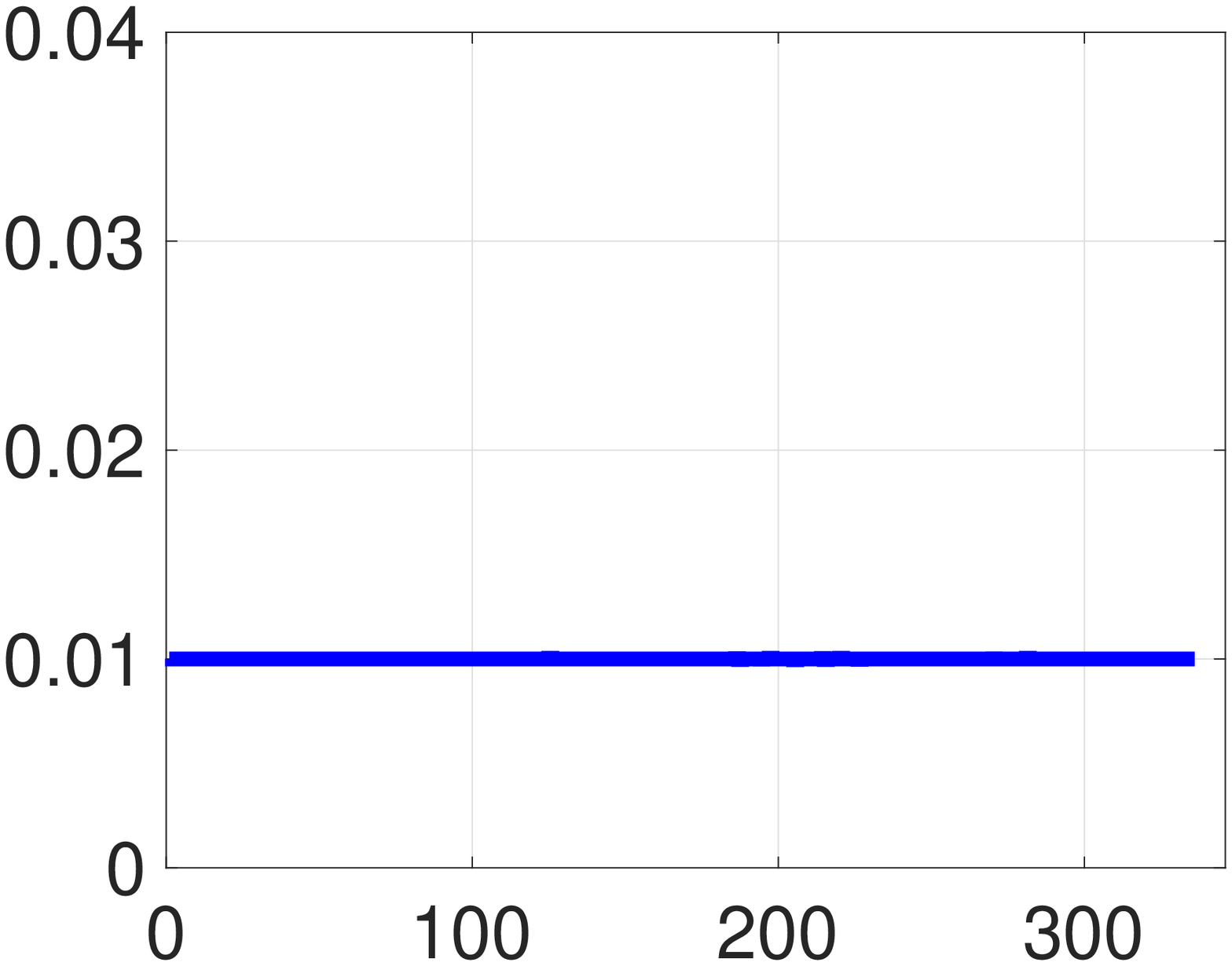}\put(-110,-7){\scriptsize \textcolor{black}{Culture period [day]}}\put(-166,12){\scriptsize \rotatebox{90}{\textcolor{black}{Unionized ammonia [$\si{mg \per L}$] }}}} \vspace{-0.35cm}
    \caption{\footnotesize \textcolor{black}{Performance results of MPC$^2$ when the fish population is equal to $10$ and $25$.}}
    \label{SpikeUIA2}
    \vspace{0.1cm}
\end{figure*}

\subsubsection*{Performance results of MPC$^2$}\label{discussion1}
\noindent Fig.~\ref{SpikeUIA2} illustrates the performance results of MPC$^2$ when the fish population growth dynamics are $10$ and $25$.  Figs.~\ref{SpikeUIA2}~(a)-(f),  Figs.~\ref{SpikeUIA2}~(b)-(g), and Figs.~\ref{SpikeUIA2}~(c)-(d)-(e) and  Figs.~\ref{SpikeUIA2}~(h)-(i)-(j) show the evolution of the fish growth trajectories, the feeding and the environmental factors for both stocking densities. Thanks to the flexibility of the model-predictive framework, MPC$^2$ achieves less food quantity and good tracking error performance for both stocking densities. Further, we notice from Table \ref{Performance_assessment_2} that the feeding and environmental control inputs ({\it i.e.} temperature, dissolved oxygen, and UIA) are well maintained around their desired references for both stocking densities, and the fish mortality remains zero.

 \begin{table}[!t]
\caption{\textcolor{black}{Performance assessment of MPC$^2$ with different fish population density}}
\begin{center}
\begin{tabular}{| c || c  | c | c | c | c |}
  \hline
 ~\bf{Fish mortality}~ & ~~~\bf{RMSE}~~~ & ~\bf{Food consumption [$\si{g}$]}~ \\ 
\hline
 0/10 & $2.723\%$ & $2487.9$\\
 0/25 & $21.335\%$ & $3311.4$\\ %\hline
\hline
\end{tabular}
\end{center}
\label{Performance_assessment_2}
\vspace{-0.6cm}
\end{table}

\section{Conclusion}\label{conclusion}
\noindent In this paper, we validated the new population dynamic fish growth model to achieve a sufficient overlap between the individual population fish growth data and the new dynamic model in which we consider the population state equal to one. Then, we designed classical and optimal control strategies, namely bang-bang, proportional-integral-derivative (PID), and model predictive control (MPCs) schemes, to track a desired fish growth trajectory and monitor the feeding and water quality. First, we focused on determining the best feeding regimen to design these controllers within the sub-optimal temperature and DO profiles under different levels of unionized ammonia (UIA) exposures. Then, we proposed an optimal algorithm that optimizes the feeding and water quality of the dynamic fish population growth process. This optimal control relies on MPC, thanks to its flexibility to incorporate constraints and inputs in the objective function. We also showed that the model predictive control simultaneously decreases fish mortality and reduces food consumption in all different cases by an average of $26.9\%$ compared to the bang-bang controller, $22.6\%$ compared to the PID controller, and $14.3\%$ compared to MPC$^1$ controller. Our findings revealed that the proposed MPC$^2$ optimizes the food consumption and enhances fish survival and growth while optimizing food consumption \textcolor{black}{while Q-learning policy provides a feeding policy and maintains a relative food consumption that potentially might underfeed the fish. For future works, we will investigate how model predictive control approach can support Q-learning framework to efficiently handle feeding constraint satisfaction and find better trajectories and policies from value-based reinforcement learning.}

%\newpage
\section{Appendix}\label{App}
\noindent The effects of temperature $\tau(T)$, unionized ammonia $v(UIA)$ and dissolved oxygen $\sigma(DO)$ on food consumption are described, respectively \cite{Yan:98}.
\begin{equation*}
\tau(T)=
\left\{\begin{array}{llll}
\displaystyle  \exp{\left\{-\kappa\Big(\dfrac{T-T_{opt}}{T_{\mbox{\tiny max}}-T_{opt}}\Big)^4\right\}} \quad \mbox{if}\quad T>T_{opt},\\
\exp{\left\{-\kappa \Big(\dfrac{T_{opt} -T}{T_{opt}-T_{\mbox{\tiny min}}}\Big)^4\right\}} \quad \mbox{if}\quad T<T_{opt},
\end{array}\right.
\end{equation*}
where $\kappa=4.6$.
\begin{equation*}
v(UIA)\!\!=\!\!
\left\{\begin{array}{llll}
\displaystyle 1 \qquad\qquad\qquad~\, \mbox{if} \quad UIA<UIA_{\mbox{\tiny crit}},\\
\!\!\dfrac{UIA_{\mbox{\tiny max}}-UIA}{UIA_{\mbox{\tiny max}} -UIA_{\mbox{\tiny crit}}} \, \mbox{if}\,  UIA_{\mbox{\tiny crit}}<  UIA< UIA_{\mbox{\tiny max}},\\
0 \qquad\qquad\qquad\quad \mbox{elsewhere}.
\end{array}\right.
\end{equation*}
\begin{equation*}
\sigma(DO)=
\left\{\begin{array}{llll}
\displaystyle 1 \qquad\qquad\qquad \mbox{if} \quad DO>DO_{\mbox{\tiny crit}},\\
\!\!\dfrac{DO - DO_{\mbox{\tiny min}}}{DO_{\mbox{\tiny crit}} - DO_{\mbox{\tiny min}}}~\, \mbox{if}\quad  DO_{\mbox{\tiny min}}\!<\!DO\!<\!DO_{\mbox{\tiny crit}},\\
0 \qquad\qquad\qquad \mbox{elsewhere}.
\end{array}\right.
\end{equation*}

\noindent The fish mortality coefficient $k_1$ depends on unionized ammonia ($UIA$) factor. It has a form of logistic regression as follows, 
\begin{equation*}
    k_1(UIA) = \frac{\mathcal{Z}}{1 + \exp\{-\beta(UIA - \eta)\}},
\end{equation*}
where $\mathcal{Z}$, $\beta$, and $\eta$ are the three tuning parameters in the logistic fitting. The fish mortality points are extracted from real-experiment in \cite{Sink2010} and the three tuning parameters turned out to be $\mathcal{Z}=99.41$, $\beta=10.36$, and $\eta=0.80$. Fig.~\ref{UIA_mortality} illustrates the extracted points and the logistic fitting.
 \begin{figure}[!h]
 \vspace{0.3cm}
\centering
      \begin{overpic}[scale =0.18]{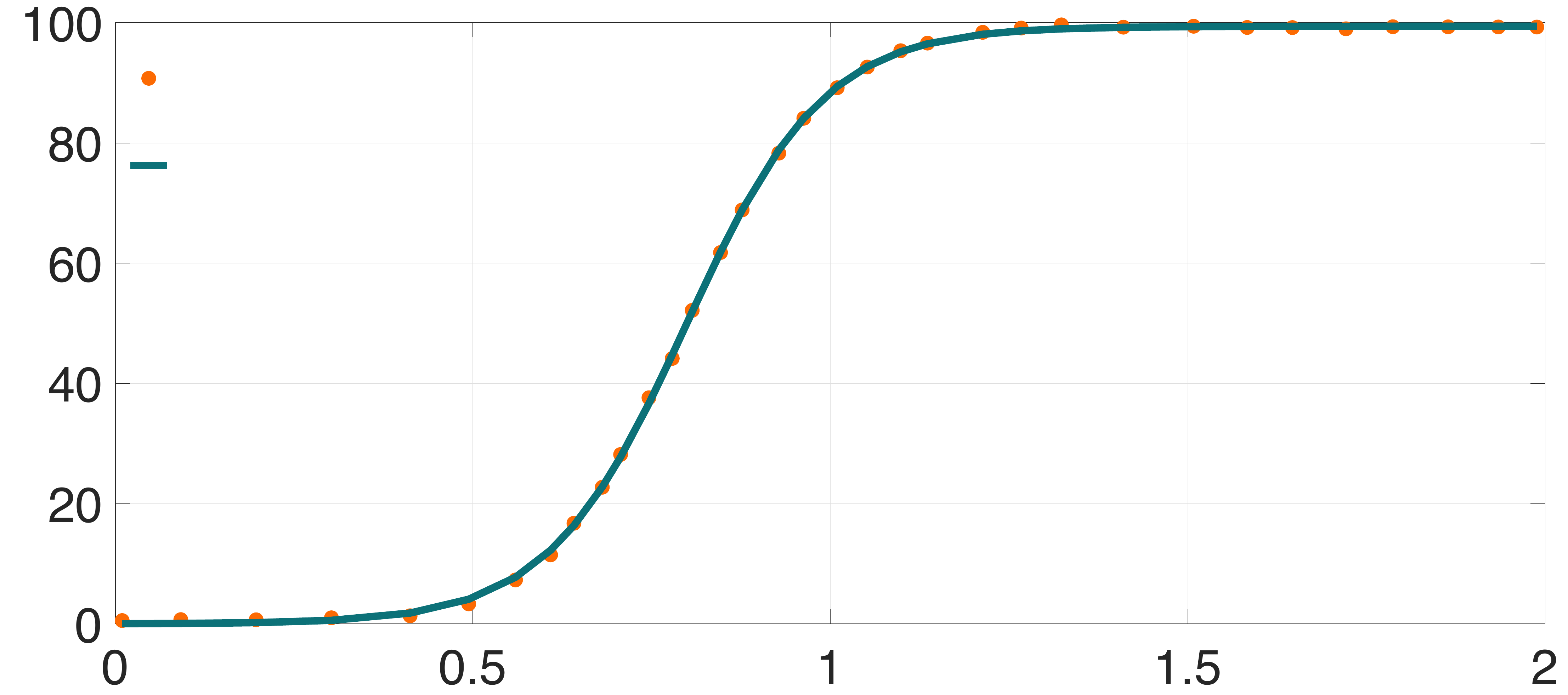}
      \end{overpic}
      \put(-235,35){\scriptsize \rotatebox{90}{Mortality $\%$}}
      \put(-200,89){\scriptsize Extracted data}
      \put(-200,76){\scriptsize Logistic model}
      \put(-130,-7){\scriptsize UIA (mg/L) }
      \caption{Relation between fish mortality coefficient and UIA}\label{UIA_mortality}  
 \end{figure}
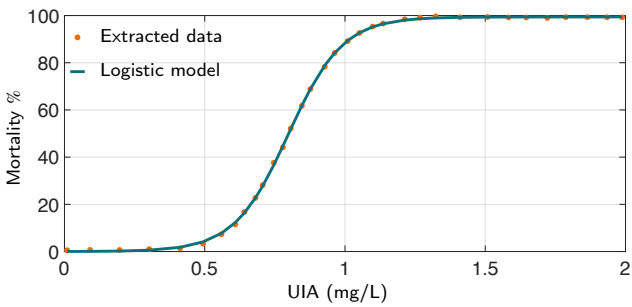
\begin{table}[!t]
\begin{scriptsize}
\caption{Nomenclature and main parameters of the growth model}
\begin{center}
\begin{tabular}{| c || c || c |}
  \hline %
\textbf{~Symbol~}  & \textbf{~~~~Description~~~~}  & \textbf{~~~~Unit~~~~}   \\ \hline
$\xi$ &  Total fish biomass  & \si{kcal \per pond} \\
$p$ &  Total fish number  & \si{fish \per pond} \\
$p_s$ &  Stocking fish number  & \si{fish \per day\per pond}\\
$\xi_i$ &  Individual fish biomass during fish stocking  & \si{kcal \per fish} \\
$\xi_a$ &  Mean fish biomass & \si{kcal \per pond}\\
$k_1$ &  Fish mortality profile & - \\
~~~$m$~~~ &Exponent of body weight for net anabolism &{$0.67$}\\
~~$t$~~ & Time  & \si{day} \\  
$n$ & Exponent of body weight for fasting catabolism  &~~~~{$0.81$}~~~~  \\    \hline
$f$ &  Relative feeding rate &$0\!<\!f\!<\!1$ \\  
$T$ &~~Temperature  &\si{^0}\si{C} \\
$DO$ &  Dissolved oxygen & \si{mg/l}   \\  
$UIA$ &~Unionized ammonia~ & \si{mg/l} \\    \hline
$b$ &  Efficiency of food assimilation & {$0.62$}  \\  
$a$ & Fraction of the food assimilated &{$0.53$}\\  
$h$ &Coefficient of food consumption &$0.8$ \si{kcal^{1\!-\!m}}\si{/day}\\   
$k_{\mbox{\tiny min}}$ &~Coefficient of fasting catabolism~ &$0.00133$\si{N}\\  
$j$ &~Coefficient of fasting catabolism~ &0.0132\si{N}\\  
$T_{opt}$ & Optimal average level of water temperature &$33^0$\si{C}\\
$T_{\mbox{\tiny min}}$& Minimum level of temperature   &$24^0$\si{C}\\
$T_{\mbox{\tiny max}}$ &  Maximum level of temperature  &$40^0$\si{C}\\ 
$UIA_{\mbox{\tiny crit}}$ &~~~Critical limit of UIA~~~  &$0.06$\si{mg/l}\\
$UIA_{\mbox{\tiny max}}$ & Maximum level of UIA &$1.4$\si{mg/l}\\
$DO_{\mbox{\tiny crit}}$ & Critical limit of DO&$0.3$\si{mg/l}\\ 
$DO_{\mbox{\tiny min}}$ & Minimum level of DO &$1$\si{mg/l}\\ 
$r$ &  Daily ration  &\si{kcal/day}\\
$R$ & Maximal daily ration & $10\%$ \si{BWD}\\
$BWD$ & Average body-weight per day &\si{kcal/day}\\ \hline  
$\tau$ &~Temperature factor~ &$0\!<\!\tau\!<\!1$ \\ 
$\sigma$ &  Dissolved oxygen factor  & $0\!<\!\sigma\!<\!1$ \\  
$v$ &  un-ionized ammonia factor & $0\!<\!v\!<\!1$ \\  
$\rho$ & Photoperiod factor & $0\!<\!\rho\!<\!2$\\  \hline  
\end{tabular}
\end{center}
\label{para}
\end{scriptsize} 
\end{table}

%\balance
\bibliography{biblio}

\begin{thebibliography}{10}
\expandafter\ifx\csname url\endcsname\relax
  \def\url#1{\texttt{#1}}\fi
\expandafter\ifx\csname urlprefix\endcsname\relax\def\urlprefix{URL }\fi
\expandafter\ifx\csname href\endcsname\relax
  \def\href#1#2{#2} \def\path#1{#1}\fi

\bibitem{Fao:18}
FAO, The state of world fisheries and aquaculture, in: Meeting the sustainable
  development goals, Rome, 2018, pp. Licence CC BY--NC--SA 3.0 IGO.

\bibitem{Seg:16}
I.~Seginer, Growth models of gilthead sea bream ({S}parus aurata {L.}) for
  aquaculture: A review, Aquacultural Engineering 70 (2016) 15--32.

\bibitem{FlM:21}
S.~A. Flinn, S.~R. Midway, Trends in growth modeling in fisheries science,
  Fishes 6~(1) (2021) 1--18.

\bibitem{SHL:16}
M.~Sun, S.~Hassan, D.~Li, Models for estimating feed intake in aquaculture: a
  review, Computers and Electronics in Agriculture 127 (2016) 425--438.

\bibitem{SHOLS:19}
D.~Sousa, D.~Hernandez, F.~Oliveira, M.~Lu\'is, S.~Sargento, A platform of
  unmanned surface vehicle swarms for real time monitoring in aquaculture
  environments, Sensors 19~(21) (2019) 4695.

\bibitem{Fo:18}
M.~Fore, K.~Frank, T.~Norton, E.~Svendsen, J.~A. Alfredsen, T.~Dempster,
  H.~Eguiraun, W.~Watson, A.~Stahl, L.~M. Sunde, C.~Schellewald, K.~R. Skoien,
  M.~O. Alver, D.~Berckmans, Precision fish farming: A new framework to improve
  production in aquaculture, Biosystems engineering 173 (2018) 176--193.

\bibitem{LEE1995205}
P.~G. Lee, A review of automated control systems for aquaculture and design
  criteria for their implementation, Aquacultural Engineering 14~(3) (1995)
  205--227.

\bibitem{Zain2013}
B.~A. Md~Zain, M.~Md~Jamal, S.~md~salleh, Modelling and control of fish feeder
  system, Applied Mechanics and Materials 465-466 (2013) 1314--1318.

\bibitem{Pedro2020}
P.~Almeida, L.~Dewasme, A.~Vande~Wouwer, Denitrification control in a
  recirculating aquaculture system—a simulation study, Processes 8~(10)
  (2020).

\bibitem{optiz1}
J.~A. Mistiaen, I.~Strand, Optimal feeding and harvest time for fish with
  {weight-dependent} prices, Marine Resource Economics 13 (1998) 231--246.

\bibitem{Optz2}
F.~R. Kazmierczak, H.~{Rex Caffey}, Management ability and the economics of
  recirculating aquaculture production systems, Marine Resource Economics
  10~(2) (1995) 187--209.

\bibitem{optiz3}
T.~Heaps, The optimal feeding of farmed fish, Marine Resource Economics 8~(2)
  (1993) 89--99.

\bibitem{Hea:95}
T.~Heaps, Density dependent growth and the culling of farmed fish, Marine
  Resource Economics 10 (1995) 285--298.

\bibitem{CNMBL:22}
A.~Chahid, I.~N'Doye, J.~E. Majoris, M.~L. Berumen, T.~M. {Laleg-Kirati}, Fish
  growth trajectory tracking using {Q}-learning in precision aquaculture,
  Aquaculture 550 (2022) 737838.

\bibitem{CNMBL:21}
A.~Chahid, I.~N'Doye, J.~E. Majoris, M.~L. Berumen, T.~M. {Laleg-Kirati}, Model
  predictive control paradigms for fish growth reference tracking in precision
  aquaculture, Journal of Process Control 105 (2021) 160--168.

\bibitem{Yan:98}
Y.~Yang, A bioenergetics growth model for {N}ile tilapia (oreochromis
  niloticus) based on limiting nutrients and fish standing crop in fertilized
  ponds, Aquacultural Engineering 18~(3) (1998) 157--173.

\bibitem{Koo:12}
B.~Kooijman, Dynamic Energy Budget Theory for Metabolic Organisation, 3rd
  Edition, Springer-Verlag, Cambridge University Press, 2012.

\bibitem{LiS:08}
S.~Libralato, C.~Solidoro, A bioenergetic growth model for comparing sparus
  aurata's feeding experiments, Ecological Modelling 214~(2--4) (2008)
  325--337.

\bibitem{MHGEJ:20}
J.~Weidner, C.~{Hakonsrud Jensen}, J.~Giske, S.~Eliassen, C.~Jorgensen,
  Hormones as adaptive control systems in juvenile fish, Biology Open 9 (2020)
  38--52.

\bibitem{VLGTH:20}
C.-T. Venolia, R.~Lavaud, L.~A. {Green-Gavrielidis}, C.~Thornber, A.~T.
  Humphries, Modeling the growth of sugar kelp (saccharina latissima) in
  aquaculture systems using dynamic energy budget theory, Ecological Modelling
  430 (2020) 109151.

\bibitem{FGCG:14}
R.~Filgueira, T.~Guyondet, L.-A. Comeau, J.~Grant, A fully-spatial
  {ecosystem-{DEB}} model of oyster ({C}rassostrea virginica) carrying capacity
  in the {R}ichibucto {E}stuary, {E}astern {C}anada, J. Mar. Syst. 136 (2014)
  42--54.

\bibitem{ChB:98}
C.~Cho, D.~Bureau, Development of bioenergetic models and the {Fish-PrFEQ}
  software to estimate production, feeding ration and waste output in
  aquaculture, Aquat. Living Resour. 11~(4) (1998) 199--210.

\bibitem{RSPFG:12}
J.~Ren, J.~{Stenton-Dozey}, D.~Plew, J.~Fang, M.~Gall, An ecosystem model for
  optimising production in integrated multitrophic aquaculture systems,
  Ecological Modelling 246~(1--2) (2012) 34--46.

\bibitem{Urs:67}
E.~Ursin, A mathematical model of some aspects of fish growth, respiration, and
  mortality, Journal of the Fisheries Research Board of Canada 24 (1967)
  2355--2453.

\bibitem{Sink2010}
T.~Sink, Influence of ph, salinity, calcium, and ammonia source on acute
  ammonia toxicity to golden shiners, notemigonus crysoleucas, Journal of the
  World Aquaculture Society 41 (2010) 411 -- 420.

\bibitem{Hiddink2008}
J.~Hiddink, B.~Mackenzie, A.~Rijnsdorp, N.~Dulvy, E.~Nielsen, D.~Bekkevold,
  M.~Heino, P.~Lorance, H.~Ojaveer, Importance of fish biodiversity for the
  management of fisheries and ecosystems, Fisheries Research 90 (2008) 6--8.
\newblock \href {https://doi.org/10.1016/j.fishres.2007.11.025}
  {\path{doi:10.1016/j.fishres.2007.11.025}}.

\bibitem{Jenkins2020}
G.~P. Jenkins, R.~A. Coleman, J.~S. Barrow, J.~R. Morrongiello, Environmental
  drivers of fish population dynamics in an estuarine ecosystem of
  south-eastern australia, Fisheries Management and Ecology 29~(5) (2022)
  693--707.

\bibitem{Francis2019}
F.~C. G.-A. Cruz, Development and modelling of an aeration control system for
  precision aquaculture, Ph.D. thesis, University of Toronto (Canada) (2019).

\bibitem{HCHL:22}
W.~Hu, L.~Chen, B.~Huang, H.~Lin, A computer vision-based intelligent fish
  feeding system using deep learning techniques for aquaculture, IEEE Sensors
  Journal 22~(7) (2022) 1--10.

\bibitem{KII:20}
H.~Kuroki, H.~Ikeoka, K.~Isawa, Development of simulator for efficient
  aquaculture of {S}illago japonica using reinforcement learning, in:
  International Conference on Image Processing and Robotics (ICIP), 2020, pp.
  1--7.

\bibitem{YCWTZ:21}
L.~Yan, X.~Chang, N.~Wang, R.~Tian, L.~Zhang, W.~Liu, Learning how to avoid
  obstacles: A numerical investigation for maneuvering of {self‐propelled}
  fish based on deep reinforcement learning, International Journal for
  Numerical Methods in Fluids 93~(10) (2021) 3073--3091.

\bibitem{YWYYZ:21}
J.~Yu, Z.~Wu, X.~Yang, Y.~Yang, P.~Zhang, Underwater target tracking control of
  an untethered robotic fish with a camera stabilizer, IEEE Transactions on
  Systems, Man, and Cybernetics: Systems 51~(10) (2021) 6523--6534.

\bibitem{VNK:18}
S.~Verma, G.~Novati, P.~Koumoutsakos, Efficient collective swimming by
  harnessing vortices through deep reinforcement learning, Proc. Natl. Acad.
  Sci. 115~(23) (2018) 5849--5854.

\bibitem{Dampin2012}
N.~Dampin, W.~Tarnchalanukit, K.~Chunkao, M.~Maleewong, Fish growth model for
  {N}ile {T}ilapia ({O}reochromis {N}iloticus) in wastewater oxidation pond,
  {T}hailand, Procedia Environmental Sciences 13 (2012) 513--524.

\bibitem{Kushner1992}
H.~J. Kushner, P.~G. Dupuis, S.~O. service), Numerical Methods for Stochastic
  Control Problems in Continuous Time, Vol.~24, Springer US, New York, NY,
  1992.

\end{thebibliography}

%\newpage

\end{document}